\documentclass[prb,aps, amsfonts,amssymb,amsmath,nofootinbib,twocolumn]{revtex4}
\usepackage{graphics}
\usepackage{hyperref}
\usepackage{epsfig}

\setcitestyle{square}

\usepackage{dcolumn,color}
\usepackage{bm}

\begin{document}

\title{Electrodynamics of correlated electron systems: \\   \large{$Based$ $on$ $lectures$ $first$ $given$ $at$ $the$ $2008$ $Boulder$ $School$ $for$ $Condensed$ $Matter$ $and$ $Materials$ $Physics;$  $revised$ $in$ 2018 $and$ 2025.} }

\author{N.\ P.\ Armitage}
\affiliation{Department of Physics and Astronomy, The Johns
Hopkins University, Baltimore, MD 21218, USA}

\date{January 22, 2025}

\maketitle

\tableofcontents

\bigskip

\textit{Note added to 2025 update:   In preparation for some summer lectures in 2025 and a class on experimental methods in condensed matter physics, I have updated these notes albeit less extensively than I did in 2018.   I have added further clarifying discussion on the use of symmetries in analyzing response functions, some discussion on the Onsager reciprocal relations (related to symmetry actually), the fluctuation-dissipation theorem, and some further physical motivation regarding Kramers-Kroning relations.   
}

\textit{Note added to 2018 update:   These notes are based on lectures I first gave at the Boulder School for Condensed Matter Physics in 2008.   They have been further updated to be distributed at the QS3 summer school at Cornell University and Princeton Summer School on Condensed Matter Physics in the summer of 2018.   I originally wrote these detailed notes 1) for students at the Boulder school and 2) for my own students and postdocs as I found myself explaining some of the same material many times and I thought it would be useful to have some of my thoughts on these subjects written down.  After the school's conclusion, I didn't expect them to have much of a life beyond my own research group although I did post a slightly cleaned up version to the arXiv in 2009.  I was then surprised and (very!) pleased to find that they've had such a big impact.  Many people (from beginning graduate students to senior faculty) have told me over the years how useful they have found them as an introduction to the electrodynamic response of materials with quantum cooperative phenomena.  This was very gratifying.}

\textit{In preparation for 2018, it is quite interesting to look back at what I thought was relevant and important for ``correlated electron systems" in 2008 and what was is absent.    Some of this reflects my own prejudices and some shows how the field has moved on.   In the original version, although  I mention the quantum Hall effect, it was only in the context of the plateau transition being a model system for quantum phase transitions (QPT) and the consequences of dynamical scaling near a QPT.    There is no reference to topological insulators or topology in the original version of these notes.   Indeed the landmark paper of Hsieh et al. \cite{Hsieh08a} had just been published and its significance was lost on me at the time.    Needless to say this has become a major topic of inquiry.   And although most experimental aspects of the work on topological materials can be understood in terms of free fermions, intellectually the field has moved in tandem with the field of strong correlations and so I include some discussion on these systems in the updated notes.  Also, note that in the 2008 version I left out all mention of nonlinear optics and nonequilbrium effects.   There was little use of the former at the time (with the exception of excellent work on magnetoelectrics and multiferroics \cite{Fiebig05a}) and I felt that -- despite some nice work  \cite{Averitt02a} -- the latter was still too immature to cover explicitly.   That has changed for both these topics.   Optics (both linear or nonlinear)  has now made important contributions in the detection of subtle broken symmetry states of matter in cuprates and iridates \cite{Xia08a,Lubashevsky14a,Zhao16a,Zhao17a}.   There is also use of non-linear spectroscopies for 2D THz, which I consider tremendously promising \cite{Lu17a}.   And there has been tremendous interest in non-equilbrium phases with -- most notably -- the possibility of driven nonequilbrium room temperature superconductivity in cuprates and K$_3$C$_{60}$ \cite{Hu14a,Mitrano16a}.   It is interesting to note that the interest in quantum criticality and scaling phenomena has faded somewhat although their importance is undiminished. }

\textit{What is clear (and has not changed one whit) that is that the field of optical effects of ``quantum correlated systems" continues to be tremendously active.   The compatibility of optical probes with many different sample environments and the increased ease of use of ultrafast laser systems means that there is a continuing vitality to the field.   it is clear that optical probes of solids will continue to make essential contributions to the field for years to come.}

\section{Preamble}

Physical and chemical systems can be characterized by their natural frequency and energy scales.  It is hardly an exaggeration that most of what we know about such systems, from the acoustics of a violin to the energy levels of atoms, comes from their response to perturbations at these natural frequencies.   For instance, chemists and biologists use  infrared spectroscopy absorptions around 1650 cm$^{-1}$  (which corresponds to frequencies of 49 THz, wavelengths of 6 $\mu$m, and energies of 0.2 eV) to identify the carbon-carbon double bond in organic compounds.  And it was the observation of light emission from atoms at discrete energies in the electron Volt (eV) range that led to the quantum theory.  

It is of course the same situation in `correlated' electron materials.  We can learn about the novel effects of strong electron-electron interactions and the properties of collective states of matter (superconductors, quantum magnets etc.) by characterizing their response to small amplitude perturbations at their natural frequencies.  In solids, these natural frequency scales span an impressively large frequency range from x-ray down to DC.  This incredibly broad range means that a blizzard of experimental techniques and analysis methods are required for the characterization of correlated systems with optical techniques.  

This short review and lecture notes attempt to lay out a brief summary of the formalism, techniques, and analysis used for `optical' spectroscopies of correlated electron systems.  They are idiosyncratic, occasionally opinionated, and - considering the breadth of the subject - incredibly brief.  Unfortunately, there is no single complete treatise that presents a complete background for these topics in the context of correlated electron materials.  However, there are a number of excellent resources that collectively give a solid background to this field.

I recommend:

 \begin{enumerate}
  \setlength{\itemsep}{0pt}
  \setlength{\parskip}{0pt}
  \setlength{\parsep}{0pt}
 \item F. Wooten, ``Optical Properties of Solids", (Academic Press, New York, 1972).
 
 \textit{The truly excellent and classic introduction to the subject of the electrodynamic response of solids   (although there is very little in it in the way of electronic correlations).  The book that multiple generations of spectroscopists learned from either directly or indirectly.  To say that it is full of only basic obvious things is akin to complaining that Macbeth is full of quotations.  Unfortunately it is out of print.... \cite{Wooten72a}}

\item M. Dressel and G. Gr\"uner, ``Electrodynamics of Solids: Optical Properties of Electrons in Matter", 
(Cambridge University Press, 2002).

\textit{Much newer and modern with many plots of relevant response functions.  Good discussion of many important techniques \cite{Dressel02a}.   Beware of typos.}

\item G. D. Mahan, ``Many-Particle Physics", (Plenum, 2nd ed., 1990).

  \textit{The go-to resource for perturbative treatments of correlations in solids \cite{Mahan90a}.  Excellent discussion of sum rules.}

\item Richard D. Mattuck,  ``A Guide to Feynman Diagrams in the Many-Body Problem", (Dover, 2nd ed., 1992). 

\textit{It has been called ``Feynman diagrams for dummies."  Well ... I like it anyways \cite{Mattuck92a}.}

\item M. Tinkham, Introduction to Superconductivity, 2nd Ed. , McGraw-Hill, NY, (1996).

 \textit{The classic exposition on superconductivity.   Excellent treatment of EM response of superconductors (and other BCS-like states) both from the phenomenological two-fluid model and the full Matthis-Bardeen formalism \cite{Tinkham}.}

 \item  David Tanner, ``Optical effects in solids"
  \textit{Excellent class lecture notes from one of the pioneers of of IR spectroscopy on correlated systems.  Available at \url{https://www.phys.ufl.edu/~tanner/notes.pdf}.}


 
 \item  D. van der Marel, ``Optical signatures of electron correlations in the cuprates", ``Strong interactions in low dimensions, Series: Physics and Chemistry of Materials with Low-Dimensional
Structures" Vol. 25,  237-276 (2004); Also available at \url{http://arxiv.org/abs/cond-mat/0301506}

 \textit{Excellent treatment of underlying theory, particularly that associated with various sum rules.  Many relevant examples \cite{vanderMarel03b}.}
 
 \item  Andrew Millis,  ``Optical conductivity and correlated electron physics'', ``Strong interactions in low dimensions, Series: Physics and Chemistry of Materials with Low-Dimensional
Structures" Vol. 25,  195-235 (2004);  Also available at \url{http://phys.columbia.edu/~millis/july20.pdf}
 
\textit{Excellent review covering many aspects of the theory of optical measurements for correlated electron physics.  Particularly good discussion on the use of sum rules.  Explicit calculations within Kubo formalism.  Many experimental examples with accompanying theoretical discussion.  Unique treatment on the applicability of the conventional descriptions of the coupling of EM radiation to solids \cite{Millis}}

\item D. Basov and T. Timusk, ``Electrodynamics of high-T$_c$ superconductors", Reviews of Modern Physics \textbf{77}, 721 (2005).  

 \textit{A thorough review of the use of optical probes in the cuprate superconductors \cite{Basov05a}.}

\item  L. Degiorgi,  ``The electrodynamic response of heavy-electron compounds", Rev. Mod. Phys. \textbf{71}, 687 (1999).

\item A. Millis and P.A. Lee, ``Large-orbital-expansion for the lattice Anderson model", Phys. Rev. B \textbf{35}, 3394 (1987).

\textit{Experimental and theoretical papers with excellent sections on the phenomenological expectations for the optical response of heavy fermion materials and optical self-energies \cite{Degiorgi99a,Millis87a}.}

\item  C.C. Homes,  ``Fourier Transform Infrared Spectroscopy", Lecture notes available here \href{http://infrared.phy.bnl.gov/pdf/homes/fir.pdf}{http://infrared.phy.bnl.gov/pdf/homes/fir.pdf}

\textit{A good introduction to various technical issues associated with Fourier Transform Infrared Reflectivity (FTIR);  the most commonly used measurement technique for optical spectra of correlated electron systems \cite{Homes}.}

\item  George B. Arfken and Hans J. Weber,  ``Mathematical Methods for Physicists"  (Academic Press,  2005).

\textit{The classic textbook of mathematical methods for physicists.  Now with Weber.  Good reference for Kramers-Kronig and Hilbert transforms \cite{Arfken05a}.}

\item John David Jackson, ``Classical Electrodynamics" (3rd ed., Wiley, 1998).

\textit{No motivation need be given \cite{Jackson98a}.}

\item  Dmitrii L Maslov and Andrey V Chubukov, ``Optical response of correlated electron systems", 2017 Rep. Prog. Phys. 80 026503.

\textit{Good treatment of many modern issues from a theoretical perspective.  Discussion of optical conductivity of non-Fermi liquids \cite{MaslovA}.}

 \end{enumerate}

Other references as cited below.

\section{Introduction}

\begin{figure*}[htb]
\begin{center}
\includegraphics[width=17cm,angle=0]{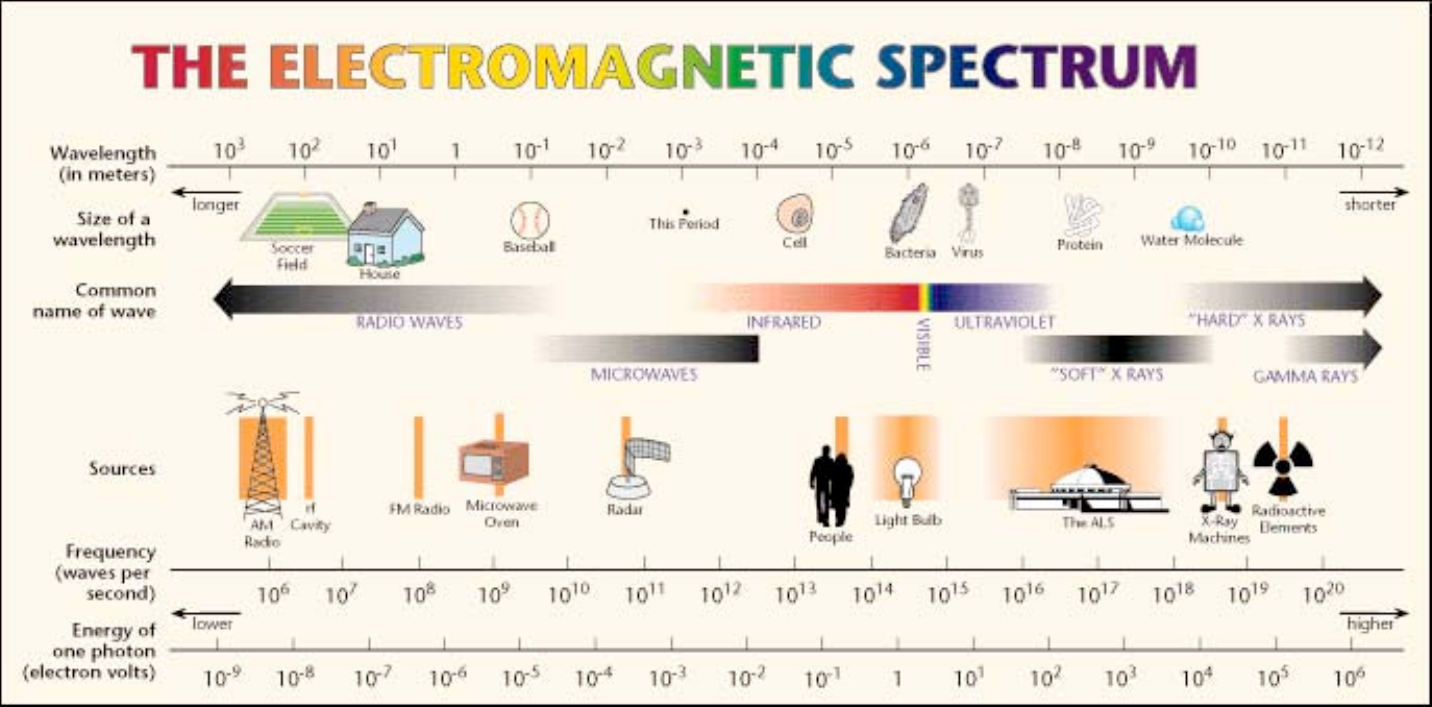}
\caption{(Color) The electromagnetic spectrum from radiofrequencies to gamma rays.  Diagram from the LBL Advanced Light Source web site  \texttt{http://www.lbl.gov/MicroWorlds/ALSTool/EMSpec/EMSpec2.html}. }\label{EMSpectrum}
\end{center}
\end{figure*}

As mentioned above, the energy and frequency scales relevant for correlated systems span the impressive range of the x-ray down to DC (Fig. \ref{EMSpectrum}).  For instance, atomic energy scales of 0.5 eV to the keV make solids possible through chemical bonding.  These energies manifest themselves explicitly in correlations by, for instance, setting the energy scale of the large on-site repulsions $U$ (1 - 10 eV) of electrons which can lead to Mott-Hubbard interactions and insulating states.  Typical overlap integrals between atomic wavefunctions in solids are at the low end of the eV energy scale and set the scale of Fermi energies in metals and hence the energy scale for electron delocalization and roughly that also of plasmon collective modes.   Various other collective modes such as phonons (lattice vibrations) and magnons (magnetic vibrations) are found at lower energy scales, typically at fractions of an eV.  At energies of order 50 meV are the superconducting gaps of optimally doped cuprates.  At even lower energies of the few meV scale the scattering rate of charges in clean metals is found.  This is also the energy scales of gaps in conventional superconductors.  Many local $f$-electron orbitals, which are relevant in Kondo materials are found at these energies.  At even lower scales (approximately 10's of $\mu$eV) can be found the width of the `Drude' peaks found in the AC conductivity and the superconducting gaps of clean heavy fermion systems.  Finally the 1 - 10 $\mu$eV energy scale corresponds to the lowest temperatures typically reached with conventional cryogenics in the solid state (10 - 100 mK).  The corresponding frequencies become important when studying crossover phenomena near quantum critical points with $\hbar \omega \approx k_B T$.    All such energy scales can be studied in a number of different contexts with various photon spectroscopies.

Although there are many different photon spectroscopies that can be discussed that span these scales, in these lectures I concentrate on `optical' spectroscopies, which I define as  techniques which involve transitions with net momentum transfer $q=0$ and whose absorption and polarization properties are governed by the optical dipole matrix element.  I do not discuss the fascinating and important work being done using other photon spectroscopies using light and charge in correlated systems with for instance, Raman spectroscopy \cite{Devereaux07a}, electron energy loss, Brillouin scattering, optical Kerr rotation\footnote{Kerr rotation is discussed below in the 2018 version},  photoemission, and fluorescence spectroscopies to give a tremendously incomplete list.   I am also not going to discuss the important work being done using THz light to look at magnetic excitations (through the magnetic dipole operator) in materials.   The emphasis here is on charge properties.

In the present case, the quantities of interest are typically the $complex$ frequency dependent conductivity $\sigma(\omega)$ or dielectric constant $\epsilon(\omega)$.  Alternatively data may be expressed in terms of one of a variety of straightforward complex parameterizations of these quantities like the index of refraction $n$ and absorption coefficient $k$, or the complex surface impedance $Z_s$.  These quantities are defined for the interaction of light with materials generically from zero frequency  to arbitrarily high frequencies.  In these lecture notes, however, I confine myself to the frequency range from microwaves, through the THz, to infrared, visible and ultraviolet.  These are the frequency ranges that correspond to various typical energy scales of solids.  Each of these regimes gives different kinds of information and requires different techniques.  For instance, microwaves are measured in cavities, striplines, or with Corbino techniques.   THz is a huge growth area with the advent of time-domain THz and the increased use of Backward Wave Oscillators (BWOs).  The infrared and visible range are measured by Fourier Transform Infrared Reflectivity (FTIR).  Visible and Ultraviolet can be measured by grating spectrometers and spectroscopic ellipsometry.  Of course there are large overlaps between all these regimes and techniques.

I caution that throughout these lecture notes, I frequently make use of the language of quasi-free electrons with well-defined masses and scattering times.  It is not clear that such a description should be automatically valid in correlated systems.  In fact, it is not even clear that such a description should be valid in `normal' materials.  Why should an ensemble of 10$^{24}$ electrons/cm$^3$ interacting with each other $via$ long range Coulomb interaction have excitations and states that resemble anything like those of free electrons?  Naively one would expect that excitations would be manifestly many-electron composite objects wholly unrelated to the individual particles which constitute the system.   It is nothing short of a miracle that, in fact, many materials are relatively well described by the assumption that interactions do not play a principle role in the explicit physics.  Low-energy experiments (e.g. DC resistivity, heat capacity, thermal conductivity etc.) on materials like sodium, gold, or silicon indicate that many aspects of such systems can be well described by free-electron physics.  The largest effect on electrons appears to be the static field of the ionic cores and the average static effect of the other electrons.  In other materials, like for instance the heavy fermion compounds, experiments seem to indicate that the charge carriers are free, but have masses many times larger than that of a bare electron.  But even here at the lowest energy scales there are no explicit signs that charger carriers are simultaneously interacting with 10$^{23}$ other charges and spins.  In general, it is surprising that considering the close proximity that many electrons have to each other that such interactions appear to be only quantitative, but not qualitative perturbations on the free-electron physics.
 
In a system of interacting fermions, the relatively weak effect that even potentially strong interactions have on the underlying physics can be understood by realizing the strong constraint that Fermi statistics and the existence of a Fermi surface provides on the scattering kinematics.  This phase space constriction gives, within the conventional treatment, a scattering rate that goes like $\sim \omega^2 + T^2$ at low frequency and temperature.  Due to this quadratic dependence, for some small frequency the particle's scattering rate will be less than its energy and one can say that the quasi-particle excitation is well-defined and scattering is a minor perturbation on the free-electron physics.  Such a perturbation renormalizes to zero in the $\omega \rightarrow 0$ limit.  The effect of interactions can be subsumed into giving quasiparticle excitations a finite lifetime and a renormalized energy parameterized by an effective mass m$^*$.
 
 Landau hypothesized that if one envisions slowly turning on an interaction potential in a gas of non-interacting electrons, some character of the original system would remain \cite{Landau56a}.  Specifically, he conjectured that there would be a one-to-one correspondence between states and excitations of the non-interacting system and those of the interacting system.  In which case there can be said to be an $adiabatic$ $continuity$ between the two and one can try to understand and model the interacting system by modeling the non-interacting one.  A system with such a mapping is termed a $Fermi$-$liquid$.  Its quasi-free electron-like excitations are termed $quasi-particles$.

The success of Fermi liquid theory comes from, as mentioned above, the constraint on scattering kinematics for low-energy excitations.  This gives the result that quasi-particle excitations are only defined at arbitrarily small energy scales.  The condition for well defined quasiparticles to exists,  $\beta \omega^2 \ll \omega$, means that as one increases the parameter $\beta$, which characterizes the strength of electron-electron excitations, the maximum energy of well-defined excitations decreases, but well-defined quasiparticles will still exist at low enough energies.  Within this view, stronger interactions mean only that Fermi liquid behavior will occur at lower-energy scales i.e. at lower temperatures and excitation energies.

Systems for which the Fermi-liquid paradigm is valid can be described at low energies and temperatures in terms of quasi-free electrons.  Although the vast majority of normal metallic systems do seem to obey the Fermi liquid phenomenology, it is unclear whether such a description is valid for `correlated' electron systems.  A violation could occur for instance in Hubbard models when the intersite hopping parameter $t$, which parameterizes the band width, is much smaller than the onsite Coulomb interaction $U$ resulting in an insulating state.  So far however the fractional quantum Hall effect is the only case where Landau's conjecture regarding a one-to-one correspondence of states has been $experimentally$ falsified \cite{Tsui82a,Laughlin83a}.  On the theoretical side, exact solutions to 1D interacting models also show definitively that the fundamental excitations in 1D are not electron-like at all, but are fractions of electrons:  spinons and holons that carry spin and charge separately \cite{Lieb68a}.  This would be another case where the electronic quasiparticle concept is not valid, but thus far in real systems the residual higher dimensionality has been found to stabilize the Fermi liquid (See for instance Ref. \onlinecite{Schwartz98a})\footnote{Carbon nanotubes, which are in principle perfect 1D systems, have been claimed to have tunneling conductances that obey a voltage - temperature scaling that is consistent with Luttinger (1D) liquid physics \cite{Bockrath99a}.  However there are serious questions whether this study showed scaling to the requisite precision in order to confirm a non-Fermi liquid nature.  It may also be possible to account for the data in terms of temperature dependent Coulomb blockade \cite{Egger01a}. }

It is true that many systems that we call strongly correlated exhibit a phenomenology inconsistent with the Fermi-liquid paradigm.  It is unclear whether this is because interactions have driven such materials truly into a non-Fermi liquid state or because the Fermi liquid phenomena is obscured at the experimentally accessible temperatures and frequencies.  For instance, it is important to keep in mind that the much heralded non-Fermi liquid behavior of high T$_c$ cuprates is actually exhibited at relatively high temperatures ($\sim$100 K) above the occurrence of superconductivity.  It may be that the inopportune occurrence of superconductivity obstructs the view of what would be the low-energy quasiparticle behavior.

At this point it is far from clear to what extent the Fermi liquid paradigm is valid in many of the materials being given as examples below.  In this regard, one must keep in mind however, that despite the murky situation much of the language tossed around in this field presupposes the validity of it and the existence of well-defined electronic excitations.  Indeed, much of the used terminology shows this bias.   The terms `density of states', `effective mass', `scattering rate', `Pauli susceptibility', `band structure', `electron-phonon' coupling all require the context of the Fermi liquid to even make sense.  Clearly such language is inappropriate if such excitations and states do not exist!  Despite this, in the literature one sees many papers discussing, for instance, optical or photoemission electronic self-energies, while at the same time the authors discuss the non-Fermi liquid aspect of these materials.  It is not clear whether it is appropriate to discuss electronic self-energies in materials where the elementary excitations are not electron-like.  It is the case that while the formalism for generating electronic self-energies from optical or photoemission data as discussed below may be followed straightforwardly, the physical signficance of such self-energies is not necessarily clear.  For instance, although the parametrizations of optical spectra in terms of the frequency dependent mass and scattering rate from the extended Drude model (see below) can always be valid as a $parameterization$, it is reasonable that one can only assign physical significance to these quantities if the quasi-particle concept is valid in the energy range of interest.

Some aspects of the below formalism are model independent and some rests on the concept of well-defined electronic Fermi-liquid excitations.  Although I will try to make the various distinctions clear, I use the language of quasi-free electrons almost entirely throughout the below, because at the very least it provides a rough intuition of the kinds of effects one expects in insulator, metals, and superconductors.   It also provides a self-contained formalism for the analysis of optical spectra.  A generalization of these ideas to strongly correlated systems does not currently exist.  In the spirit of learning to walk before one learns to run, I use the language of quasi-free electrons in these lecture notes throughout, but I caution on the naive application of these ideas, which are only formally true for non-interacting systems to strongly interacting ones!  The generalization to non-Fermi liquid, correlated, strongly interacting etc. etc.  systems is left as an exercise for the reader!!!

\section{Formalism}

\subsection{Response functions} \label{Responsefunctions}

The principle quantity that we will use to characterize materials are response functions, which are the constants of proportionality between an applied field like an electric field and a response like a polarization or a current.   Generally we can expand the response in powers of the driving field e.g. 

\begin{equation}
\bf{P}_i = \chi_{ij}^{(1)} E_j +  \chi_{ijk}^{(2)}E_j E_k + \chi_{ijkl}^{(3)}E_j E_k E_l,
\label{PolarizationExpansion}
\end{equation}
 or
\begin{equation}
\bf{J}_i = \sigma_{ij}^{(1)} E_j +  \sigma_{ijk}^{(2)}E_j E_k + \sigma_{ijkl}^{(3)}E_j E_k E_l.
\label{OhmsLawExpansion}
\end{equation}

Of course there are other relations which include ones with magnetic fields and crossed magnetic and electric fields.  The lowest order term is the linear response and will be the principle subject of these notes.   (I briefly discuss the use of nonlinear responses in Sec. \ref{Symmetries}.).  Linear response functions are very familiar to us with Hooke's law $\mathbf{F} = - k\mathbf{x}  $ for a spring being perhaps our earliest example in our training, but Ohm's Law $V=IR$ should be equally familiar.    Eq. \ref{OhmsLawExpansion} is a generalization of Ohm's law to higher order nonlinearities.

It is remarkable that quite general considerations such as material symmetries, Onsager reciprocity, and causality constrain the form of the response functions quite stringently.  These constraints are widely used, particularly for the lowest order linear response tensorial forms of $\chi$.

\subsubsection{Constraints from symmetries} 

Symmetries constrain the forms of the susceptibilities.  Formally this is codified as Neumann's principle that states that the symmetry transformations of any intrinsic physical property of a crystal (such as repsonse functions or a scattering matrix) must include $at$ $least$ the symmetry transformations of the point group of that crystal \cite{Nye85a}.  The symmetries of a material typically manifest in the form of an algebra relating matrix elements of the response functions or overall constraints (transposition, unitarity, hermiticity, normality, etc.) on the form of the matrix.  I discuss the role of symmetries further in Sec.\ref{Symmetries}.

\subsubsection{Fluctuation-dissipation theorem}

The fluctuation-dissipation theorem is a fundamental concept in statistical physics that establishes a direct connection between the spontaneous thermodynamic fluctuations in a physical variable and the system's response functions, as quantified by measures like susceptibility or conductivity. These measures apply broadly, beyond their electromagnetic context, to variables such as voltage and temperature differences. The theorem is applicable to both classical and quantum mechanical systems.

First proven by Callen and Welton~\cite{Callen}, the fluctuation-dissipation theorem was later extended by Kubo~\cite{Kubo54,Kubo57}. Precursors to this general theorem include Einstein's explanation of Brownian motion~\cite{Einstein} during his annus mirabilis and  Nyquist's 1928 analysis of Johnson noise in electrical resistors~\cite{Johnson,Nyquist}.  For instance, Einstein observed that the random forces responsible for the erratic motion of a particle in a fluid also give rise to drag when the particle is pulled through the fluid.   He derived the  Einstein–Smoluchowski relation which is 

\begin{equation}
D = \mu k T,
\label{diffusion}
\end{equation}
where the left side of the equation is the diffusion constant $D$ that quantifies how far a particle will drift spontaneously under random thermal motion (e.g.fluctuations) in a time via for instance Fick's law.

\begin{equation}
\frac{d \phi}{ d t} = D \frac{d^2 \phi}{d x^2}.
\end{equation}
On the right side of the Eq. \ref{diffusion} is particle mobility $\mu$, which is the ratio of the particle's terminal drift velocity to an applied force.  For an electron it is defined as

\begin{equation}
\mathbf{v} = \mu \mathbf{E}.
\end{equation}

Above $k$ is the Boltzmann constant, and $T$ is the absolute temperature.

The thermal fluctuations causing the particle’s random motion are fundamentally linked to the dissipative frictional force encountered when an external force perturbs the system.  The fluctuation–dissipation theorem states that any process dissipating energy by converting it into heat (e.g., friction) has a corresponding reverse process driven by thermal fluctuations.   It makes intuitive sense.   A system that can have spontaneous displacements in some fashion is also very susceptible to a field that will drive the same displacement  and ultimately the equilibrium thermal fluctuations are indistinguishable from a non equilibrium fluctuation produced by an infinitesimal external perturbation

Consider some system with a dynamical variable $x(t)$ that is at finite temperature.  $x(t)$  will fluctuate around its mean value with fluctuations characterized by a power spectrum $S(\omega) = \langle x(\omega) x^*(\omega) \rangle$.   The power spectrum is the Fourier transform of the time correlation function $ S(t-t') = \langle x(t) x(t') \rangle $.  The fluctuation–dissipation theorem relates the power spectrum at both positive and negative frequencies of the spontaneous fluctuations of a system to the imaginary part (dissipative) of the susceptibility.  

\begin{equation}
S(\omega) = - \frac{2kT}{\omega} \mathrm{Im} \chi(\omega).
\label{FluctDiss}
\end{equation}
The right side is the imaginary part of the susceptibility which is related to the rate that energy is dissipated with a driving field.  This is the classical form of the theorem; quantum fluctuations are taken into account by replacing $\frac{2kT}{\omega} $ by $\hbar \; \mathrm{ coth} (\hbar \omega/kT)$.  This expression is equivalent to the classical one in the  $T \rightarrow 0 $ limit.  It is interesting to note that scattering experiments (Raman, inelastic neutron scattering, electron energy loss spectroscopy), measure (different) correlation functions.   Experiments like optics measure susceptibility.   It is an analog of Eq. \ref{FluctDiss} that allows one to connect THz measurements of the magnetic susceptibility to neutron scattering as the latter can be rendered into an effective measure of the magnetic susceptibility.

\subsubsection{Onsager reciprocal relations}

The Onsager reciprocal relations express equalities between the coefficients in response functions that relate currents and properly defined forces in thermodynamic systems-out-of equilibrium, but where a notion of local equilibrium exists.  It was derived by Onsager~\cite{Onsager,Casimir,Miller} to relate different transport processes like heat and electrical conduction (he was originally thinking about transport in electrolytes), where there were different kinds of simultaneously existing currents and forces.  They also provide a theoretical framework to analyze and predict the behavior of systems far from equilibrium and are often described as the ``fourth law of thermodynamics".   Although often used to relate the transport coefficient of charge in a temperature gradient or the transport coefficient of energy in an electric field, they are also determine aspects of fundamental symmetries of the response functions.

An example of the reciprocal relations for thermoelectric response are as follows.

\begin{align}
\frac{\partial S_c}{\partial t} & =  \mathbf{J}_u \cdot \nabla \frac{1}{T} +  \mathbf{J}_n \cdot \nabla \frac{- \mu}{T}  \\ \nonumber
& = \sum_\alpha  \mathbf{J}_\alpha \cdot \nabla f_\alpha \\ \nonumber
& = \sum_\alpha \sum_\beta L_{\alpha \beta} \nabla f_\alpha \cdot \nabla f_\beta
\label{reciprocal}
\end{align}
Here $S_c$ is the entropy from irreversible processes, $\mathbf{J}_u $ is the heat current, and $\mathbf{J}_n$ is the particle current.  In going from the 2nd to the 3rd line we have used  a generic form the response functions

\begin{align}
\mathbf{J}_\alpha = \sum_\beta L_{\alpha \beta} \nabla f_\beta,
\end{align}
where $\nabla f_\beta$ are entropic forces, which in this case are $\nabla f_u = \nabla \frac{1}{T}$ and $\nabla f_\rho = \nabla \frac{-\mu}{T}$ that are the forces conjugate to the ``displacements" $u$ and $n$ and $L_{\alpha \beta}$ is the Onsager matrix of transport coefficients.  
Since entropy production given in Eq. \ref{reciprocal} must always be non-negative the matrix of transport coefficients must be positive semi-definite.  Onsager's further contribution was to demonstrate if time-reversal symmetry is not broken and forces are written in the proper form then the matrix is symmetric e.g. $L_{\alpha \beta}= L_{\beta \alpha  }$.

The aforementioned fluctuation dissipation theorem means that these coefficients are related to the correlation function $  S_{\alpha \beta   }(s) =  \langle \dot{x}_\alpha(t) \dot{x}_\beta(t+s) \rangle  $.  The correlation function is invariant to translations of the absolute value of time, which means that it is unchanged by the substitution $t \rightarrow t- s $ so that

\begin{align}
 \langle \dot{x}_\alpha(t) \dot{x}_\beta(t+s) \rangle  =  \langle \dot{x}_\alpha(t-& s) \dot{x}_\beta(t)\rangle  =  \langle \dot{x}_\beta(t) \dot{x}_\alpha(t - s)\rangle \nonumber \\ 
 S_{\alpha \beta   }(s) = S_{\beta \alpha  }(-s).
 \label{Translation}
\end{align}

Furthermore imagine reversing the direction of time altogether e.g. one reverses the particle velocities AND the direction of any applied magnetic field.  As the microscopic equations of motion are invariant under global reversal of the direction of time the correlation function will obey this relation 

\begin{align}
 \langle \dot{x}_\alpha(0) \dot{x}_\beta(s) \rangle  = & \langle \dot{x}_\alpha(0) \dot{x}_\beta( - s)\rangle^\dagger \nonumber \\ 
 S_{\alpha \beta   }(s) = &S^\dagger_{ \alpha \beta  }(-s)
  \label{Reversal}
\end{align}
where the dagger denotes time-reversal.  Combining Eqs. \ref{Translation} and  \ref{Reversal} and changing the sense of the forward direction of time one gets 

\begin{align}
 S_{\alpha \beta   }(s) = &S^\dagger_{ \beta \alpha  }(s)
  \label{Reversal}
\end{align}

In the case where $\dot{x}_\alpha$ and $\dot{x}_\beta$ are velocities or currents (they do not need to be; in the above example they were energy and density) and going to back to the response functions and noting the reversing the direction of the magnetic field is the same as reversing the direction of time one has 

\begin{align}
 L_{\alpha \beta   }(\mathbf{B}) = &L_{ \beta \alpha  }(-\mathbf{B}).
\end{align}.

This constraint on the response coefficients  is important in considering charge currents in magnetic fields.   For instance, the Hall effect obeys this relation $\sigma_{xy}(\mathbf{B}) = \sigma_{yx}(\mathbf{-B})$.

\subsection{Optical Constants of Solids} \label{OpticalConstants}

Any discussion of the interaction of light with matter starts with Maxwell's equations

 \begin{eqnarray}
\nabla \cdot \mathbf{E}(r,t) = 4 \pi \rho(r,t), \\
\nabla \times \mathbf{E}(r,t) = - \frac{1}{c} \frac{\partial}{\partial t} \mathbf{B}(r,t), \\
\nabla \cdot \mathbf{B}(r,t) = 0, \\
\nabla \times \mathbf{B}(r,t) =   \frac{1}{c} \frac{\partial}{\partial t} \mathbf{E}(r,t) +  \frac{4 \pi}{c} \mathbf{J}(r,t),
\label{Maxequations}
\end{eqnarray}

\noindent  where $E$ and $B$ represent the electric and magnetic fields averaged over some suitable microscopic length (typically the incident light wavelength).  As usual we introduce auxiliary fields as 

 \begin{eqnarray} 
\mathbf{D} = \mathbf{E} + 4 \pi \mathbf{P}, \\
\mathbf{H} = \mathbf{B} - 4 \pi \mathbf{M},
\label{DandH}
\end{eqnarray}

\noindent  where $D$ and $H$ have their conventional definitions and $M$ and $P$ are magnetization and polarization.  Using the continuity equation for the electric current $\nabla \cdot J = - \frac{\partial \rho}{\partial t} $ and accounting for all sources for free conduction, polarization, and magnetization currents  $J_{tot}=J_{cond} + \frac{\partial P}{\partial t} + c \nabla \times M $ as well as the external and induced charges $\rho_{total} = \rho_{ext} + \rho_{ind}$ we get the additional equations (please refer to Jackson \cite{Jackson98a} for a more extended discussion).

 \begin{eqnarray}
\nabla \cdot \mathbf{D} = 4 \pi \rho_{ext},\\
\nabla \times \mathbf{H} =   \frac{1}{c} \frac{\partial  \mathbf{D} }{\partial t} +  \frac{4 \pi}{c} \mathbf{J}_{ext}+  \frac{4 \pi}{c} \mathbf{J}_{cond}.
\label{Maxwell}
\end{eqnarray}

Here we assume that we are in the linear regime and the response of polarization, or magnetization of current is linear in the applied field.  We therefore write

 \begin{eqnarray}
\mathbf{P} = \chi_e \mathbf{E},  \nonumber \\
\mathbf{M} = \chi_m \mathbf{H}, \nonumber \\
\mathbf{J} = \sigma \mathbf{E}.
\label{linear}
\end{eqnarray}

Typically we express the electric and magnetic susceptibilities in terms of dielectric functions $\epsilon = 1 +4 \pi \chi_e$ and magnetic permittivity  $\mu= 1 +4 \pi \chi_m$.  Except for explicitly magnetic materials, $\chi_m =0$.  The dielectric function (or its Fourier transform) is a response function that connects the field $E$ at some time $t$ and position $r$ with the field $D$ at some later time and position.  Generally we first define it in the time and position domain $via$ the relation

 \begin{eqnarray}
\mathbf{D}(r,t) = \int_{- \infty}^t \int\hat{ \epsilon}(r,r',t,t')\mathbf{E}(r',t') d^3r' dt'.
\label{Dresponse}
\end{eqnarray}

One could also describe the system's response in terms of current and conductivity $\hat{\sigma}(\omega)$

 \begin{eqnarray}
\mathbf{J}(r,t) = \int_{- \infty}^t \int \hat{\sigma}(r,r',t,t')\mathbf{E}(r',t') d^3r' dt'.
\label{Jresponse}
\end{eqnarray}

In this context, the real space and time dependent $\hat{\epsilon}$ and $\hat{\sigma}$ are usually referred to as $memory$ functions for obvious reasons.  Note that the principle of causality is implied here;  effects cannot precede their causes.  For analysis of optical spectra we are typically more interested in their Fourier transforms $\sigma(q,\omega)$ and $\epsilon(q,\omega)$.  This is because these are the quantities which are actually measured, but also because the mathematical complications of convolution in Eqs. \ref{Dresponse} and \ref{Jresponse} is avoided;  convolution in the timed-domain becomes multiplication in the frequency domain.  The conductivity and dielectric function are related by\footnote{One frequently sees the complex response function written using the real part of the conductivity and the real part of the dielectric function e.g. $\epsilon = \epsilon_1 + i 4 \pi \sigma_1/ \omega$.}

 \begin{eqnarray}
\sigma = \frac{i \omega}{4 \pi} (1 - \epsilon).
\end{eqnarray}

Given the very general form of Eqs. \ref{Dresponse} and \ref{Jresponse}, simple physical considerations allow a number of general statements to be made.   First, due to the vast mismatch between the velocity of light and the typical velocity of electrons in solids, we are typically concerned with the $\mathbf{q}=0$ limit of their Fourier transforms\footnote{Alternatively one may say it is the $q=0$ limit is relevant because of the large difference between the momentum carried by an optical photon and typical lattice momenta}.  This means that except in a few circumstances, where one must take into account non-local electrodynamics,  (very pure metals  clean superconductors or chiral systems for example), one can assume that down to the scale of some microscopic length there is a local relationship between the quantities given in Eqs.  \ref{linear}.  This means that while these quantities may have spatial dependence (for instance the current $\mathbf{J}$ is confined to surfaces in metals), the proportionality expressed in Eqs. \ref{linear} holds.  One can use the above expressions to rewrite Maxwell's equations explicitly in terms of $\epsilon$, $\mu$ and $\sigma$.  (Please see Jackson \cite{Jackson98a} for further details.)

As we noted, it is frequently more useful to transform the response functions from the time domain to the frequency domain.   Starting with the Fourier transform of the current

 \begin{eqnarray}
\mathbf{J}(\omega) = \int  \mathbf{J}(t) e^{-i \omega t} dt
\end{eqnarray}
and substituting into it a simpler form of Eq. \ref{Jresponse}

 \begin{eqnarray}
\mathbf{J}(t) = \int_{- \infty}^t \hat{\sigma}(t-t')\mathbf{E}(t') dt',
\end{eqnarray}
one has
  \begin{eqnarray}
\mathbf{J}(\omega) = \int  dt  e^{-i \omega t}  \Big[  \int   \hat{\sigma}(t-t')\mathbf{E}(t') dt'      \Big]  ,   \\
= \int  dt' \mathbf{E}(t')     \Big[  \int  \hat{\sigma}(t-t') e^{-i \omega t}  dt       \Big] ,  \\
= \int  dt' \mathbf{E}(t')  e^{-i \omega t'}   \Big[     \int   \hat{\sigma}(t-t') e^{-i \omega ( t  - t')}  dt        \Big] .
\end{eqnarray}
In the expression above the 1st and 2nd integrals are equal to the Fourier transform of the electric field and the conductivity separately.   Therefore one can write down the frequency domain relations that we typically  directly measure

 \begin{eqnarray}
\mathbf{J}(\omega) = \sigma(\omega) \mathbf{E}(\omega).
\end{eqnarray}


%

%

The principle of causality - that effects can not proceed their causes - demands strict temporal considerations regarding the integrals in Eqs. \ref{Dresponse} and \ref{Jresponse}.  This leads to the powerful Kramers-Kronig relations relating the real and imaginary parts of the Fourier transform of such response functions.  The typical proof of the Kramers-Kronig relations begins with an application of Cauchy's residue theorem for complex integration. 
The causality condition implies that the Fourier transform of the response functions are analytic in the upper half complex plane.   See Wooten \cite{Wooten72a}, Jackson \cite{Jackson98a} or Arfken \cite{Arfken05a} for details on the usual integration and derivation of the Kramers-Kronig relations based on integration in the complex plane.   I give here an alternative derivation based on only this essential principle of causality and the convolution theorem that may be more intuitive.  First consider the Fourier transform of $\sigma(\omega)$

 \begin{eqnarray}
\sigma(\omega) = \int_{- \infty}^  \infty \sigma(t) e^{ - i \omega t} dt.
\end{eqnarray}
Causality mandates that $\sigma(t<0) = 0$ and hence one can write 

 \begin{eqnarray}
\sigma(\omega) =  \int_{0 }^  \infty \sigma(t) e^{-i \omega t} dt.
\end{eqnarray}

This is equivalent to 

 \begin{eqnarray}
\sigma(\omega) = \int_{- \infty}^  \infty \sigma(t) h(t) e^{-i \omega t} dt,
\end{eqnarray}
where $h(t)$ is the Heaviside step function.   Via the convolution theorem, the multiplication in the time domain is equivalent to convolution in the frequency domain.  Therefore we can replace the Fourier transform of the product $ \sigma(t) h(t)$ in the above with the convolution of $\sigma(\omega)$ and the Fourier transform of the Heaviside function e.g.

 \begin{eqnarray}
\sigma(\omega) =  \frac{1}{2 \pi }   \int_{- \infty}^  \infty  d \omega'     \sigma(\omega')      \Big[      \pi \delta(\omega - \omega')     - \frac{i}{ \omega - \omega'}          \Big].
\end{eqnarray}

This expression demonstrates an invariance of the response function after convolution with the Fourier transform of the Heaviside step function.  Evaluation of the $\delta$ function and solving for $\sigma(\omega)$ one gets

 \begin{eqnarray}
\sigma(\omega) =  \frac{1}{ \pi }   \int_{- \infty}^  \infty  d \omega'  \frac{  i \sigma(\omega')     }{\omega- \omega'}.
\end{eqnarray}



%
%


%
%


%
%


%
%


%
%

From this the Kramers-Kronig relations can be inferred.  The fact that the time-domain signal is real means that the real part of  $\sigma(\omega)$ is an even function of $\omega$ and the imaginary part of  $\sigma(\omega)$ is an odd function of $\omega$.   This allows one to write a simpler form of the Kramers-Kronig relations using positive frequencies only as

 \begin{eqnarray}
\sigma_1(\omega) = \frac{2}{\pi}  \int_{0}^{\infty} d \omega '  \frac{\omega' \sigma_2(\omega')}{\omega'^2 - \omega^2}, \\
\sigma_2 (\omega)= - \frac{2 \omega}{\pi} \int_{0}^{\infty} d \omega '  \frac{\sigma_1(\omega')}{\omega'^2 - \omega^2}.
\end{eqnarray}


Although the above derivation is perhaps more intuitive than the conventional one based on analyticity in the upper half complex plane, it still perhaps fails to convey the {\it why} of why a structure in the dissipative part of the response function must have consequences in the reactive part.  This can be illustrated as follows.  Consider a thought experiment regarding a ``notch filter", which in our idealized case is an optical filter that is opaque for one frequency and transparent for a wide band of frequencies around it.  In our idealized case, the transmission of this filter is 100$\%$ except at a specific frequency.

Consider an incident pulse, which is given in its Fourier decomposition by a broad band of frequencies resulting in a pulse that is quite limited in time e.g. 

\begin{equation}
\mathbf{E}^i(t) = \frac{1}{2 \pi}  \int_{-\infty}^{\infty}  d \omega \mathbf{E}^i(\omega) e^{i \omega t},
\end{equation}
where $\mathbf{E}^i(\omega)$ is the incident pulse spectral decomposition.  Now, imagine that I let the pulse be incident on the filter. The filter removes a single frequency from $\mathbf{E}^i(\omega)$.   Suppose the amplitude of the $\omega_0$ component in $\mathbf{E}^i(\omega)$ is $A$.  If it has been completely absorbed upon transmission, one may guess that the spectrum after after transmission is

\begin{equation}
\mathbf{E}^t(\omega) =   \mathbf{E}^i(\omega) - 2 \pi  A  \delta(\omega - \omega_0).
\label{InitialTry}
\end{equation}

The inverse Fourier transform of this is 

\begin{align}
\mathbf{E}^t(t) = & \frac{1}{2 \pi}  \int_{-\infty}^{\infty}  d \omega \mathbf{E}^t(\omega) e^{i \omega t},  \\ 
\mathbf{E}^t(t) =& \frac{1}{2 \pi}  \int_{-\infty}^{\infty}  d \omega  [ \mathbf{E}^i(\omega) - 2 \pi A  \delta(\omega - \omega_0)    ]   e^{i \omega t}, \\
\mathbf{E}^t(t) =&  \mathbf{E}^i(t) - A \mathrm{cos} \omega_0 t.
\end{align}

Initially this might look quite reasonable.   The filter has completely removed the $\omega_0$ component and  transmitted the rest.  However, upon inspection a problem is revealed that the signal persists for all times including large negative times even before the original pulse arrives at the filter.  This appearance of a signal at early times is a violation of causality and is obviously unphysical.  The resolution to this is that another inescapable effect of even a perfect filter is to introduce large phase shifts $\phi(\omega)$ to frequencies near $\omega_0$ in such a fashion so that they interfere destructively so as to cancel the signal at times before the pulse arrived, but do not cancel the signal at long times after the pulse. 

Eq. \ref{InitialTry} can be fixed by introducing a frequency dependent phase shift to frequencies near $\omega_0$.   This can be done via the model of a Lorentz oscillator discussed below.  The transmitted signal will have the general form 

\begin{equation}
\mathbf{E}^t(\omega) =   e^{i\phi (\omega) } [ \mathbf{E}^i(\omega) - 2 \pi A  \delta(\omega - \omega_0) ].
\label{FixedTry}
\end{equation}

The filter does produce a signal at long times after the end of the original pulse.   This ``ringing" is a consequence of having a narrow band filter.

Another example of the necessity of the interrelation between real and imaginary parts is as follows.  Consider the conductivity of a perfect metal.  At higher frequencies the
electron velocity has a harder time following the driving field due to finite mass of the electrons.  Moreover the electron velocity is 90$^\circ$ out of phase with the driving field.  For our current purposes this can be expressed as a purely imaginary conductivity that is $\sigma = i \frac{n e^2}{m \omega} $.  This form can be easily derived from the Drude model in Sec. \ref{SectionDL} by letting the current relaxation rate go to zero.  This functional form works fine for finite frequency monochromatic waves that by definition exist with the same amplitude for all times.   The electric field of such a wave is $ \mathbf{E}(\omega) = 2 \pi \mathbf{E}_0 \delta(\omega - \omega_0) $.  So one can write

\begin{align}
\mathbf{J}(t) = & \frac{1}{2 \pi}  \int_{-\infty}^{\infty} d \omega \; \mathbf{J}(\omega) e^{i \omega t} \nonumber \\ = & \frac{1}{2 \pi}  \int_{-\infty}^{\infty} d \omega \; \sigma(\omega) \mathbf{E}(\omega) e^{i \omega t}  \nonumber \\  = & i \frac{Ne^2}{m \omega_0} \mathbf{E}_0  e^{ i \omega_0 t}  ,
\end{align}
where we understand that we should consider the physical response to be the real part of $\mathbf{J}$.

Now consider the opposite limit where the drive field is an impulse e.g. $\mathbf{E}(t) = \mathbf{E}_0 \delta(t)$.  An impulsive field as such has a frequency spectrum that is infinitely wide with amplitude $\mathbf{E}_0$.  The time dependence of a current coming from this drive field follows from the inverse Fourier transform of $\mathbf{J}(\omega) = \sigma(\omega) \mathbf{E}(\omega) $ e.g.

\begin{equation}
\mathbf{J}(t) = \frac{1}{2 \pi}  \int_{-\infty}^{\infty} d\omega \; \sigma(\omega) \mathbf{E}(\omega) e^{i \omega t}.
\end{equation}

We substitute in the above form for $\sigma_2 $ and get 

\begin{equation}
\mathbf{J}(t) = i\frac{1}{2 \pi} \frac{N e^2}{ m } E_0  \int_{-\infty}^{\infty} d \omega \frac{ e^{i \omega t} } {\omega} . 
\end{equation}

This integral depends on the sign of $t$.  One finds that 

\begin{eqnarray}
\mathbf{J}(t) = - \frac{N e^2}{2m} E_0  \; \mathrm{for} \; t<0, \\
\mathbf{J}(t) = + \frac{N e^2}{2m}  E_0 \; \mathrm{for} \; t>0.
\end{eqnarray}

We may get a sense of {\it d\'ej\`a vu} with another non-causal result, as the current flows in the direction of the electric field polarization for $t>0$, but flows in the negative direction for all early times before the pulse arises.   What has gone wrong?   Similar to before there is another component of the complex response that serves to cancel current contributions for $t<0$.   It turns out that there is actually a zero frequency contribution to the conductivity (this is what would be derived in a Drude model derivation where the scattering rate goes to zero).  The conductivity is actually $\sigma(\omega) = \frac{n e^2 \pi } {m} \delta(\omega) + i \frac{n e^2}{m \omega} $.  If we inverse Fourier transform the delta function part we find that 

\begin{eqnarray}
\mathbf{J}_\delta(t) =   \frac{1}{2 \pi  } E_0  \int_{-\infty}^{\infty} d \omega   \frac{N e^2 \pi } {m}  \delta(\omega), \\
\mathbf{J}_\delta(t) = + \frac{N e^2}{2m}  E_0 . 
\end{eqnarray}
This contribution exists for all times, but serves to exactly cancel the current from the imaginary part of the conductivity for $t<0$.   Putting it all together one gets

\begin{align}
\mathbf{J}(t) =  0  \; \mathrm{for} \; t<0, \\
\mathbf{J}(t) =  \frac{N e^2}{m}  E_0  \; \mathrm{for} \; t>0.
\end{align}
In the absence of scattering the impulse starts a current that continues forever.


Other forms of the  Kramers-Kronig relations can derived for any response function that is causal including the dielectric constant, index of refraction, loss function, magnetic susceptibility and (as mentioned below) reflectivity.  They are tremendously useful in the analysis and determination of optical spectra.  When one knows one component of a response function for all frequencies, the other component automatically follows. As will be discussed below, they are used extensively in Fourier Transform Infrared Reflectivity (FTIR) measurements to determine the complex reflectivity (amplitude and phase) when only measuring reflected power.

\begin{figure}[htb]
\begin{center}
\includegraphics[width=8.5cm,angle=0]{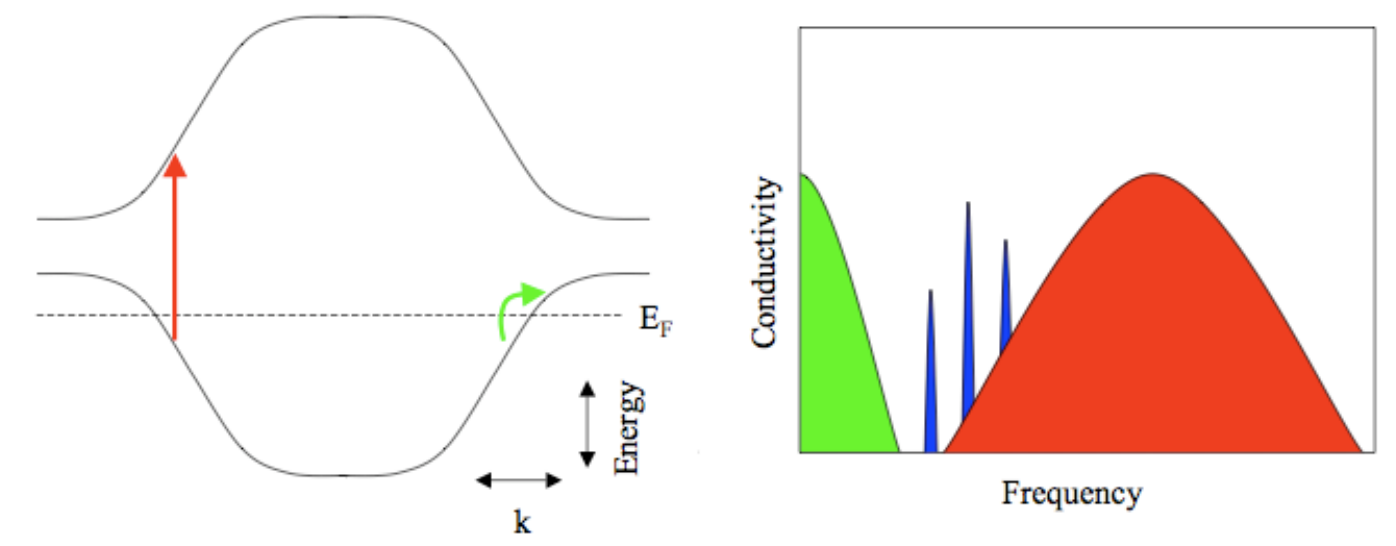}
\caption{(Color) Various different optical absorptions that satisfy the $q=0$ constraint of optical spectroscopy can appear in the optical conductivity of simple metals.  Near E$_F$ intraband absorptions [green].  Interband absorptions [red].  Lattice vibrations (phonons) [blue] .}\label{OpticalConduct}
\end{center}
\end{figure}

The above discussion was based on some generalized frequency dependent conductivity $\sigma(\omega)$ and its model independent properties.  In real materials, one has a zoo of different possible contributions to the electromagnetic response.  For instance in simple metals with periodic translational symmetry (i.e. a crystal), one can identify a number of absorptions that satisfy the $q=0$ constraint.  For the schematic band structure shown in Fig. \ref{OpticalConduct} (left), one expects a broad feature at finite energy (in red) in the optical conductivity (Fig. \ref{OpticalConduct} (right)), which comes from the sum over all possible direct interband band absorptions (red arrow) in which electrons are promoted from below E$_F$ across an energy gap to a higher lying band, with zero net momentum change.  These are so-called vertical transitions.  Near E$_F$ absorptions with low but finite $\omega$ (green arrow) are only possible if strict translational symmetry has been broken by, for instance, disorder.  Electrons moving in Bloch waves with mean free path $\ell$, can violate strict momentum conservation in optical absorption at momentum scales on the order of $2 \pi/ \ell$.   One can think of this heuristically as smearing out the band structure on this scale to allow vertical transitions within the same band. This gives a peak centered at zero frequency (in green).  In the limit of perfect translational symmetry this peak would be a delta function centered at $\omega = 0$.  One can also have excitations of harmonic waves of the lattice (phonons) if such phonons posses a net dipole moment in the unit cell.  They appear as distinct and frequently very sharp peaks in the optical conductivity (in blue).  Many different other excitations exist as well that can potentially be created $via$ optical absorption.

Having introduced the general idea of optical response functions, I now discuss the derivation of them, both classically and quantum mechanically.

\subsection{Classical Treatments:  Drude-Lorentz} \label{SectionDL}

Almost the simplest model of charge conduction we can conceive of is of a single charge $e$ of mass $m$, driven by an electric field $E$, and subject to a viscous damping force that relaxes momentum on a time scale $\tau$.   Consider the force equation describing this situation

 \begin{eqnarray}
mx'' = - eE - m x'/ \tau .
\label{DrudeForces}
\end{eqnarray}

If we assume harmonic motion then $x = x_0 e^{-i \omega t}$ and $E = E_0 e^{-i \omega t}$.  Substituting in for $x$ and $E$ and solving for $x'_0$ we get

 \begin{eqnarray}
x'_0 =  \frac{e\tau E_0}{m} \frac{1}{1 - i \omega \tau}.
\end{eqnarray}

If we then consider an ensemble of such charges with density N, and realize that the maximum current density is $J_0 = Ne x'_0$, we get the relation

 \begin{eqnarray}
J_0 =  \frac{Ne^2 \tau E_0}{m} \frac{1}{1 - i \omega \tau}.
\end{eqnarray}

Using the previously defined relation $J = \sigma E$ we find the relation for the frequency dependent `Drude' conductivity is

 \begin{eqnarray}
\sigma(\omega) = \frac{Ne^2 \tau }{m} \frac{1}{1 - i \omega \tau} = \frac{Ne^2 \tau}{m} \frac{1+ i \omega \tau}{1 + \omega^2 \tau^2}.
\label{DrudeCond}
\end{eqnarray}

Interestingly, this classical model, first conceived of by Paul Drude, is actually of great use even in the analysis of particles obeying quantum mechanical statistics.  As discussed below one finds the same functional form to leading order in that case as well.

\begin{figure}[htb]
\begin{center}
\includegraphics[width=6.5cm,angle=0]{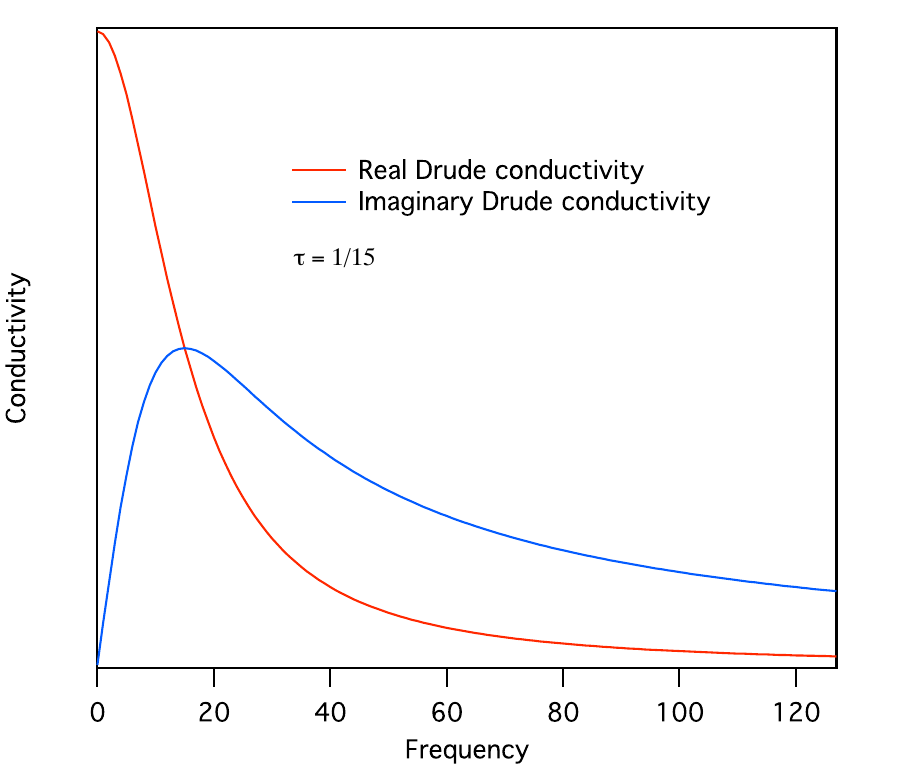} 
\includegraphics[width=6.5cm,angle=0]{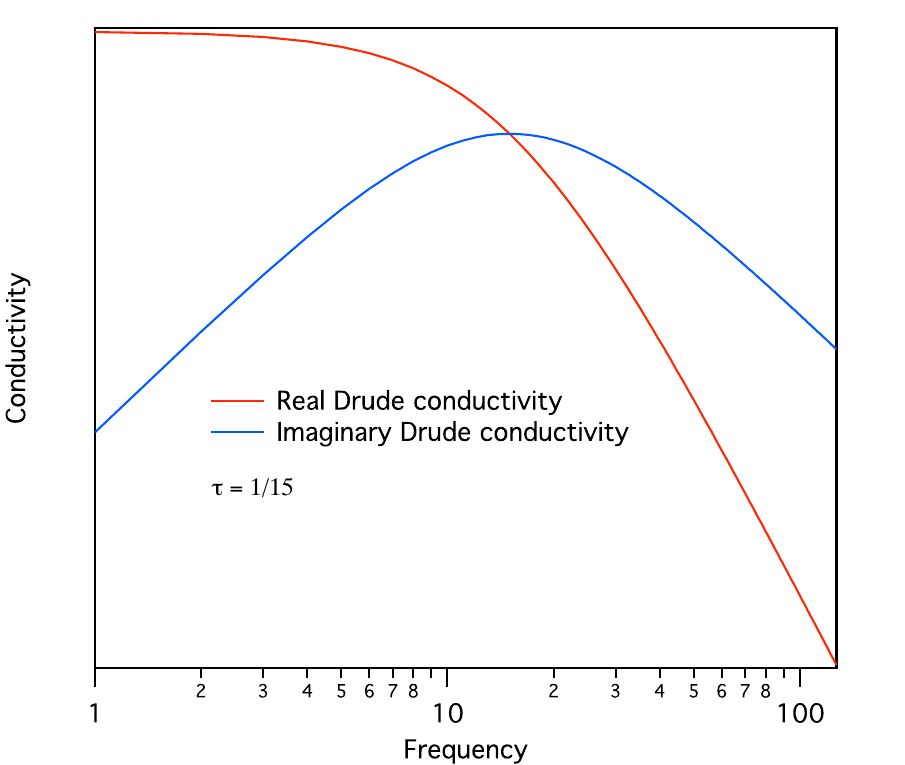}
\caption{(Color) Frequency dependent Drude conductivity with scattering rate $1/\tau =15$ on linear scale  (top)  log-log scale (bottom).  }\label{Drude}
\end{center}
\end{figure}

The Drude model (plotted in Fig. \ref{Drude}) demonstrates a number of important features that are found generally in response functions.  Its limit at $\omega \rightarrow 0$ is $Ne^2 \tau/m $ is well behaved mathematically and equivalent to the DC value.  As shown in Fig. \ref{Drude}, it has real and imaginary components that differ from each other.  At low frequency the current's response is in phase with the driving field and purely dissipative (real).  At intermediate frequencies (when the driving frequency is of order the scattering rate $1/\tau$), the real conductivity falls to half its DC value and imaginary conductivity peaks and equals the real value.  At higher frequencies, two things happen.  Not only does the current begin to lag the driving field, but its overall magnitude decreases as the response has a harder time keeping up with the driving electric field.    The conductivity also exhibits distinct power laws in its various frequency limits: linear for the imaginary part at low $\omega$, $1/\omega$ at high $\omega$, and  $1/\omega^2$ for the real part at high frequency.

\begin{figure}[htb]
\begin{center}
\includegraphics[width=8cm,angle=0]{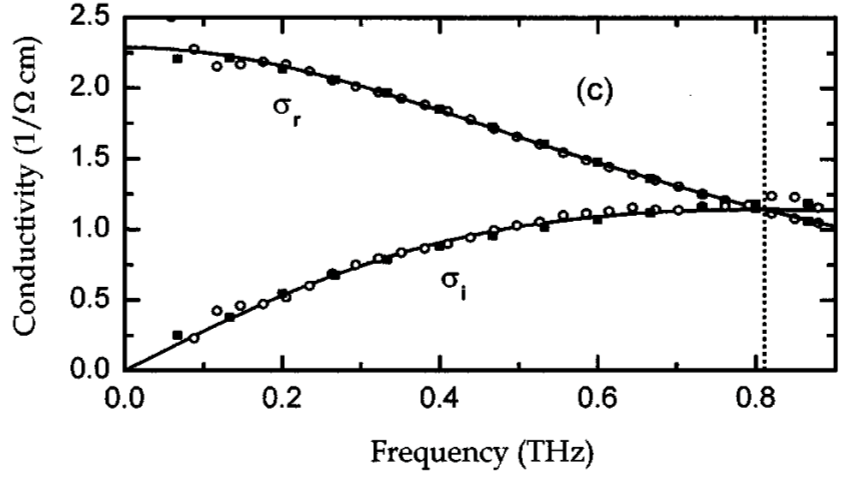} 
\caption{ Complex electric conductivity of unintentionally doped n-type GaN.  The dotted vertical line gives a scattering rate of 0.81 THz.  Lines are fits to the Drude expression of Eq. \ref{DrudeCond}.  One can see the almost perfect agreement with the Drude form.  From Ref. \onlinecite{Zhang03a}.  }\label{Drude2}
\end{center}
\end{figure}

The Drude functional form can be used to capture the finite frequency response of simple metals.   See Fig. \ref{Drude2} for measurements using time-domain THz spectroscopy of unintentionally doped n-type GaN.   For high enough dopings and at room temperature such simple semiconductors can be modeled as simple metals.   One can see the almost perfect agreement with the Drude form.

In the limit of zero dissipation $1/\tau \rightarrow 0$ Equation \ref{DrudeCond} has the limiting form

 \begin{eqnarray}
\sigma_1 = \frac{\pi}{2} \frac{Ne^2}{m} \delta(\omega = 0) ,\\
\sigma_2 = \frac{Ne^2}{m \omega} .
\label{zerodissp}
\end{eqnarray}

$\sigma_1$ is zero everywhere, but at $\omega = 0$.  As $\sigma_1$ is proportional to dissipation, this demonstrates that a gas of collisionless electrons cannot absorb photons at finite frequency.  This can be shown directly by demonstrating that the Hamiltonian that describes this interaction, commutes with the momentum operator $p$ and hence it  has no time dependence.  With the exception of umklapp scattering, electron-electron interactions do not change this situation, as such interactions cannot degrade the total system momentum.

\begin{figure}[htb]
\begin{center}
\includegraphics[width=8cm,angle=0]{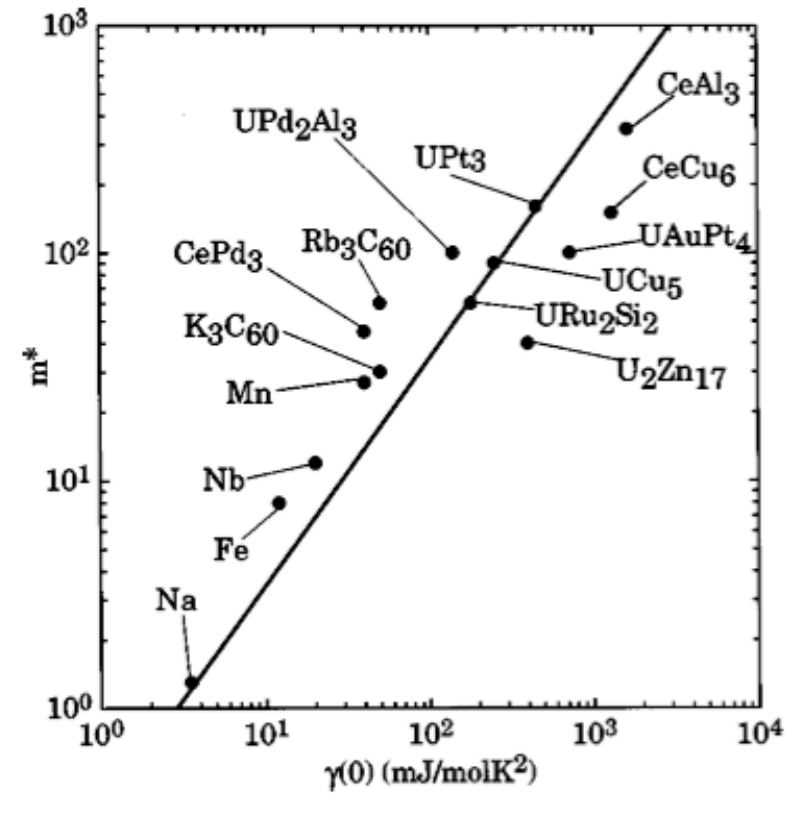}
\caption{(Color) Specific-heat values $\gamma$ vs effective mass m$^*$, evaluated from the optical data using spectral-weight arguments \cite{Degiorgi97a} }\label{SpecificHeatOpticalMass}
\end{center}
\end{figure}

The dependence of the Drude conductivity on $\tau$, $N$, and $m$, shows that the optical mass of charge carriers in metals can be obtained if for instance the charge density is known from other parameters like the Hall coefficient.  For instance, in Fig. \ref{SpecificHeatOpticalMass} the mass determined by optical conductivity is plotted against the linear coefficient in the specific heat (which is proportional to the electronic density of states and hence the mass).  From simple metals to exotic heavy fermion materials, it shows a dramatic linear dependence over 3 orders of magnitude.

It is convenient to express the prefactors of the Drude conductivity in terms of the so-called $plasma$ $frequency$  $\omega_p = \sqrt{\frac{4 \pi N e^2}{m}}$.  The plasma frequency is equivalent to the frequency of the free longitudinal oscillations of the electron gas\footnote{In actual materials, high energy excitations give a contribution to the dielectric function $\epsilon_\infty$ that renormalizes the observed plasma frequency downward to the $screened$ plasma frequency $\sqrt{\frac{4 \pi N e^2}{m \epsilon_\infty}}$.  This is the actual frequency of plasmon oscillations in a real material.}.  The conductivity can be expressed

 \begin{eqnarray}
\sigma(\omega) = \frac{\omega_p^2}{4 \pi}\frac{\tau}{1 - i \omega \tau} .%
\label{Drudeplasmafreq}
\end{eqnarray}

Therefore within the Drude model, the optical response is fully determined by two frequencies:  the plasma frequency $\omega_p$ and the scattering rate $1/\tau$.  Since $\omega_p$ is many orders of magnitude greater than $1/\tau$, this allows us to define three distinctly different regimes for the optical response.

At low frequencies $\omega \ll 1/\tau$, in the so-called $Hagen-Rubens$ regime, the conductivity is almost purely real, frequency independent, and approximately equal to the DC value.    At intermediate frequencies, in the $relaxation$ $regime$ where $\omega$ is on the order of $1/\tau$, one must explicitly take into account the $\omega \tau$ factor in the denominator of  Eq. \ref{DrudeCond}.  As mentioned above when $\omega = 1/\tau$ the real conductivity falls to half its DC value and the imaginary conductivity peaks and is equal to the real value\footnote{Of course in real materials, there are interband excitations which can also contribute to the conductivity.  The Drude model as presented here is an idealized case.}.  The significance of the high frequency regime $\omega>\omega_p$ can be seen in rewriting the conductivity as the dielectric function.  It is 

 \begin{eqnarray}
\epsilon(\omega) = 1-  \frac{\omega_p^2}{\omega^2 - i \omega / \tau}. %
\end{eqnarray}

\begin{figure}[htb]
\begin{center}
\includegraphics[width=7cm,angle=0]{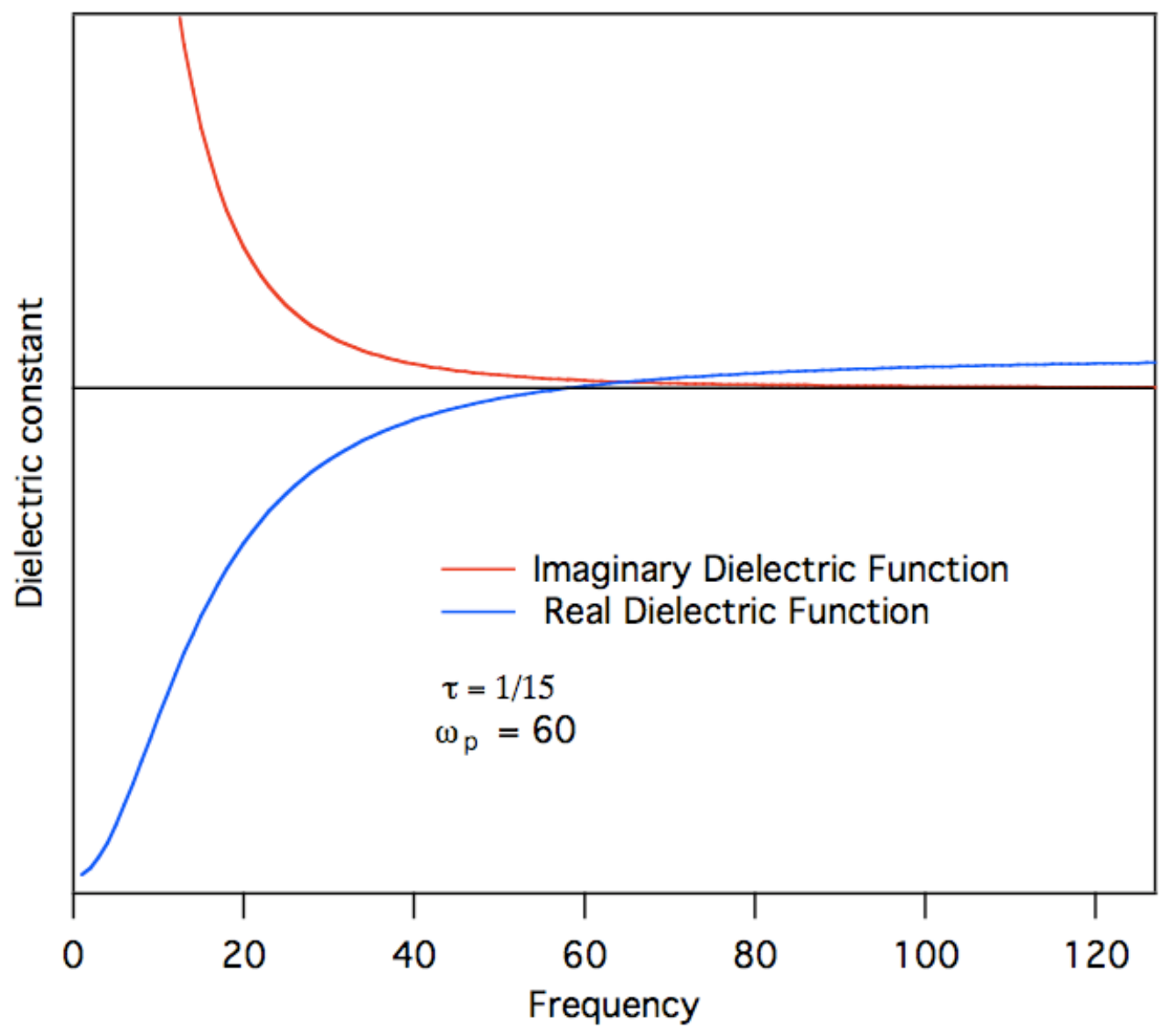} 
\caption{(Color) Frequency dependent Drude dielectric function with scattering rate $1/\tau =15$ and plasma frequency $\omega_p = 60$.   The real part of the dielectric function changes sign at $\omega_p$.  }\label{DielectricDrude}
\end{center}
\end{figure}

The real and imaginary parts of this expression are given in Fig. \ref{DielectricDrude}.  One sees that for the conventional case where $\omega_p \gg 1/\tau$, the plasma frequency is the frequency at which the real part of the dielectric function changes sign from negative to positive i.e.  $\omega_p$ sets the scale for the zero crossing of $\epsilon_1$.   An analysis of the reflection and transmission using the $Fresnel$ equations (see Jackson \cite{Jackson98a} ), shows that above the zero crossing, metals described by the Drude model become transparent.  Hence this high frequency limit is called the $transparent$ regime.

Despite its classical nature, the Drude model describes the gross features of many metals quite well at low frequencies.   We give a number of examples of its use and extensions below.  Of course, our interest in the electrodynamics of solids extends far beyond the case of simple metals.  We will find the Drude model lacking in this regard.  And even for `simple' metals, one has the interesting aspects of finite frequency absorptions that parameterize band structure and, for instance, give copper and gold their beautiful colors.  The Drude model is wholly inadequate to describe such finite frequency absorptions.  For such processes in more complicated metals, semiconductors, and insulators, one can model them quantum mechanically in a number of ways, but we can also gain predictive power and intuition from an extension to the classical Drude model called $Drude$-$Lorentz$.

Here we envision the electrons are additionally also subject to a simple harmonic restoring force  $-Kx$.  Of course, such a model is $physically$ inadequate to describe finite frequency absorption in most band semiconductors and insulators, as their insulating nature is due to the properties of filled bands and not electron localization.  Nevertheless, numerous aspects of absorption at `band-edges' in semiconductors can be modeled phenomenologically with Drude-Lorentz.  Of course its applicability to the modeling of harmonic phonon absorptions is obvious.

 With $ \omega_0^2 = K/m$, we  extend Eq. \ref{DrudeForces} as

 \begin{eqnarray}
mx'' = - eE - m x'/ \tau -K  x ,
\label{DrudeForces2}
\end{eqnarray}

Proceeding in exactly the same fashion as for the simple Drude model, we obtain for the Drude-Lorentz conductivity

 \begin{eqnarray}
\sigma(\omega) = \frac{Ne^2  }{m} \frac{\omega}{i(\omega_0^2 - \omega^2) + \omega / \tau} ,
\label{DLmodel}
\end{eqnarray}
One can see that in the limit $\omega_0 \rightarrow 0$ the Drude relation is obtained.  

An important limiting case of the Drude-Lorentz model is obtained in the limit of large damping and large spring force.   First consider Eq. \ref{DLmodel} rewritten in terms of the plasma frequency and expressed as  $\epsilon$.

 \begin{eqnarray}
\epsilon(\omega) =  1 +   \frac{\omega_p^2  }{\omega_0^2 } \frac{1}{ 1- \frac{i  \omega}{\tau \omega_0^2}  -  \frac{ \omega^2}{\omega_0^2 } } ,
\label{Factored}
\end{eqnarray}

Here we have factored the resonance frequency $\omega_0$ out of the denominator.   Note that Eq. \ref{Factored} written in this fashion expresses the fact that the denominator of the dielectric function can be written as an expansion in powers of $-i\omega$.   By inspecting the frequency dependence of a dielectric response, one can determine what terms are dominating in the classical equations of motion.  Taking the limit of this expression where $\omega_0^2 \rightarrow \infty $ and $1/\tau \rightarrow \infty $ but in such a fashion that quantity $ \tau \omega_0^2 $ stays finite, one has the expression, 

 \begin{eqnarray}
\epsilon(\omega) =  1 +   \frac{\omega_p^2  }{\omega_0^2 } \frac{1}{ 1-  \frac{i  \omega}{\tau \omega_0^2}  } .
\end{eqnarray}

This expression defines a new frequency scale $\Gamma =  \tau \omega_0^2 $ that we can write as 

 \begin{eqnarray}
\epsilon(\omega) =  1 +   \frac{N e^2  }{4 \pi K } \frac{1}{ 1-  i  \omega / \Gamma  } .
\label{Debye}
\end{eqnarray}

Eq. \ref{Debye} is equivalent to the expression of Debye relaxation, which is a common form for polarization relaxation phenomena in insulators, particularly glasses.   It describes the situation where a particle subject to a strong restoring force moves through a medium that is so viscous so as to longer have inertial forces be relevant.  In this regard, note that in the form given above, the particle mass has dropped out of the expression.  An uncommonly broad ``resonance" peak is produced.  An equivalent form to this expression is found commonly for magnetic dipole relaxation as well.   Also note that although Eq. \ref{Debye} can be found to match large classes of experimental data quite well, it is not strictly speaking a fully mathematically consistent response function.   For instance, it doesn't satisfy the sum rules discussed below (Sec. \ref{sectionsumrules}) due to the neglect of the $\omega^2/\omega_0^2$ term in the denominator of Eq. \ref{Factored}.   Typically the Debye model is derived for relaxing electric dipoles in a time dependent electric field, with a neglect of their moment of inertia \cite{Frohlich58a}.  The above derivation is different, but results in the equivalent mathematical form of Eq. \ref{Debye}.    Extension of the model exist \cite{Onodera93a} for situations where inertial effects are important (high frequencies and for fulfillment of the sum rules).   We found that such effects were important in the analysis of optical data for quantum spin-ice materials and were interpreted as being consistent with inertial effects of the effective magnetic monopoles \cite{Pan16a,Armitage17a} in those systems.

\begin{figure}[htb]
\begin{center}
\includegraphics[width=7cm,angle=0]{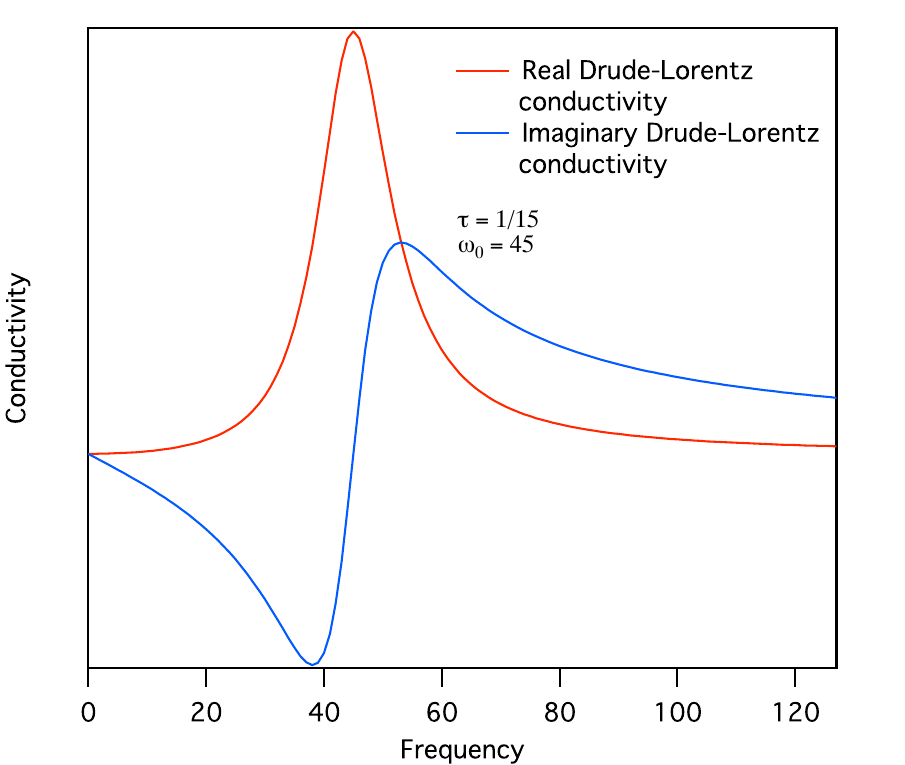} 
\caption{(Color) Frequency dependent Drude-Lorentz complex conductivity with scattering rate $1/\tau =15$ and oscillator resonant frequency $\omega_0 = 45$.   The real conductivity is zero at $\omega = 0$ and positive for all frequencies.  }\label{DrudeLorentz}
\end{center}
\end{figure}

As mentioned above, despite its classical nature and the inapplicability of the underlying physical picture for some situations, in some cases the Drude-Lorentz model can be used to quantify finite frequency absorptions.  In the case of the band edge absorption  in semiconductors, a number of oscillators of the form of Eq. \ref{DLmodel} with different weights can be used to capture the decidely non-Lorentzian line shape.  One can convince oneself that since any lineshape can be fit arbitrarily well using an arbitrarily large number of oscillators and moreover since the imaginary part of the response follows from a Kramers-Kroning transform of the real part, it is always perfectly feasible to parametrize the response using Drude-Lorentz.

\subsection{The Quantum Case}

The above Drude and Drude-Lorentz models are classical models used to describe intrinsically quantum mechanical phenomena and the fact that they are useful at all to describe solids is surprising.  It is clear that a quantum mechanical treatment is desired.  The most commonly used method for the quantum mechanical calculation of electromagnetic response of materials is the $Kubo$ formalism \cite{Mahan90a,Kubo54,Kubo57}.  It is based on the fluctuation-dissipation theorem, which relates the spontaneous fluctuations of a system described $via$ its correlation functions to its driven linear response.  The derivation here loosely follows the one of Wooten \cite{Wooten72a}, which is mirrored in Dressel and Gr\"uner \cite{Dressel02a}.

We calculate conductivity, by calculating the absorption rate of a system based on the power dissipated $P = \sigma_1 E^2$.  We first use Fermi's Golden rule to calculate the scattering probablility that incident radiation excites an electron from one state $|s \rangle$ to another $|s' \rangle$

 \begin{eqnarray}
W_{s \rightarrow s'} = \frac{2 \pi}{\hbar^2} |\langle s'|H_{i}|s \rangle|^2 \delta(\omega  - \omega_{s'} + \omega_{s}).
\end{eqnarray}

 $|s \rangle$ and $|s' \rangle$ are the system's eigenstates, which may include the effects of many body physics.  The interaction Hamiltonian for charge with the electromagnetic field to first-order in field is

 \begin{eqnarray}
H_{i} = \frac{e}{2mc} \sum_{i=1}^N [p_i \cdot A(r_i) +A(r_i) \cdot  p_i ] - e \sum_{i=1}^N \Phi(r_i) ,
\end{eqnarray}

\noindent where $p$, $A$, and $\Phi$ are the quantum mechanical operators for momentum, vector potential, and scalar potential and $c$ is the speed of light.  Since the vector potential $A$ depends on position, it does not in general commute with $p$.  However since $p = - i \hbar \nabla$, one can show that $p \cdot A - A \cdot p = -i \hbar \nabla \cdot A$ and since within the Coulomb gauge $\nabla \cdot A = 0$, $p$ and $A$ can in fact commute.  This means that for purely transverse waves with $\Phi = 0$ the interaction Hamiltonian simplifies to 

 \begin{eqnarray}
H_{i}^T = \frac{e}{mc} \sum_{i=1}^N p_i \cdot A(r_i).
\end{eqnarray}

By substituting in for the canonical momentum $p = mv - eA/c$, dropping all terms that are higher than linear in $A$, and replacing the summations by an integral one $H_{i}$ takes the form

 \begin{eqnarray}
H_{i}^T = - \frac{1}{c} \int dr J^T(r) \cdot A^T(r).
\end{eqnarray}

This gives a Fourier transformed quantity of 

\ \begin{eqnarray}
H_{i}^T = - \frac{1}{c}  J^T(q) \cdot A^T(q).
\end{eqnarray}

This means we can substitute in for the form of $H_i$ in the expression for $W_{s \rightarrow s'}$ to get

 \begin{eqnarray}
W_{s \rightarrow s'} = \frac{2 \pi}{\hbar^2} |\langle s'|H_{i}^T|s \rangle|^2 \delta(\omega  - \omega_{s'} + \omega_{s}).
\end{eqnarray}

The matrix element in this expression follows from the form of $H_{i}^T$.  It is 

 \begin{eqnarray}
\langle s'|H_{i}^T|s \rangle =  - \frac{1}{c} \langle s'|J^T(q)|s \rangle A^T(q).
\end{eqnarray}

which leads to 

 \begin{eqnarray}
W_{s \rightarrow s'}  = \frac{2 \pi}{\hbar^2 c^2}   \langle s'|J^T(q)|s \rangle   \langle s|J^{T*}(q)|s' \rangle ,\\ \nonumber
 |A^T(q)|^2 \delta(\omega  - \omega_{s'} + \omega_{s}).
\end{eqnarray}

\noindent where $J^{T*}(q)= J^{T}(-q)$.  One then sums over all occupied initial and all empty final states $W = \sum _{s,s'} W_{s \rightarrow s'} $.    The dissipated power per unit time and volume at a particular photon frequency $\omega$ follows after a few mathematical steps as

 \begin{eqnarray}
P = \hbar \omega W =  \hspace{1cm} \\ \nonumber
 |A^T(q)|^2 \sum_{s} \frac{\omega}{\hbar c^2} \int dt \langle s|J^{T}(q,0)J^{T*}(q,t)|s \rangle e^{-i\omega t}.
\end{eqnarray}

The electric field is related to the vector potential as $E^T = i \omega A^T/c$ for transverse waves.  Upon substitution we get the absorbed power per unit volume per unit time expressed as a current-current correlation function

 \begin{eqnarray}
P =  |E^T(q)|^2 \sum_{s} \frac{1}{\hbar \omega} \int dt \langle s|J^{T}(q,0)J^{T*}(q,t)|s \rangle e^{-i\omega t}.
\end{eqnarray}

Using our previously given relation $P = \sigma_1 E^2$, the expression for the real part of the conductivity is

 \begin{eqnarray}
\sigma_1^T(q,\omega) =  \sum_{s} \frac{1}{\hbar \omega} \int dt \langle s|J^{T}(q,0)J^{T*}(q,t)|s \rangle e^{-i\omega t}.
\label{Kubo}
\end{eqnarray}

The imaginary part of $\sigma(\omega)$ follows from the Kramers-Kronig relation.  This is the $Kubo$ formula.  It has the form expected for the fluctuation-dissipation theorem, that of a correlation function (here current-current) averaged over all the states  $|s \rangle$ and describes fluctuations of the current in the ground state.  The conductivity depends on the time correlation between current operators integrated over all times.

The above formalism is general and applies to any set of system states $|s \rangle$.  An extension can be made in cases where Fermi statistics applies.  The end result is (see Dressel and Gr\"uner \cite{Dressel02a} for the full derivation)

 \begin{eqnarray}
\sigma_1^T(\omega) = \frac{\pi e^2}{m^2 \omega} \frac{2}{(2 \pi)^3} |\langle s' |p|s \rangle|^2 D_{s's}( \hbar \omega).
\end{eqnarray}

Here $\langle s'|p|s \rangle$ is known as the dipole matrix element.  $D_{s's}$ is the so-called joint density of states, defined as $\frac{2}{(2\pi)^3} \int \delta(\hbar \omega - \hbar \omega _{s's}) dk$.  This equation, which is often referred to as the $Kubo$ - $Greenwood$ formula, is extremely useful for the calculation of higher lying interband transitions in metals and semiconductors when the states $|s \rangle$ and $|s' \rangle$ belong to different bands (as for instance illustrated by the red arrow in Fig. \ref{OpticalConduct}).  It also has the form that one may naively expect; it is proportional to the joint density of states and a matrix element which incorporates aspects like symmetry constraints and selection rules for allowed transitions.

We can use the Kubo formalism (Eq. \ref{Kubo}) to calculate the leading order metallic conductivity in a slightly more rigorous fashion than was done in the classical approach above.    We start with Eq.  \ref{Kubo} and then posit as an ansatz that finite currents relax as 

 \begin{eqnarray}
J(q,t) = J(q,0) e^{-t/\tau}.
\end{eqnarray}

\noindent and assume that the current correlation time $\tau$ has no dependence on $q$.  Inserting this into the Kubo formula, one gets that

 \begin{eqnarray}
\sigma_1(q,\omega) = \frac{1}{\hbar \omega} \sum_{s} \int d t e^{-i \omega t - |t|/\tau} \langle s|J^2(q)|s \rangle.
\end{eqnarray}

One now needs an expression for the current fluctuations at finite $q$.  The expectation value for $J$ is zero, but fluctuations lead to a finite value of $\langle J^2 \rangle$.   For small $q$, one can show in the `dipole approximation' that $J(q) = -\frac{e}{m} \sum_i p_i$ where $i$ labels the individual charges.  With the reinsertion of a complete set of states $s'$ one gets the expression 

 \begin{eqnarray}
\sigma_1(q,\omega) = \frac{e^2}{m^2 \hbar \omega}  \int d t e^{-i \omega t - |t|/\tau} \sum_{s,s',i}|\langle s'|p_i|s \rangle|^2.
\label{QDrude}
\end{eqnarray}

The quantity $2 \sum_{s,s',i}| \frac{  |\langle s'|p_i|s \rangle|^2}{m \hbar \omega_{s,s'}}$ found in Eq. \ref{QDrude} is called the $oscillator$ $strength$ $f_{s,s'} $.  Here $\omega$ in Eq. \ref{QDrude} is the energy difference between states $|s \rangle$ and  $|s' \rangle$ as is  $ \omega_{s,s'}$ in the definition of the oscillator strength.  Below, we pay special attention to the oscillator strength when we calculate the optical sum rules.

The oscillator strength can be easily calculated for the case of free electrons where the energy is $\hbar \omega = \frac{\hbar^2 k^2}{2m}$ and the average momentum  is $ |\langle s'|p_i|s \rangle|^2 = \frac{\hbar^2 k^2}{4}$.  In such a case  $f_{s,s'} = N$ the total number of free charge carriers.  After performing the integral in Eq. \ref{QDrude}  the final result for the leading order conductivity calculated $via$ the Kubo formalism is  

 \begin{eqnarray}
\sigma_1^{Kubo}(\omega) = \frac{Ne^2 \tau}{m} \frac{1}{1 - i\omega \tau},
\end{eqnarray}

\noindent which is the exact same result as calculated classically $via$ the Drude model above. For more details see the chapters in Wooten \cite{Wooten72a} and Mahan \cite{Mahan90a}.

\section{Techniques}

As mentioned above, the incredibly large spectral range spanned by the typical energy scales of solids means that many different techniques must be used.  Below I discuss the most used techniques of microwave measurements, THz spectroscopies, optical reflectivity, and ellipsometry that are used to span almost six orders of magnitude in measurement frequency.    These various regimes are loosely defined below, but of course there are overlaps between them\footnote{Although I don't discuss them, one might naturally include various radio frequency techniques in this list (which would increase the frequency range by another three orders of magnitude at least at the low end).  At typical experimental temperatures, the low frequency available in techniques like mutual inductance measurements allows almost the DC limit of quantities like the superfluid density in superconductors to be measured.  Interested readers are directed to Ref. \onlinecite{Turneaure98a} and reference therein.}.

\subsection{Microwaves}

A standard measurement tool in the microwave range (100 MHz - 100 GHz) is microwave resonant cavity techniques.  At lower frequencies (kHz and MHz), measurements can be performed by attaching contacts to samples and the complex conductivity can be measured by lock-in techniques, network analyzers, and impedance analyzers.  Such methods become problematic in the GHz range however, as wavelengths become comparable to various measurement dimensions (sample size, microwave connector dimensions, cable lengths), and capacitive and inductive effects become appreciable.  In microwave cavity resonance techniques, one is less sensitive to such considerations as the sample forms part of a resonance circuit, which dominates the measurement configuration.  Typically one measures transmission through a cavity, which is only possible when at resonance.  The technique is widely used for the study of dielectric and magnetic properties of materials in the GHz range, because of its high sensitivity and relative simplicity.

A microwave resonator is typically an enclosed hollow space of cylindrical or rectangular shape machined from high-conductivity metal (typically copper or superconducting metals) with interior dimensions comparable to the free-space wavelength.  On resonance, an electromagnetic standing wave pattern is set up inside the cavity.  One characterizes the resonance characteristics of the cavity, both with and without a sample inside.  Upon sample introduction the resonance's center frequency shifts and it broadens.  From the shifting and broadening, one can define a complex frequency shift, which with knowledge of the sample size and shape allows one to quantify the complex conductivity directly.  For instance, in the limit of very thin films, the center frequency shift is proportional to $\sigma_2$ and the broadening is proportional to $\sigma_1$.   For samples much thicker than the skin or penetration depth, different relations apply \cite{Klein93a}.  The fact that the measurements are performed ``on resonance" means that a very high sensitivity is achieved.  The technique does have the considerable disadvantage that only discrete frequencies can be accesses, as one is limited to the standing wave resonance frequencies of the cavity, which means that true spectroscopy is not possible.  In such measurements one typically determines a sample's complex conductivity as a function of temperature at some finite fixed frequency (See Figs.  \ref{NbSurface}, \ref{YBCO2} and \ref{YBCO1}).  

In contrast, the Corbino geometry is a measurement configuration that is capable of $broadband$ microwave spectroscopy.  It is compatible with low $T$ and high field cryogenic environments and capable of broadband microwave spectroscopy measurements from 10
MHz to 40 GHz.  It is a relatively new technique, but one which has been used with great success by a number of groups in the correlated electron physics community \cite{ScheffNature,ScheffRSI,Anlage1,Anlage2,MarkLeeRSI,MarkLeePRL}.  Although powerful, it has the disadvantage that its use is confined to samples which have typical 2D resistance within a few orders of magnitude of 50 Ohms. 

\begin{figure}[htb]
\begin{center}
\includegraphics[width=3.5cm,angle=90]{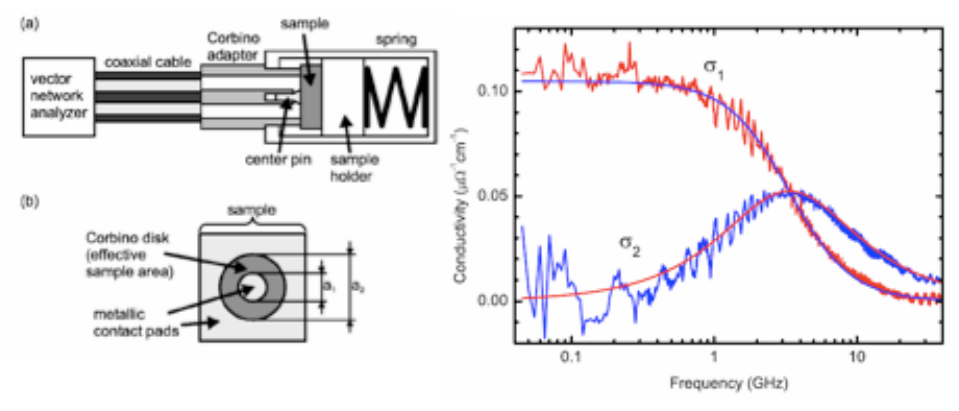}
\caption{(Color) (left) Experimental geometry for a Corbino spectrometer.  (right) The
real and imaginary parts $\sigma_1 + i \sigma_2$ of the optical
conductivity spectrum of UPd$_2$Al$_3$ at temperature T = 2.75 K.
The data are fit to $\sigma_{dc} (1 - i \omega \tau)^{-1}$ with
$\sigma_{dc} = 1.05 \times 10^5 (\Omega -cm)^{-1}$ and $\tau = 4.8
\times 10^{-11} s$ and show excellent agreement with the Drude
prediction.  Adapted from Refs \cite{ScheffNature,ScheffRSI}} \label{SCInsAC2}
\end{center}
\end{figure}

In a Corbino geometry spectrometer a microwave signal from a
network analyzer is fed into a coaxial transmission line.  A
schematic is shown in Fig.~\ref{SCInsAC2} (left).  The signal
propagates down the coaxial line and is reflected from a sample
that terminates an open ended coaxial connector.  The network
analyzer determines the complex reflection coefficient $S_{11}$ of
the sample which can be related to the complex sample impedance
$Z_L$ $via$ the relation $S_{11} = \frac{Z_L - Z_0}{Z_L + Z_0} $
where $Z_0$ is the coaxial line impedance (nominally 50 $
\Omega$). Because the sample geometry is well defined, knowledge
of the sample impedance yields intrinsic quantities like the
complex conductivity. In the typical case of a thin
film where the skin or penetration depth is much
larger than the sample thickness $d$, the sample impedance is related in
a straightforward fashion to the complex conductivity as $\sigma =
\frac{\textrm{ln }(r_2/r_1)} {2 \pi d Z_L}$  where $r_1$ and $r_2$
are the inner and outer conductor radii respectively.

An experimental challenge in Corbino geometry measurements is that
the coaxial cables and other parts of the transmission lines can
have strongly temperature dependent transmission characteristics.  Errors in the intrinsic reflection coefficient, coming from standing wave reflections or phase shifts and damping in the
transmission lines, are accounted for by performing a number of
calibration measurements.  It is imperative that the same cryogenics conditions are
reproduced between calibration and each subsequent measurement.
In general this is easily done by ensuring the same starting
conditions and using a computer controlled cool-down cycle.  It has been demonstrated that
a three sample calibration  (open, short, and a standard 50 ohm resistor) at all temperatures dramatically increases the precision of the technique
over previous single sample calibrations \cite{ScheffNature}.

As an example of the power of the technique we show in
Fig.~\ref{SCInsAC2} (right)  measurements taken on films of
the heavy-fermion compound UPd$_2$Al$_3$ at $T$ = 2.75 K. The data
show almost ideal Drude behavior with a real conductivity of
Lorentzian shape and an imaginary conductivity that peaks at the
same $\omega$ as the real conductivity has decreased by half. The
anomalous aspect is the Drude peak is seen to be of remarkably
narrow width - almost 500 times narrower than that of a good metal
like copper. This is a consequence of the mass renormalizations in the heavy fermion
compounds and a feature that would have been completely undetectable
with a conventional spectrometer.

There also been the very interesting development of a bolometric technique for high-resolution broadband microwave measurements of ultra-low-loss samples, like superconductors and good metals \cite{Turner04a}.  This technique is a non-resonant one where the sample itself is actually used as a detector in a bolometric fashion.  Microwaves are fed through a rectangular coaxial transmission line into which the sample under test and a reference sample are mounted.   Small changes in the sample and reference's temperature are monitored as they absorb microwave radiation.  The key to the success of this technique is the $in$ $situ$ use of this normal metal reference sample which calibrates the absolute microwave incident power.  As the absorbed power is proportional to the surface resistance $R_s(\omega)$, the independent measure of the incident power allows one to measure the surface resistance precisely.

In such an apparatus, the sample temperature can be controlled independently of the 1.2 K liquid-helium bath, allowing for measurements of the temperature evolution of the absorption.  The minimum detectable power of this method at 1.3 K is 1.5 pW, which corresponds to a surface impedance sensitivity of $\approx 1 \mu \Omega$ for a typical $1 mm \times 1 mm $ platelet sample.  The technique allows very sensitive measurements of the microwave surface resistance over a continuous frequency range  on highly conducting samples.  This is a region of sample impedances generally inaccessible with the Corbino technique and only available at discrete frequencies in microwave cavities.  

A disadvantage of the technique is that although one measures $R_s(\omega)$ the quantity of interest is typically the complex conductivity $\sigma(\omega)$.  One must make various assumptions to get this from the relation for surface impedance

 \begin{eqnarray}
Z_s = R_s + i X_s  = \sqrt{\frac{i \omega \mu_0}{\sigma_1 + i \sigma_2}}.
\end{eqnarray}

\noindent as the surface reactance $X_s$ is not measured.   At low temperatures and frequencies in the superconducting state  $\sigma_2$  is mostly determined by the superconducting response and can be related to the independently measured penetration depth $\lambda$.  $\sigma_1$ then follows from the relation $R_s = \frac{1}{2} \mu_0^2 \omega^2 \lambda^3 \sigma_1$.  At higher temperatures and frequencies a more complicated iterative procedure must be used that is based on the Kramers-Kronig transform  \cite{Turner04a}.

\begin{figure}[htb]
\begin{center}
\includegraphics[width=7cm,angle=0]{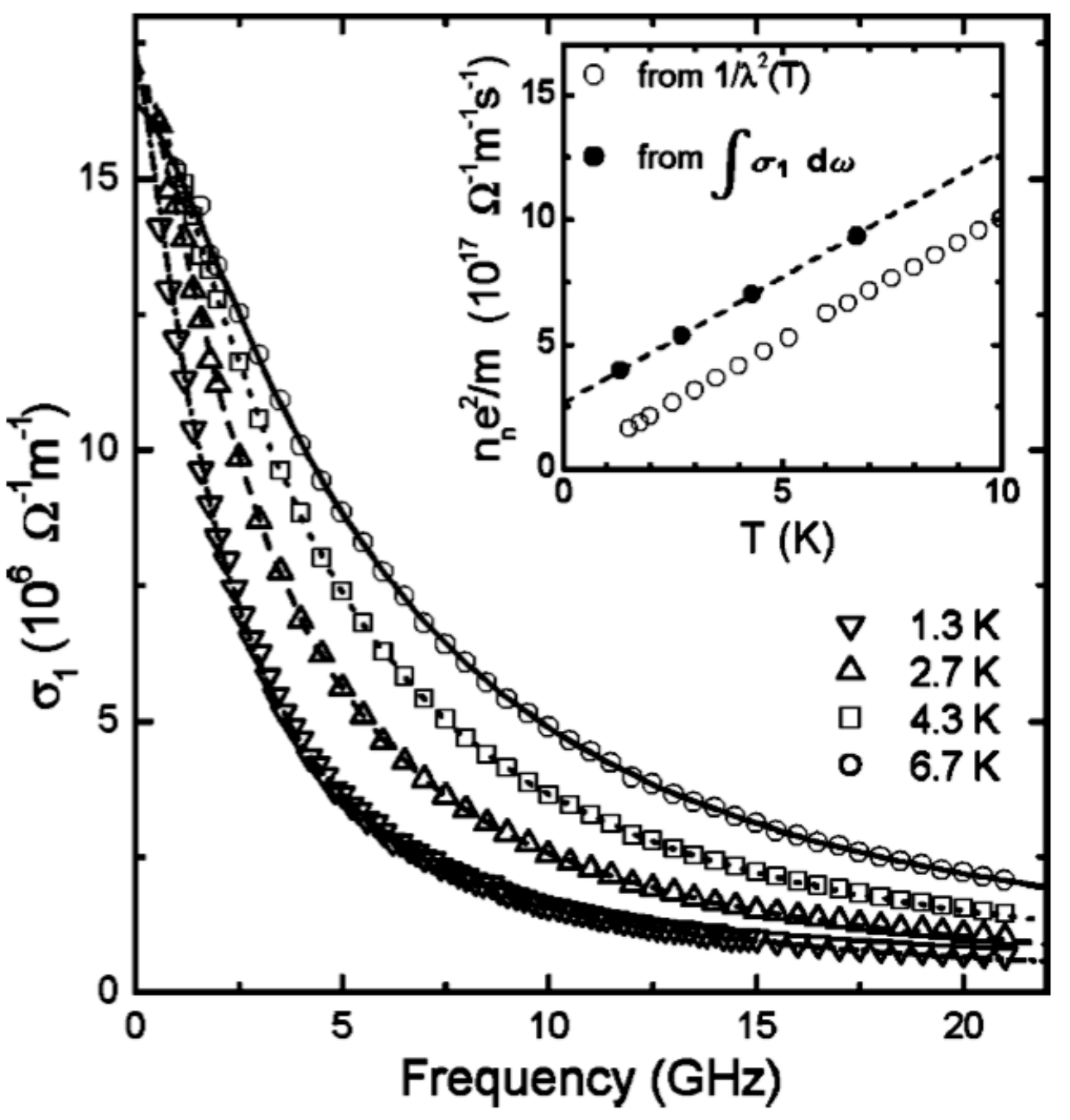}
\caption{The real part of the microwave conductivity $\sigma_1(\omega)$ extracted from measurements of $R_s(\omega)$ as described in the text and Ref.  \onlinecite{Turner04a}.  Figure adapted from \cite{Turner04a}.} \label{BroadBandYBCO}
\end{center}
\end{figure}

This technique has been used to measure the real part of the conductivity of YBa$_2$Cu$_3$O$_{6.5}$ single crystals across a wide frequency range as shown in Fig. \ref{BroadBandYBCO}.  One can see a quasi-Drude like peak at low frequencies that comes from quasi-particle excitation at the nodes of the $d$-wave superconductor.  The peak has a narrow width of the order of 5 GHz, signifying a collapse of quasiparticle scattering in the superconducting state and  very long quasiparticle scattering lengths.  This is indicative of the very high quality of these YBa$_2$Cu$_3$O$_{6.5}$ single crystals.  Measurements of a broadband nature on these kind of materials are not possible with other techniques.  As powerful as it is, the technique is hampered by the fact that various assumption must be made about $\sigma_2$.  This is problematic for materials where its dependence is not known $a$ $priori$.

In addition to the ones detailed above, there are many other techniques which have been used for materials characterization in the microwave range.   These include microstrip and stripline resonators, transmission line wave bridges, radio frequency bridges, parallel-plate resonators, and helical resonators.  See Ref. \onlinecite{Dressel02a,Gruner98a,Engel93a} for further details.

\subsection{THz spectroscopy}

Time scales in the picosecond ($10^{-12}$ sec) range are among the most ubiquitous in condensed matter systems. For example, the resonant period of electrons in semiconductors and their nanostructures, the  scattering times of electrons in metals, vibrational frequencies of molecular crystals, superconducting gap energies, the lifetime of biologically important collective vibrations of proteins, and - now - even the transit time for an electron in Intel's new THz transistor  - these are all picosecond phenomena.  This ubiquity means that experimental probes employing Terahertz (THz) electromagnetic radiation are potentially quite powerful.  It is unfortunate then that this spectral range lies in the so-called `Terahertz Gap' - above the capabilities of traditional electronics and the microwave range, but below that of typical optical instrumentation.  The THz spectral region has been a traditionally difficult part of the electromagnetic spectrum to work in because of a number of reasons including weak sources, long wavelengths, and contamination by ambient room temperature black body radiation.  There have been a number of developments that allow measurements in the THz range in a manner that was not previously possible.  These come in the form of time domain THz spectroscopy and the increasing prevalence of Backward Wave Oscillator (BWO) based spectrometers.    A number of dramatic technical advances such as \textit{time-domain} THz spectroscopy (TDTS) using so-called `Auston' switch generators and detectors have enabled measurements that span this gap in measurement possibilities.

THz spectroscopy has become a tremendous growth field \cite{DOEreport}, finding use in a multitude of areas including characterization for novel solid-state materials \cite{Kaindl,Heyman}, optimization of the electromagnetic response of new coatings \cite{coatings}, probes of superconductor properties \cite{Dodge,Corson}, security applications for explosives and biohazard detection \cite{THzBiohazard}, detection of protein conformational changes \cite{THzproteins}, and non-invasive structural and medical imaging \cite{THzimaging0,THzimaging1,THzimaging3}.

\begin{figure}[htb]
\begin{center}
\includegraphics[width=8.5cm,angle=0]{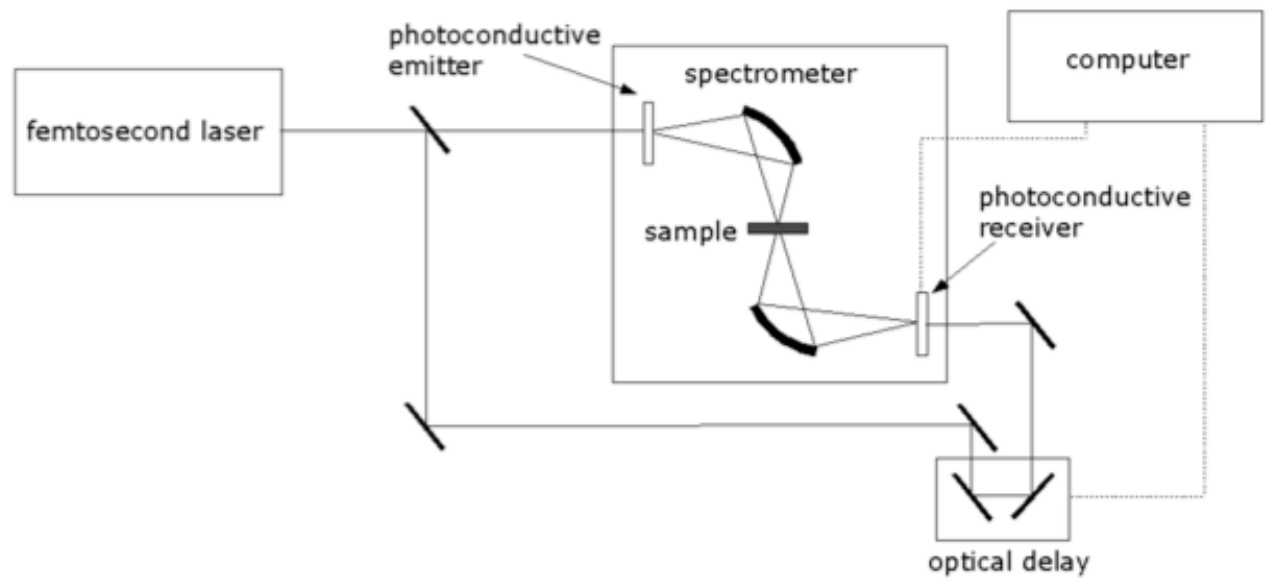}
\caption{ A diagram of a typical experimental layout used in TDTS.   } \label{Timedomain}
\end{center}
\end{figure}

\begin{figure}[htb]
\begin{center}
\includegraphics[width=7cm,angle=0]{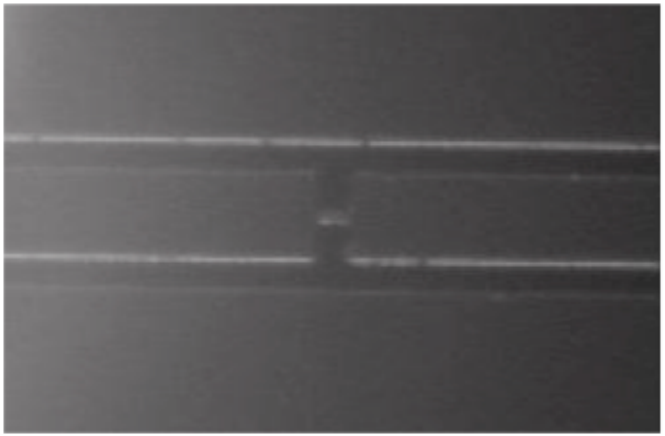}
\caption{An optical image of the two-contact photoconductive `Auston' switch antenna structure used in TDTS.  Two electrodes are grown on top of an insulating high defect semiconductor, like low temperature grown GaAs or radiation damaged silicon.  Electrode spacing is typically on the order of 10 microns.} \label{Austonswitch}
\end{center}
\end{figure}

\begin{figure}[tb]
\includegraphics[width=3.5cm,angle=90]{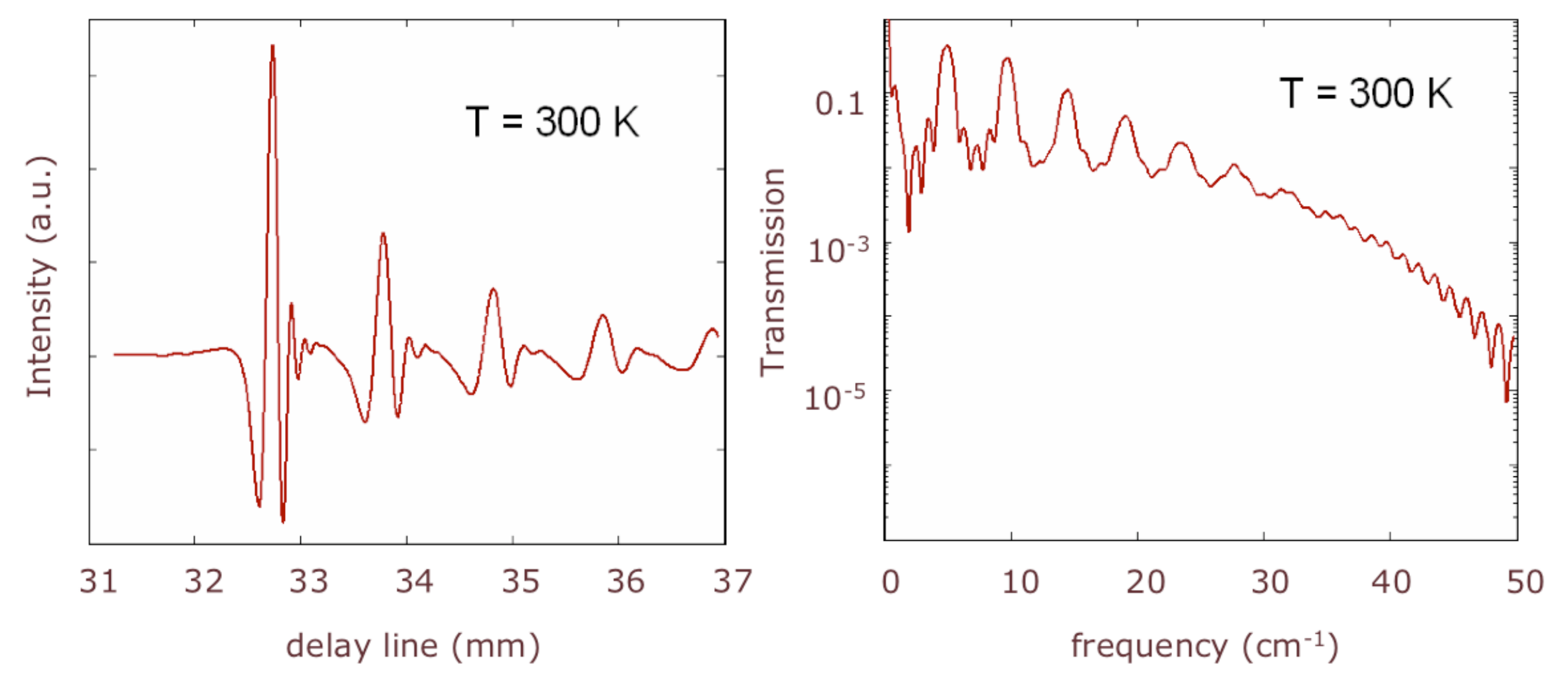}
\caption{(Color) (left) Time-domain trace of electric field
transmission through 60 $\mu$m thick SrTiO$_3$ subtrate.  Multiple
reflections from front and back surfaces are readily visible.
(right) Power spectrum of the transmission through the same 60 $\mu$m
thick SrTiO$_3$ subtrate.  The data is obtained from the squared
Fourier transform of the data in (left) and ratioed to a
reference aperature signal.  From Ref. \onlinecite{DookSTO}}
\label{DookSTO}\end{figure}

TDTS works (See Ref. \onlinecite{timedomain} for an additional excellent short summary) by the excitation of a source and activation of a detector by ultrafast femtosecond laser (typically Ti-sapphire) pulses.  A basic schematic is shown in Fig. \ref{Timedomain}. A femtosecond laser has, $via$ a beam splitter, its radiation split off to fall on source and detector Auston switches, which are typically pieces of low temperature-GaAs or radiation damaged silicon with two electrodes grown on top in a dipole arrangement and separated by approximately 20 $\mu m$ (Fig. \ref{Austonswitch}). Before laser illumination, the switch has a resistance of a few megaohms. After the fast illumination by the femtosecond laser pulse, the source switch's resistance falls to a few hundred ohms and with a bias by a few tens of volts, charge carriers are accelerated across the gap on a time scale of a few picoseconds. Their acceleration produces a pulse of almost single-cycle radiation, which then propagates through free space and - collimated by mirrors and lenses - interacts with the sample. Measurements have been typically performed in transmission.  After passing through the sample, THz radiation falls on the second Auston switch, which is also activated by the femtosecond pulse.  Whereas the first Auston switch was DC biased, the 2nd Auston switch is biased by the electric field of the THz radiation falling upon it.  Current can flow only across the second switch with a direction and magnitude proportional to the transient electric field at the instant the short femtosecond pulse impinges on it. The experimental signal is proportional to this current with a magnitude and polarity that reflects the THz electric field at the switch.  A delay line is then advanced and the THz electric field at a different relative times can be measured.  In this way, the entire electric field profile as a function of time can be mapped out as shown in Fig. \ref{DookSTO}(left).  The time-domain pulse is then Fourier transformed to get the complex electric field as a function of frequency.  A reference measurement is also performed on an aperture with no sample. The measured transmission function is the ratio of the signal transmission to the reference [Fig. \ref{DookSTO}(right)].

There are a number of unique aspects to TDTS that allow it to work exceptionally. Since the detected signal is proportional to the instantaneous electric field, and not the power, the measured transmission function is the $complex$ transmission coefficient for the $electric$ $field$, which has both an amplitude and a phase. This allows one to invert the data directly to get the real and imaginary optical constants (e.g. the complex conductivity) of the material. This is essential for a thorough characterization of material's properties.  Typical optical measurements discussed below measure only reflected or transmitted power, quantities in which the complex optical constants of interest are mixed into in a non trivial fashion.  In such measurements one has to measure over the largest frequency range possible (even if one is only interested in a limited spectral range), extrapolate to both DC and infinite frequency, and then Kramers-Kronig transform to get phase information.  In TDTS one can get both components by direct inversion.  

TDTS is also capable of unprecedentedly high signal-to-noise ratios.  Typically efforts in the THz and far infrared spectral range are complicated by a very large ambient black body radiation background.  In the case of TDTS the fact that the signal is `time-gated' and coherent, whereas black-body radiation is incoherent, allows a very high detection efficiency.  Additionally, since the detected quantity is actually electric field and not power (proportional to electric field squared), the noise in the spectral power is reduced by a factor proportional to the electric field itself. These aspects allow detection of transmission signals approaching one part in 10$^6$. This is an incredibly large dynamic range and is essentially unprecedented in optically based spectroscopies where one part in 10$^3$ is typically considered extremely good.

\begin{figure}[tb]
\includegraphics[width=7cm,angle=0]{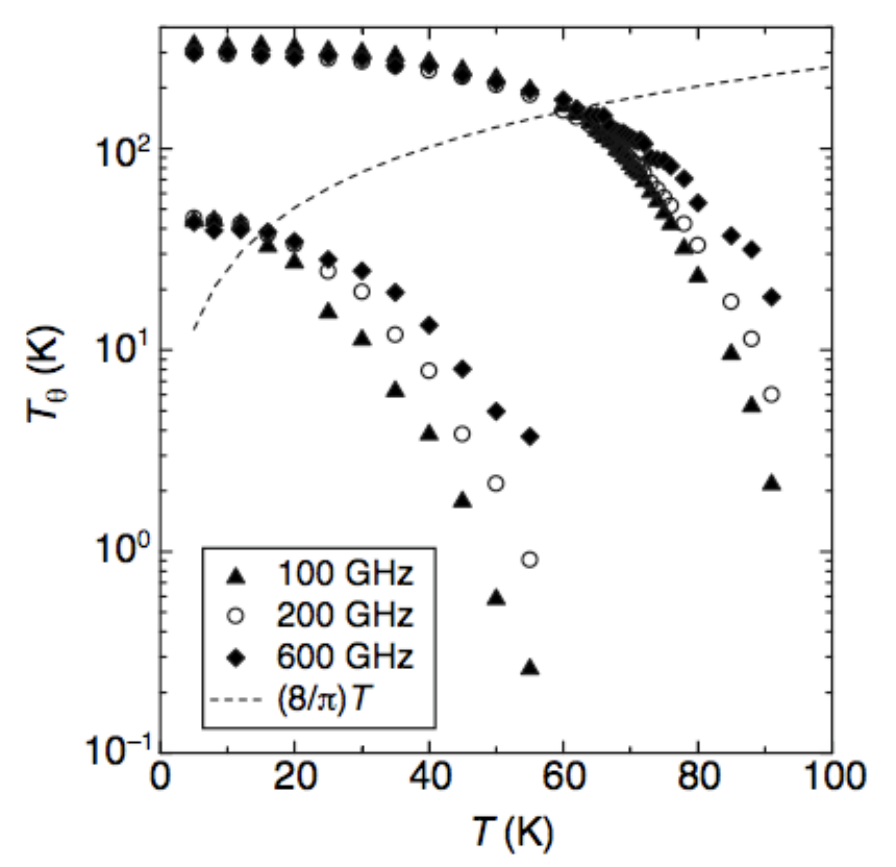}
\caption{The dynamic (frequency dependent) phase stiffness T$_{\theta}(\omega)$
as a function of temperature T.  Data are shown for two samples, one with T$_c$ 33 K.  The data for each sample correspond to measurement frequencies of 100, 200 and 600 GHz.  Each set of curves identify a crossover from frequency-independent to frequency-dependent phase stiffness. The dashed line shows that the crossover is set by the Kosterlitz-Thouless-Berenzkii condition for two-dimensional melting.  From Ref. \onlinecite{Corson}}
\label{Corson}\end{figure}

TDTS was used for instance in the measurement of thin Bi2212 cuprate superconducting films to show evidence for phase fluctuations above T$_c$.  Corson \textit{et al. }measured transmission through thin films and using the unique phase sensitivity of TDTS could show that there was an enhanced $\sigma_2$ (Fig. \ref{Corson}) even at temperatures well above T$_c$ \cite{Corson}.  This was interpreted as a finite phase stiffness on short length and time scales.  Superconducting fluctuations as such should manifest themselves as a finite superfluid stiffness measured at finite frequency.  The imaginary part of the optical conductivity is sensitive to superfluid stiffness through the relation  $\sigma_2 = \sigma_Q \frac{k_B T_{\theta}}{\hbar \omega}$, where $\sigma_Q = \frac{4e^2}{hd} $ is the quantum of conductance for Cooper pairs divided by a length scale (typically the c-axis lattice constant) and $T_{\theta}$ is the generalized frequency dependent superfluid stiffness.  Here, the superfluid stiffness $T_{\theta}$ is an energy scale expressed in temperature units.    The stiffness is the energy scale to introduce phase twists in the superfluid order parameter $\Psi = \Delta e^{i \phi}$.

The other major development for THz spectroscopy is the increased prominence of Backward Wave Oscillators (BWOs)  Although developed in the 60's, BWOs are gaining increasing prominence in the investigation of correlated systems in the important THz range \cite{Kozlov98a}.  These BWOs are traveling wave guide tubes, capable of producing very monochromatic THz range radiation over a relatively broad range per device.  With a number of tubes, it is powerful method to cover the 0.03 THz to 1.5 THz range in a continuous wave configuration.

The longer wavelengths in this spectral region give the capabilities to measure phase of transmitted waveforms through Mach-Zender two-beam polarization interferometry.  This method also
allows direct precision measurements of both real and imaginary components of the complex optical response without resort to Kramers-Kronig transforms and their associated ambiguities.  A spectrometer based on BWOs is easily integrated with a low temperature cryostat and magnetic field system as well as glove boxes for environmental control.

\subsection{Infrared}

\begin{figure}[tb]
\includegraphics[width=7.5cm,angle=0]{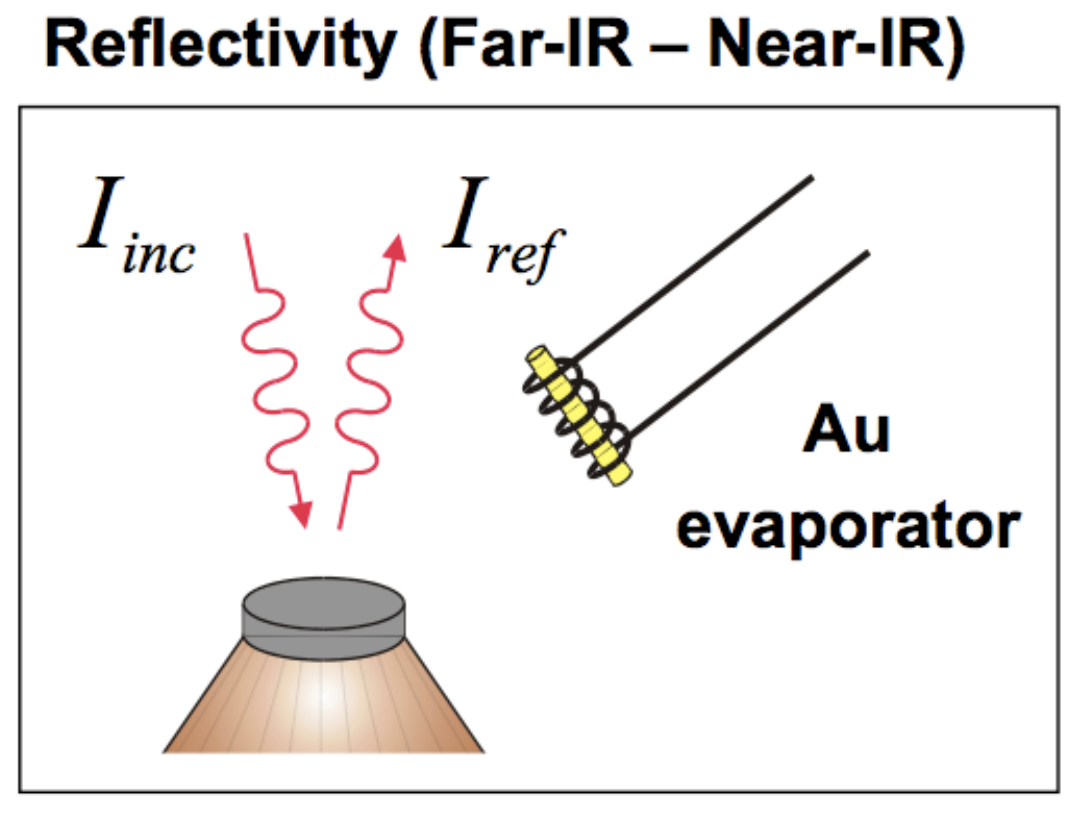}
\caption{(Color) Absolute reflectivity $R(\omega)$ in FTIR is obtained by measuring the reflected intensity and ratioing with the intensity after gold evaporation or from a reference mirror.  Figure courtesy of A. Kuzmenko. }
\label{GoldReference}\end{figure}

\begin{figure}[htb]
\begin{center}
\includegraphics[width=9cm,angle=0]{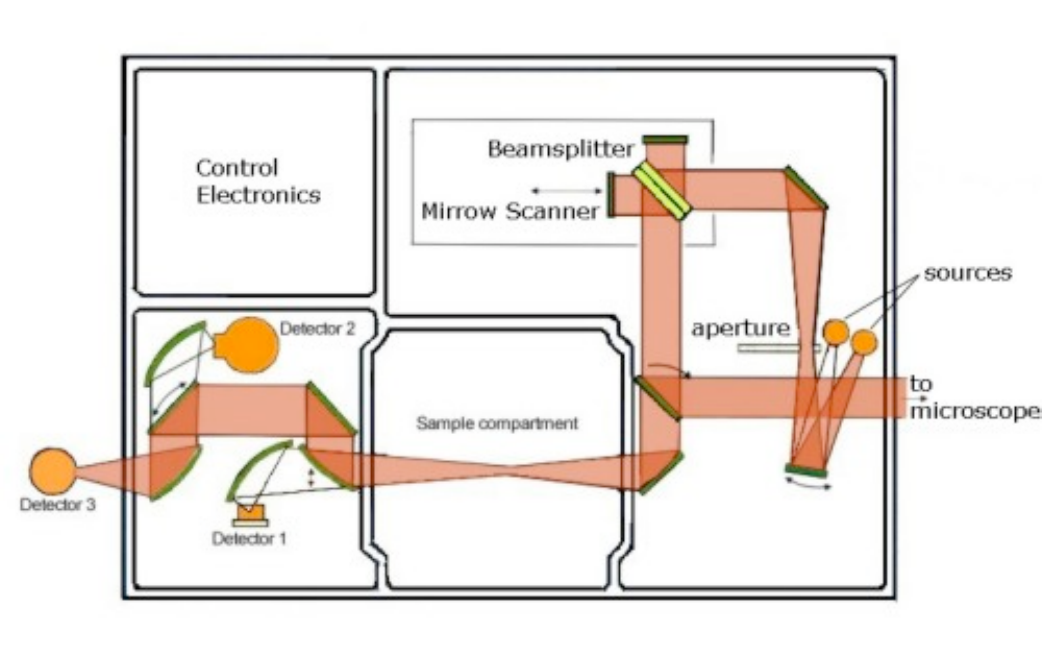}
\caption{(Color) Modified Fourier Transform Spectrometer Bruker 66.  As shown, the spectrometer is setup to perform reflection, but with additional mirrors near-normal incidence reflection is possible.  The apparatus works by shining broad band light from a variety of sources on a sample.  Reflected or transmitted intensity is measured with a few different detectors.  A beamsplitter/interferometer apparatus allows the resolution of distinct frequency components.  Figure from \texttt{http://www.pi1.physik.uni-stuttgart.de/research\\/Methoden/FTIR\_e.php}.  Figure courtesy of N. Drichko.} \label{Bruker113}
\end{center}
\end{figure}

Fourier transform infrared reflectivity (FTIR) is the workhorse spectroscopy for optical characterization of solids (Fig. \ref{Bruker113}).  Measurements are possible from $\sim$ 10 $cm^{-1}$ to approximately 25,000 $cm^{-1}$, although measurement get increasingly difficult below  50 $cm^{-1}$.  Heroic efforts can push the lower end of this range slightly below 10 $cm^{-1}$ \cite{BasovRevSci}.  Here the quantity of interest is typically the power reflectivity (transmission measurements are possible as well).  One shines broadband light, sometimes from several different sources, on a sample surface and $via$ an interference technique measures the reflected intensity.  As the source spectrum and detector response can have all kinds of frequency structure, it is necessary to reference the reflected signal against a standard sample to obtain the absolute reflectivity.  Typically the reference spectrum is chosen to be a noble metal like gold whose reflectance can to good approximation be taken to be unity over a broad frequency region below the material's plasma frequency (see below).  Typically gold is evaporated onto a sample as shown in Fig. \ref{GoldReference} or the sample is replaced with a mirror to perform this referencing.

As the materials we are interested in typically have interesting experimental signatures in both the real and imaginary response,  it is not sufficient to to compare to theory by simply measuring quantities like $absorbence$.    We want quantities like the complex conductivity or dielectric function.  It is not necessarily straightforward to obtain a complex quantity from the measured scalar magnitude of the reflectivity $R(\omega)$.  Unlike TDTS, phase is not measured in FTIR measurements.   Therefore typically use is made of the above discussed Kramers-Kronig transforms, which apply to any causal response function.  If one knows one component of the response (real or imaginary) for \textit{all} frequencies then one can determine the other component.  For reflected power, one can obtain the reflected phase as

 \begin{eqnarray}
\phi(\omega) = - \frac{1}{\pi} P \int^{\infty}_{-\infty} d \omega \frac{ln | R(\omega)|}{\omega ' -\omega}.
\end{eqnarray}

Of course the problem is that one doesn't measure over an infinite frequency range.  In typical FTIR spectroscopy, the usual mode of operation is therefore to measure over as large a frequency range as possible and then extrapolate with various schemes to $\omega \rightarrow 0$ and $\omega \rightarrow \infty$.  This method works quite well for some materials, although it can generate large errors when, for instance, reflectivities approach unity as they do in good conductors at low frequencies\footnote{See Ref. \onlinecite{Kamal06a} for a good comparison between FTIR and time-domain THz data at low frequencies on a good metal.}.   It can also be rather inefficient and cumbersome as one must measure out to the multi-eV range even if one is interested in meV level phenomena.

As mentioned, FTIR spectroscopy is a standard spectroscopic technique for the characterization of materials and chemicals.  It is used routinely to identify the presence of various chemical bonds in chemistry as in, for instance, $C=C$ or $C=O$, which all have distinct frequencies and absorptive strengths.  It is also the standard optical tool for probing materials like high-T$_c$ superconductors \cite{Basov05a}.  It has the advantage of being relatively easy to perform and possessing a very large spectra range.  It has the above mentioned disadvantage of only measuring a scalar - the power - and having a lower bound on the spectral range that is at the limit of that explicitly relevant for many correlated materials.

\subsection{Visible and Ultraviolet}

The technique of optical reflection as outlined above begins to become more difficult at frequencies on the order of the plasma frequencies of metals.  To do reflectivity measurements, one always needs a standard sample.  Standard metal references cannot be used above their plasma frequencies as their reflectivity changes quickly with frequency.  Quartz or silicon which have an approximately flat (but low) reflectivity can be used as a reference in a $grating$ $spectrometer$ \cite{Grating}.   However, more common nowadays is to use techniques such as ellipsometry to determine complex optical coefficients in this spectral range as one can measure complex response functions directly.

Ellipsometry is an optical technique for determining properties of
surfaces and thin films. Although the method was originally used
as far back as 1887 by Paul Drude to determine the dielectric
function of various metals and dielectrics, it only gained regular
use as a characterization tool in the 1970s
 \cite{Aspnes1,Aspnes2}. It is now widely used in the near infrared
(NIR) through ultraviolet (UV) frequency ranges in semiconductor
processing for dielectric and thickness characterization.

\begin{figure}[htb]
\includegraphics[width=6cm,angle=270]{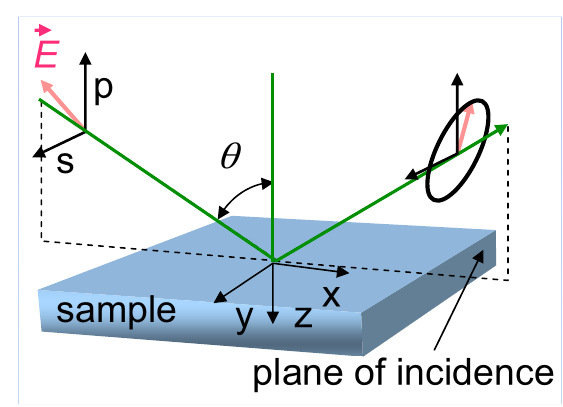}
\caption{(Color) Schematic showing the basic principle of
ellipsometry.  Glancing incidence light linearly polarized with
both $s$ and $p$ components is reflected with different
intensities and a relative phase shift. A characterization of the
resulting elliptically polarized reflected light's minor and major
axes, as well as tilt angle gives a unique contribution of the
complex optical constants of the material.}
\label{THzAPL2}\end{figure}

When linearly polarized light is reflected from a surface at
glancing incidence the in- ($p$) and out-of-plane  ($s$) of incidence
light is reflected at different intensities as well as
suffering a relative phase shift. The reflected light becomes
elliptically polarized as shown in  Fig. \ref{THzAPL2}.  The shape
and orientation of the ellipse depend on the angle of incidence,
the initial polarization direction, and of course the reflection
properties of the surface. An ellipsometer measures the change in
the light's polarization state and characterizes the $complex$
ratio $\rho$ of the in- ($r_p$) and out-of-plane  ($r_s$) of incidence
reflectivities.  The Fresnel equations allow this quantity
to be directly related to various intrinsic material parameters,
such as complex dielectric constant or layer thicknesses.

The ellipsometer itself is designed to measure the change in
polarization state of the light reflected from a surface at
glancing incidence. From a knowledge of the orientation and
polarization direction of the incident light one can calculate the
relative phase difference, $\Delta$, and the relative amplitude
difference ($\textrm{tan} \Psi$) between the two polarization
components that are introduced by reflection from the surface.

Given the $complex$ ratio $\rho = r_p/r_s = e^{i \Delta} tan \Psi $ of the in- ($r_p$) and
out-of-plane reflectivities ($r_s$), various intrinsic material
parameters can be generated $via$ the Fresnel equations.  In a homogeneous sample the complex dielectric constant $\epsilon = \epsilon_1 + i \epsilon_2$ is related to
$\rho$ as given in Eqs. \ref{Eqrho}, \ref{EqP} and \ref{EqS} below. Here as usual $\epsilon_1$
parameterizes the polarizability of a material, whereas
$\epsilon_2$ parameterizes the dissipative properties.  $\epsilon_z$ is the complex dielectric constant perpendicular to the reflection surface and $\epsilon_x$ is the complex
dielectric constant in the plane of the reflection surface, and
$\theta$ is the angle of incidence as shown in Fig. \ref{THzAPL2}.

\begin{eqnarray}
\rho = r_p / r_s = \textrm{tan} \Psi e^{i \Delta} , \label{Eqrho} \\
 r_p = \frac{\sqrt{1 - \epsilon_z^{-1} \textrm{sin} ^2 \theta}-
\sqrt{\epsilon_x} \textrm{cos} \theta }{\sqrt{1 -
\epsilon_z^{-1} \textrm{sin} ^2 \theta}+ \sqrt{\epsilon_x}
\textrm{cos} \theta  },  \label{EqP} \\ 
 r_s = \frac{\textrm{cos}
\theta - \sqrt{\epsilon_y - \textrm{sin} ^2
\theta}}{\textrm{cos} \theta + \sqrt{\epsilon_y - \textrm{sin}
^2 \theta}}. \label{EqS} 
\end{eqnarray}

\begin{figure}[htb]
\includegraphics[width=6cm,angle=90]{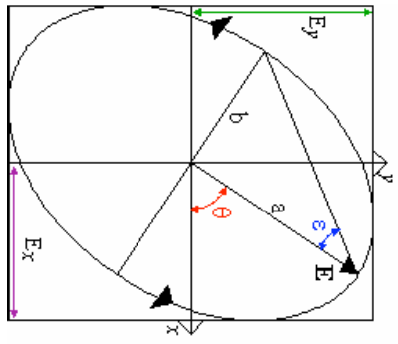}
\caption{(Color) To completely characterize the electric field
vectors and their phase, it is insufficient to measure the
projection of the ellipse along two orthogonal directions.  The
complete ellipse must be mapped out to get its major and minor
axes and tilt angle.} \label{Ellipse}\end{figure}

In a typical configuration, monochromatic light is incident on a
glancing trajectory $\theta$ close to the Brewster angle (usually
$\sim 65-85 ^\circ$ for bad metals at high frequencies) with a linear polarization state
of $45^\circ$.  As the typical detector measures power and not
electric field, it is essential that the light's elliptical
polarization is measured over at least $180^\circ$ (and more
typically  $360^\circ$).  To completely characterize the phase, it
is clearly $not$ sufficient to simply measure the projection of
the ellipse along two orthogonal directions, as can be seen by
construction in Fig. \ref{Ellipse}.  In order to get the phase
information, the ellipse's orientation and major and minor axes
must be known. Generically, the light originally linearly polarized
at $45^\circ$ is changed to an elliptical polarization with its
major axis displaced from $45^\circ$. A characterization of this
ellipse's major and minor axes, as well as tilt angle is a direct
measure of the complex amplitude reflection coefficients, which
can then be related to intrinsic material quantities.

The fact that ellipsometry measures the ratio of two
simultaneously measured values gives it several distinct
advantages as a characterization tool over simple reflectivity. It
is highly accurate and reproducible even at low intensities as
many systematic errors are divided out. Moreover, unlike
reflectivity measurements, the technique is self-normalizing and a
reference sample is not necessary. This is important as the choice
of a reference can be problematic in the optical and ultraviolet
spectral range or when the surface is of unknown quality.   The method is also particularly insusceptible
to source intensity fluctuations as they are also divided out.
Perhaps most important for the overall utility of the technique is
that the method also measures a phase, which gives additional
information regarding materials properties.  The phase information
can be used to generate the $complex$ optical constants (the complex dielectric constant for instance, or the complex index of refraction $n$ and $k$ or the complex
conductivity $\sigma$) which is essential
information for a complete characterization of a material's
optical response.  If the optical constants are known, the phase
information can be used to sensitively measure films thicknesses.   If conventional ellipsometry has a disadvantage it is that with typical grating monochromator data acquisition
(i.e. one frequency stepped at a time) it is relatively slow as
both the grating and polarizers must continue to be moved. In such
a setup one loses the rapid acquisition time of the multiple
frequency multiplexing in, for instance, Fourier transform
infrared reflectivity (FTIR)\footnote{In FIR ellipsometers one typically uses the interferometer of an FTIR spectrometer, which retains this multiplexing.}.  As mentioned above, the technique is well established and widely
used in the NIR through ultraviolet (UV) frequency
ranges \cite{Roseler,Barth}.  There have been a number of attempts to extend
ellipsometry to lower infrared frequencies
 \cite{Roseler,Barth,Bremer}.  Unfortunately the lack of sufficiently intense sources  (among other reasons) has meant that such efforts have met with limited success, although synchrotron based efforts have had some success in this regard \cite{Bernhard}. 

The energy scales of the visible and ultraviolet, which are measured by ellipsometry are not usually explicitly relevant for strongly correlated systems.  Typically we are interested in much lower energies that are on the order of the temperatures that phenomena are exhibited.    However, such energies are relevant for the determination of important band structure parameters on, say, 3D materials that photoemission cannot be performed on.  Excitations in this range also determine the high frequency dielectric constant $\epsilon_\infty$, which can have a significant (but indirect) effect of the low energy physics.  Characterization of optical constants in the visible and UV range is also imperative to constrain the phase information at low energies when performing Kramers-Kronig with reflectivity data.   An extremely powerful method for combining different data sets using a Kramers-Kronig consistent variational fitting procedure was developed by A. Kuzmenko \cite{Kuzmenko05a}.   It allows a general procedure for combining say, DC resistance, cavity microwave measurements at a few distinct frequencies, IR power reflectivity, and ellipsometry to extract over a broad energy range of all the significant frequency and temperature dependent optical properties.

\section{Examples}

\subsection{Simple Metals}

\begin{figure}[htb]
\begin{center}
\includegraphics[width=8.5cm,angle=0]{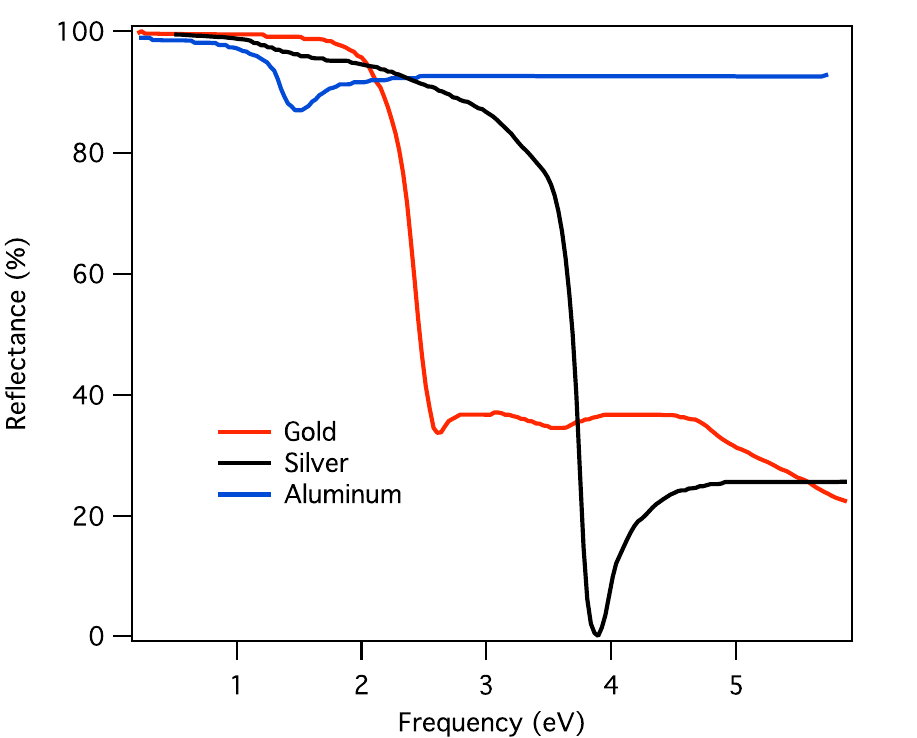}
\caption{(Color) Reflectance curves for aluminium (Al), silver (Ag), and gold (Au) metal at normal incidence as a function of incident energy.  The plasma frequencies of silver and gold are clearly seen.  The small dip in aluminum's curve around 1.65 eV is caused by an interband absorption and is not its plasma frequency.  Aluminum's plasma frequency lies at much higher energies at around 15.5 eV.  Adapted from Ref. \onlinecite{Bass94a}. }\label{MetalsRefl}
\end{center}
\end{figure}

Noble metals like silver, copper, and gold provide a good first step for the understanding the electrodynamics of solids.  Compare the reflection spectra of silver and gold (and aluminum) given in Fig. \ref{MetalsRefl} as a function of energy.  The reflection spectra of silver shows a sharp plasma edge around 3.8 eV   Gold has a much lower plasma frequency with its plasma edge at a wavelength of around 2.5 eV.  Also shown is the reflection spectra for aluminum which has a much larger plasma frequency than any of these at around 15.5 eV.  The small dip in Al around 1.65 eV is caused by an interband absorption (see below) and is not its plasma frequency.

\begin{figure}[htb]
\begin{center}
\includegraphics[width=5.5cm,angle=0]{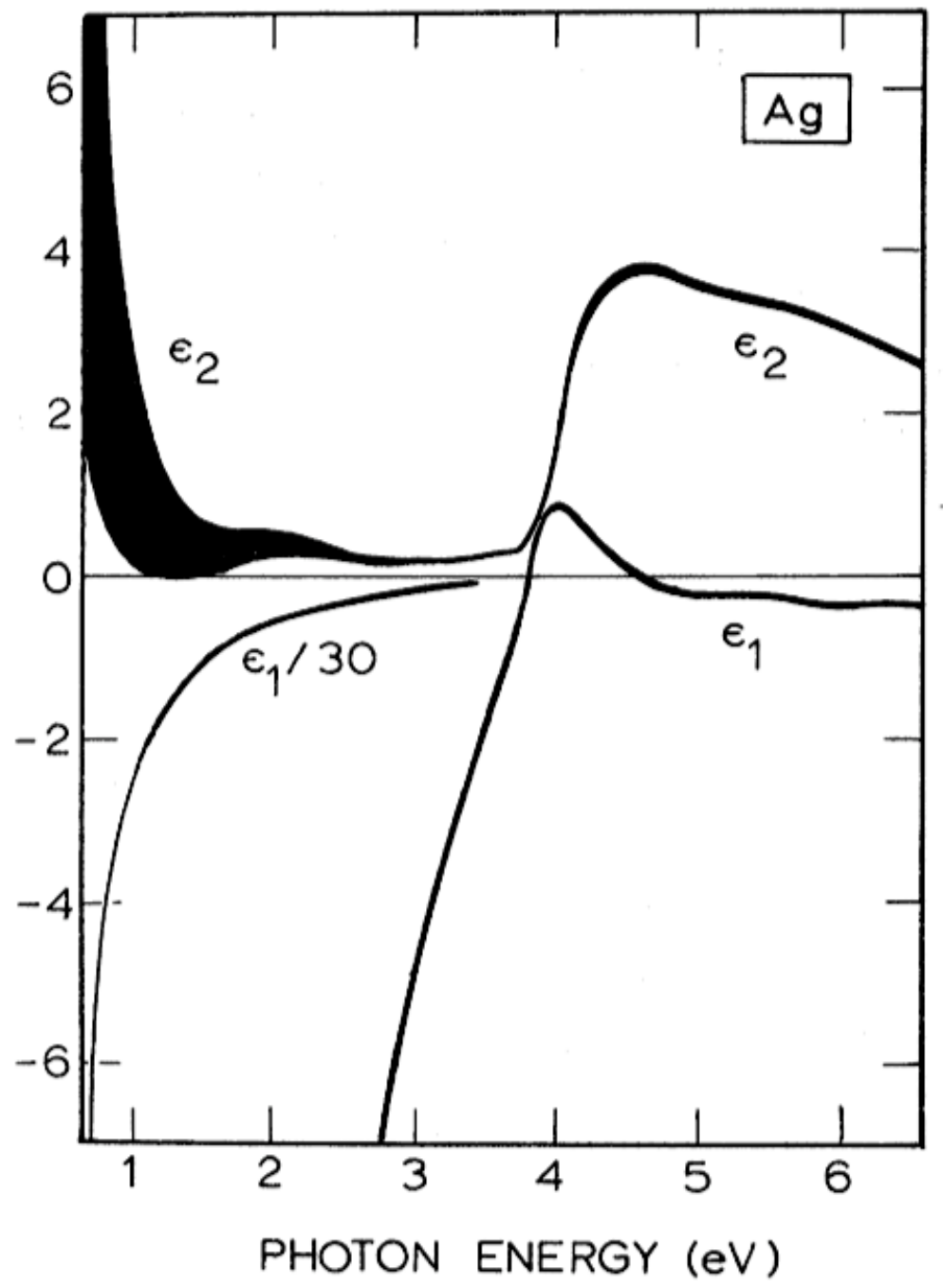}
\includegraphics[width=5.5cm,angle=0]{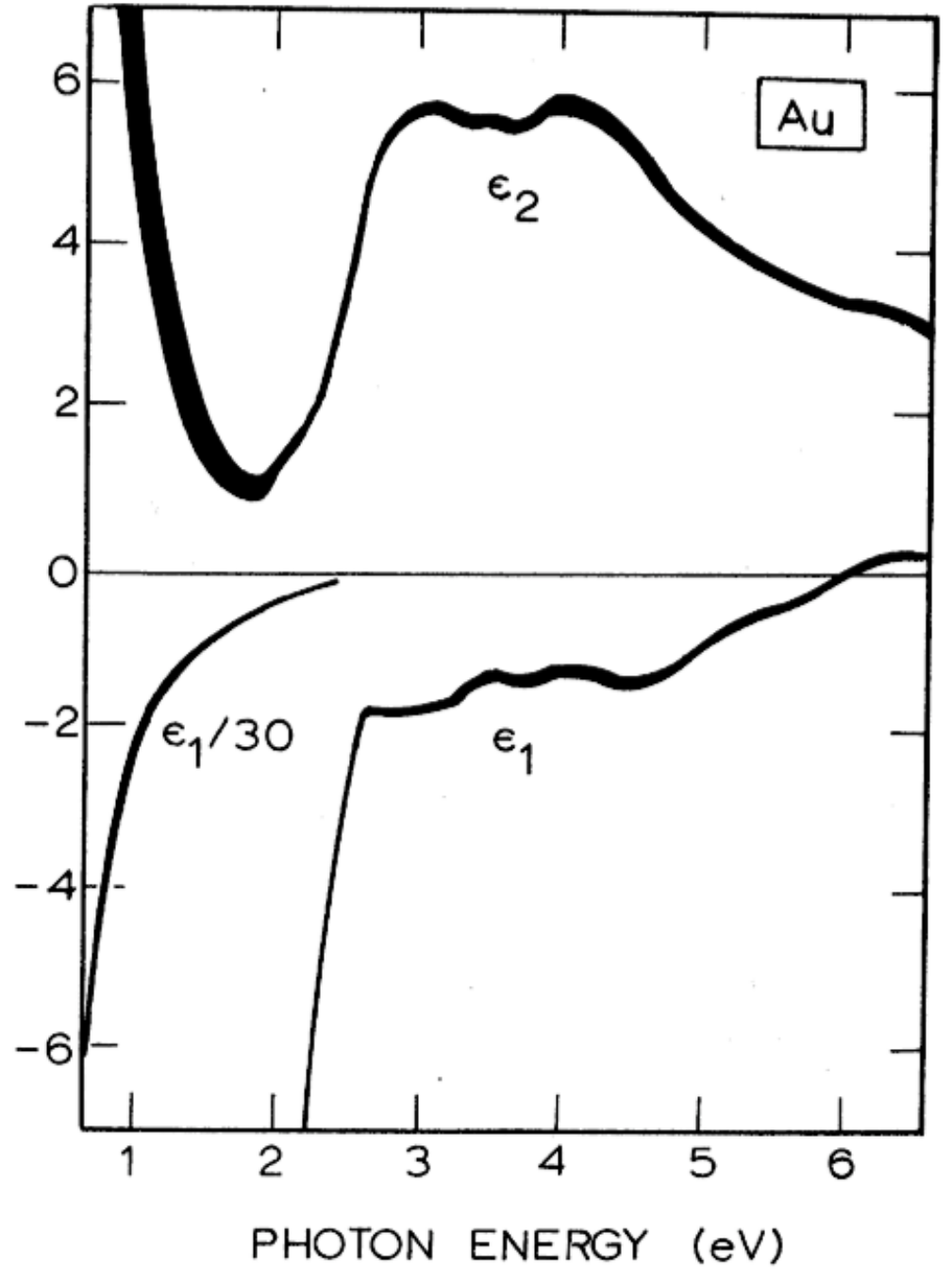}
\caption{Real and imaginary dielectric functions for silver (top) and gold (bottom).  The thickness of the curve indicates the experimental uncertainty, which originates mainly in the Kramers-Kroning transform.  From Ref. \onlinecite{Johnson72a}. }\label{NobleDielectric}
\end{center}
\end{figure}

From these spectra we can immediately see the origin for the visual difference between silver and gold.  As mentioned above, the reflection/transmission properties of metals greatly change above their plasma frequencies.  If not for the presence of interband transitions, metals would be transparent in this regime.  The lower plasma frequency of gold is what determines its yellowish color in reflection.  It reflects less in the blue/violet portion of the spectra and hence looks yellowish.  In contrast the plasma frequency of silver is in the UV range and so the reflection of silver is almost constant at constant throughout the infrared to visible range.  Aluminum's plasma frequency is higher yet still and so the major differences in its optical properties as compared to silver are invisible to the human eye.

As discussed above, within the Drude model with only free carriers the plasma frequency is set by $\omega_p = \sqrt{\frac{4 \pi N e^2}{m}}$.  This is frequency of the dip in the reflectivity, the zero crossing in $\epsilon_1$, and the frequency of free longitudinal charge oscillations (the $plasmon$ frequency).  However, the presence of interband excitations, which contribute a high frequency dielectric constant $\epsilon_\infty$ changes the situation.   In real materials, these features are exhibited at the so-called $screened$ plasma frequency set by  $\tilde{\omega_p} = \sqrt{\frac{4 \pi N e^2} {\epsilon_{\infty} m}}$.   It is interesting to note that the carrier density and masses of silver and gold are almost identical and so the difference in their plasma frequencies comes almost entirely from the renormalization effects of different interband transitions $\epsilon_\infty$.  
 
\begin{figure}[htb]
\begin{center}
\includegraphics[width=6cm,angle=0]{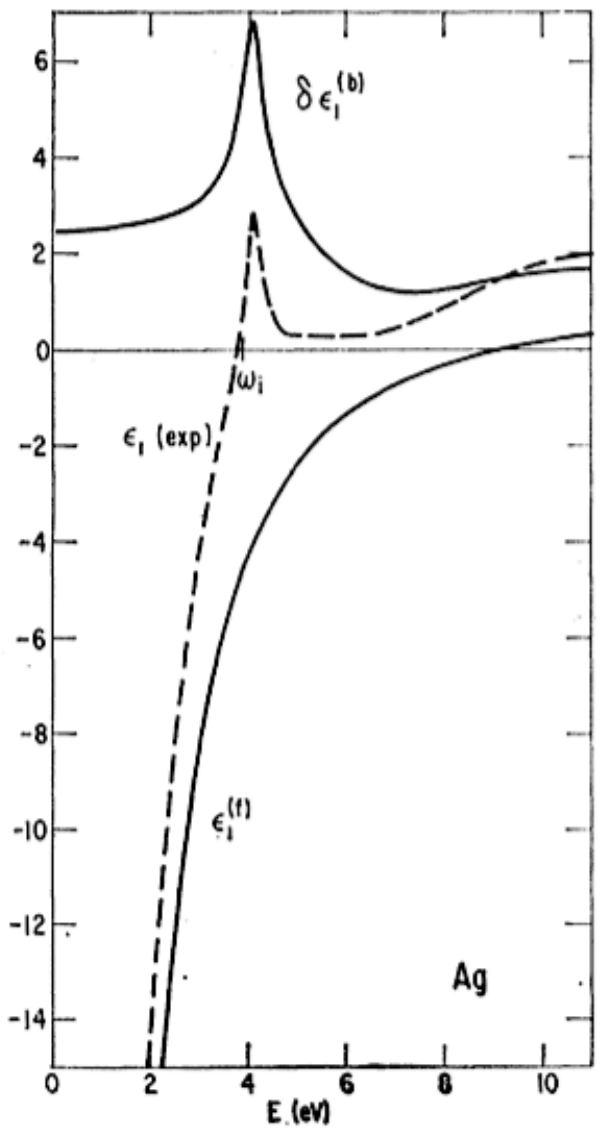}
\caption{ Decomposition of the dielectric function of silver into Drude and interband contributions. The threshold absorption for interband excitations is indicated by $\omega_i$.  $\epsilon_1^{(1)}$ is the Drude contribution.  $\delta \epsilon_1$ is the interband piece.  From Ref. \onlinecite{Ehrenreich62a}. }\label{AuDecomp}
\end{center}
\end{figure}

 Data like that shown in  Fig. \ref{MetalsRefl} can, with appropriate extrapolations to high and low frequency, be Kramers-Kronig transformed to get the real and imaginary dielectric function as shown in Fig.  \ref{NobleDielectric} for silver and gold.   As expected they both show strong almost perfect $\omega = 0$ Lorentzian Drude peaks of almost equal weight, but gold shows stronger interband absorptions, gives it a larger $\epsilon_\infty$ that renormalizes its plasma frequency\footnote{Recall from section \ref{DLsection} that a low frequency oscillator (absorption) can give a large $\epsilon_\infty$.  All other things being equal, the lower the frequency of the oscillator the higher the $\epsilon_\infty$.}.     Such spectra can be decomposed into various contributions as shown in Fig. \ref{AuDecomp} for silver.  As expected there is a strong zero frequency Drude peak contribution as well as interband contributions.  Although the interband piece shows some resemblance to the Drude-Lorentz plots in Fig. \ref{DrudeLorentz}, the shape is a more rounded cusp, which is characteristic of transitions within the manifold of d-electron states.
 
\begin{figure}[htb]
\begin{center}
\includegraphics[width=7cm,angle=0]{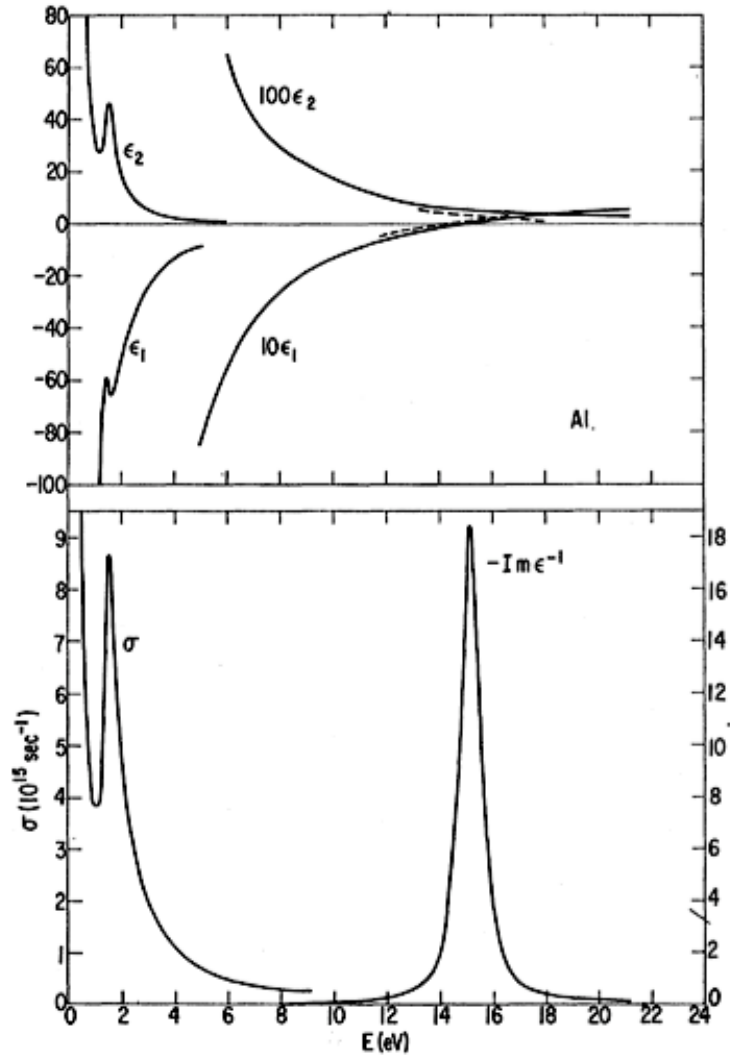}
\caption{ (top) Real and imaginary dielectric functions for aluminum.  (bottom) Real part of the conductivity and the loss function.  In addition to the Drude peak, a prominent absorption is found around 1.65.  This feature gives the dip in the reflectivity curve around the same energy.  The loss function shows a strong peak at the screened plasma frequency 15.5 eV.   From Ref. \onlinecite{Ehrenreich63a}. }\label{Aluminum}
\end{center}
\end{figure}

In Fig. \ref{Aluminum} a similar plot of the complex response functions for aluminum is shown.    In addition to the metallic $\omega = 0$ Lorentzian Drude peak,  the prominent absorption that gives the dip in the reflectivity curve around 1.65 eV can be clearly seen.  That this dip in the reflectivity does not reflect the plasma frequency can be see in the plot of the loss function Im $(- \frac{1}{\epsilon})$, which shows a strong peak only at the `screened' plasma frequency 15.5 eV (i.e. the zero cross of $\epsilon_1$.

\subsection{Semiconductors and Band Insulators}

\begin{figure}[htb]
\begin{center}
\includegraphics[width=6cm,angle=0]{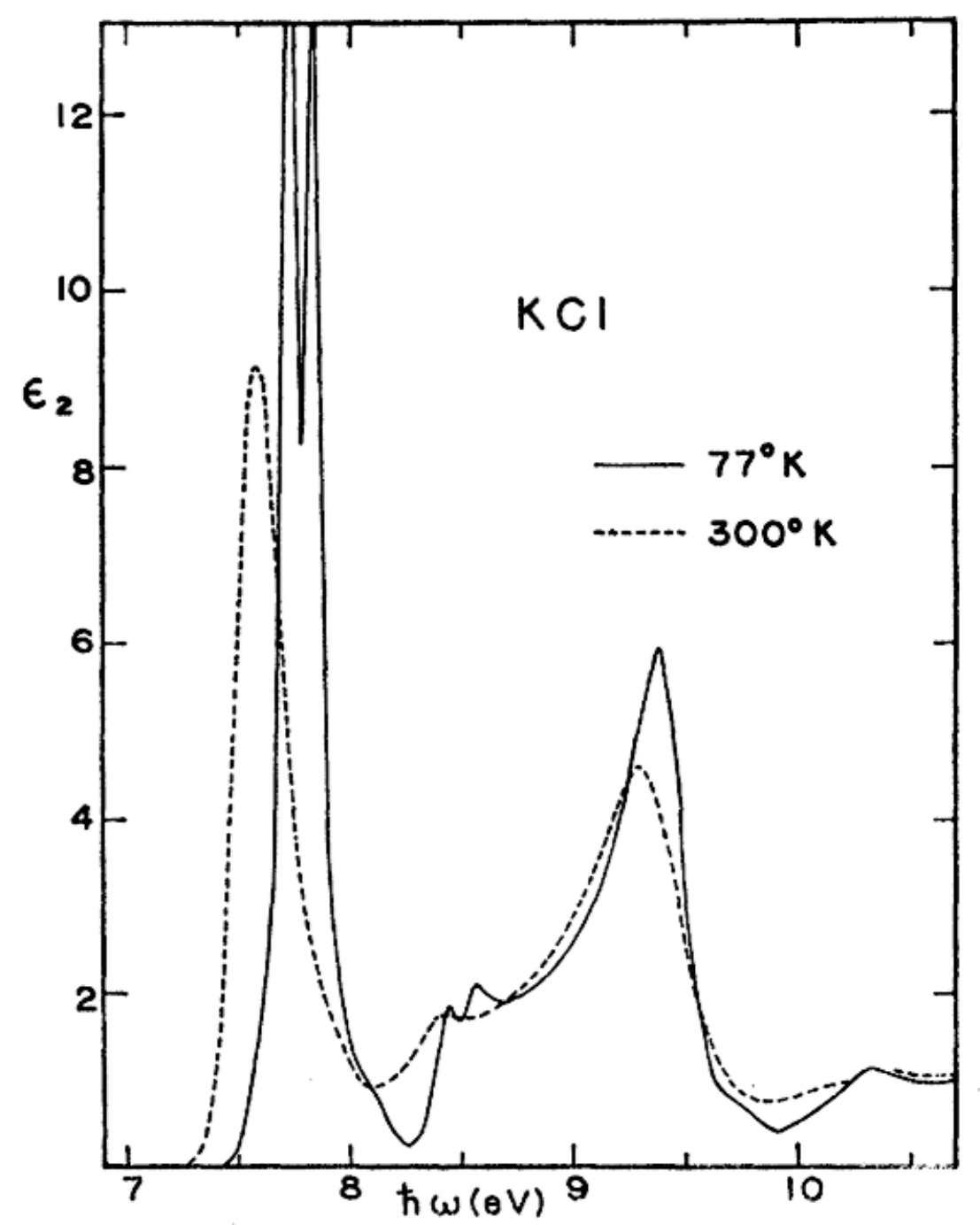}
\caption{Imaginary dielectric function of KCl. The absorption around 9 eV is the bad gap.  The series of sharp features at energies below the band gap are due to excitons.  From Ref. \onlinecite{Roessler68a}} \label{KClExcitons}
\end{center}
\end{figure}

\begin{figure}[htb]
\begin{center}
\includegraphics[width=7cm,angle=0]{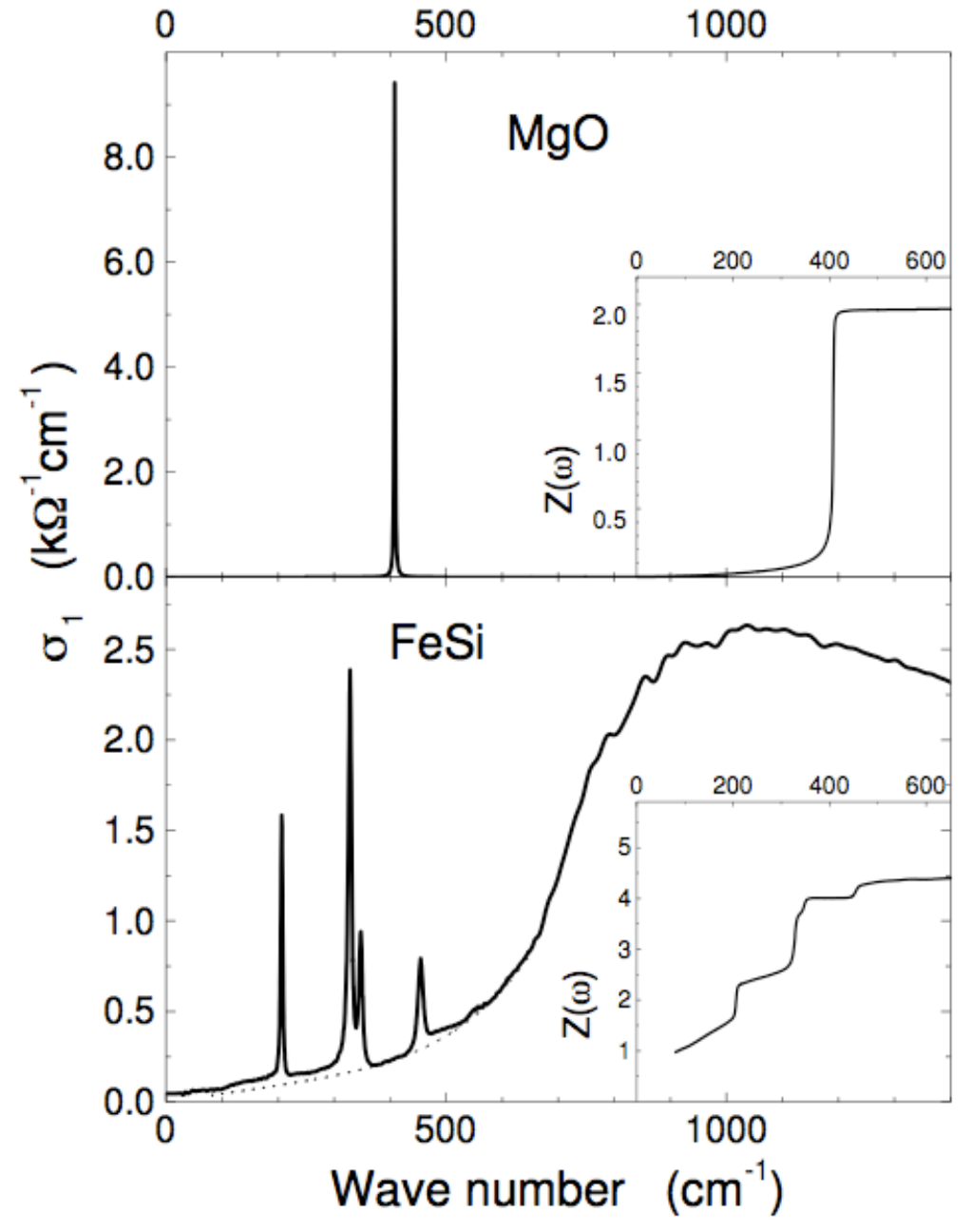}
\caption{Optical conductivityof MgO (top panel) and FeSi at T=4K (bottom panel) showing the strong phonon absorption found in such insulating compounds.  In the insets are the integrated spectral expressed in units of effective charge (See Sec. \ref{sectionsumrules}), with the electronic contribution removed.   For MgO the effective measured $transverse$ $charge$ is 1.99, which is in good agreement with the formal valences of the Mg$^{+2}$ and O$^{-2}$ ions.  From Ref. \onlinecite{Damascelli97a}.  For further discussion on effective charge see Ref. \onlinecite{vanderMarel03b}.} \label{Phonons}
\end{center}
\end{figure}

In semiconductors and band insulators, the real part of the conductivity is dominated by interband transitions (red in Fig. \ref{OpticalConduct}).  At low temperatures in a clean undoped insulator, naively one expects to see a gap with no conductivity and then a sudden onset at the gap edge.  In fact,  various other contributions are also possible inside the gap.  Electron-hole pair bound states can form as excitons and give dissipative response and be seen as sharp peaks  below the gap.   This is found for instance in Fig. \ref{KClExcitons} for the ionic insulator KCl, where  in addition to the band edge excitation there are a series of below gap exciton absorptions.  Even lower measurement frequencies than those shown would reveal a series of sharp peaks from optical phonons down in the 50 meV range.   Such behavior can be seen in Fig. \ref{Phonons} for MgO and FeSi.

\begin{figure}[htb]
\begin{center}
\includegraphics[width=7cm,angle=0]{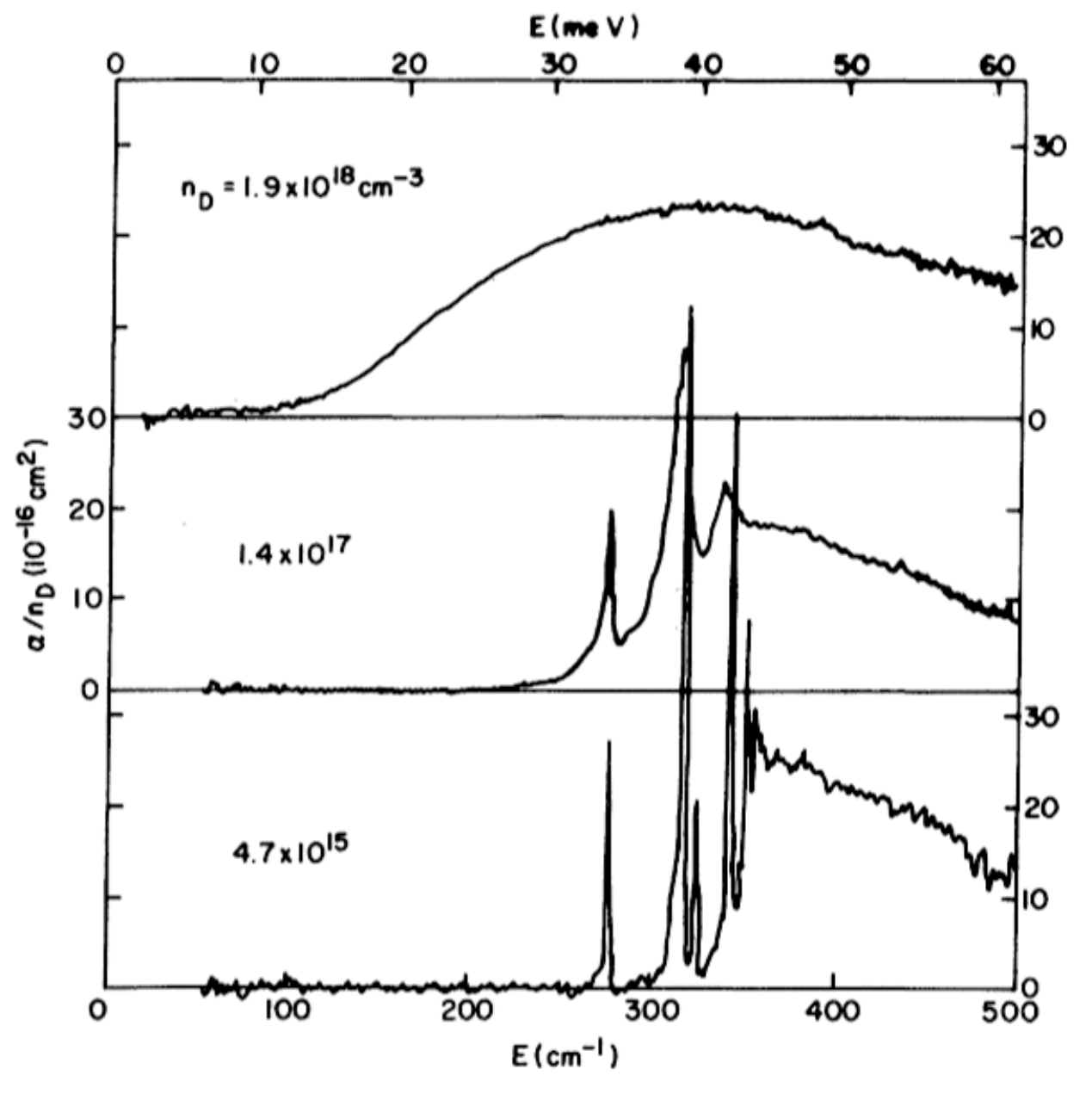}
\caption{Absorption coefficient in P doped Si normalized to the number of donors $n_D$ for the three different labeled doping levels at 2 K.  At the lowest doping, absorptions due to individual donor atoms can be seen.  These absorption broaden into a impurity band at higher dopings.  The direct gap is at much higher energies as shown in Fig. \ref{SiPLargeE}.  From Ref. \onlinecite{Thomas81a}} \label{Thomas}
\end{center}
\end{figure}

\begin{figure}[htb]
\begin{center}
\includegraphics[width=7cm,angle=0]{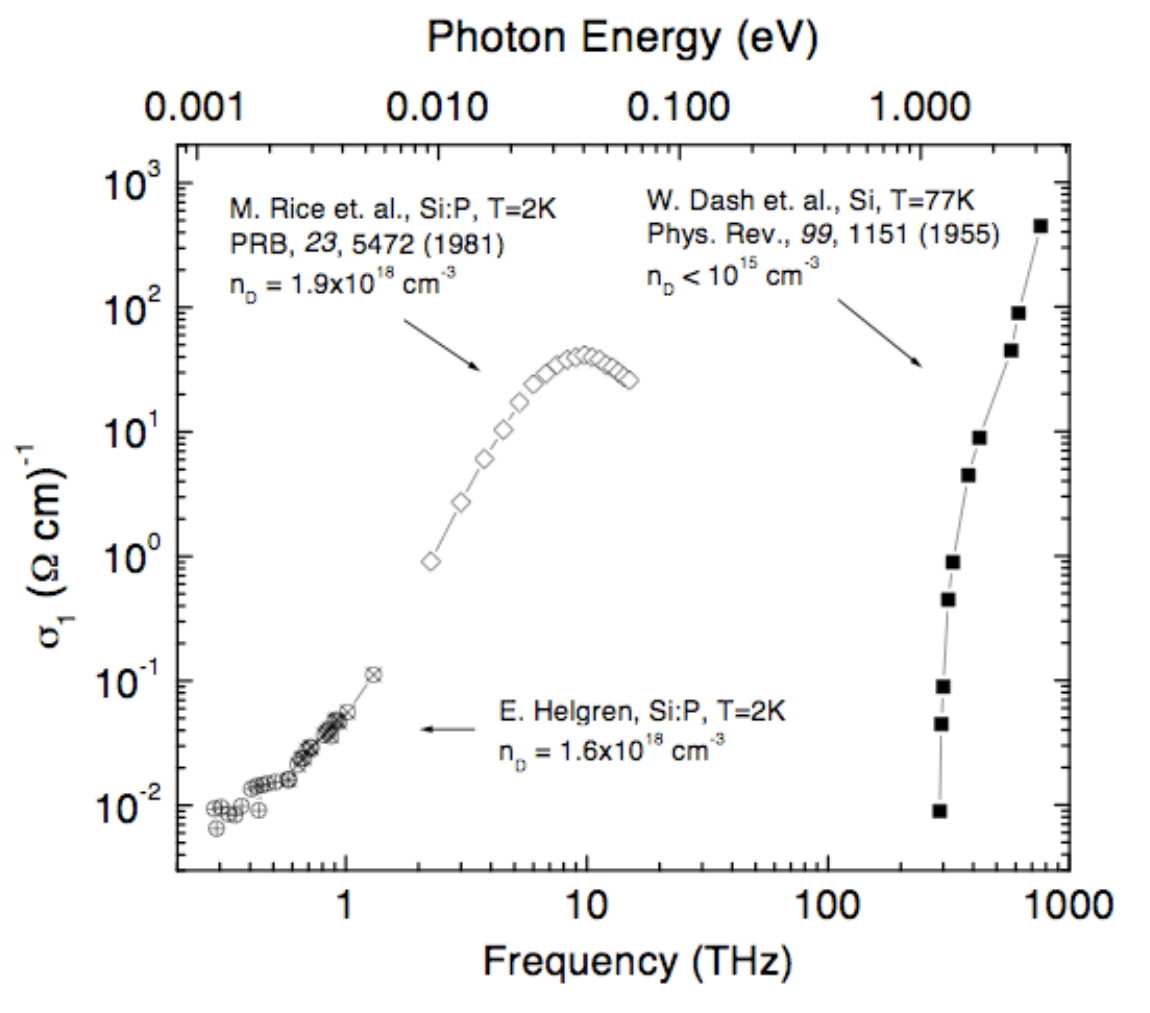}
\caption{Optical conductivity of silicon doped close to the 3D metal-insulator transition. The optical band gap is visible $via$ the high energy onset at 1.12eV. The intermediate energy conductivity is associated with excitations from the dopant band to the conduction band. The conductivity at the lowest energies, i.e. photon assisted hopping conductivity, is due to intra-dopant band excitations.  From Ref. \onlinecite{Helgren04a}.} \label{SiPLargeE}
\end{center}
\end{figure}

If one starts to dope into a typical semiconductor like phosphorus into silicon, massive changes in the optical spectra occur.  At very low doping levels one can discern the sharp absorptions of individual donors as seen in the bottom of Fig. \ref{Thomas}.  As the dopant density is increased inter-pair and inter-cluster quantum tunneling broadens the spectra until a prominent a prominent impurity band forms in the gap as shown in Fig. \ref{SiPLargeE}.  Charge transitions can occur in this manifold of impurity states.  Below a certain critical doping such a doped state retains its insulating character and is termed an $electronic$ $glass$.  I discuss this state of matter below.

\subsection{Electron glasses}

Many crystalline and amorphous semiconductors show a metal-insulator transition (MIT) as a function of dopant concentration or stoichiometry.  On the insulating side of the transition, the materials are insulators not principally for the reason of completely filled bands (as in the case of a band insulator) or interactions (as in the case of a Mott insulator), but because electronic states at the chemical potential are localized due to disorder.  Such materials which have a random spatial distribution of localized charges have been called - in analogy with structural glasses - electron glasses and are more generally termed Anderson insulators.

At finite temperatures these materials conduct under DC bias $via$ thermally activated transport of charges hopping between localized states in an impurity band.  Hence, the DC conductivity tends to zero as the temperature tends to zero.    However, because their insulating nature derives from localization of the electronic orbitals and not necessarily a vanishing density of states at E$_F$, such materials can have appreciable conductivity at low frequency deriving from transitions between localized states inside the impurity band.  Although electron transport in electronic glasses is still not completely understood, different regimes for AC transport can be distinguished depending on the strength of electron-electron interactions \footnote{There has been a lot of interest in the subject of many-body localization.   The proposal of Basko, Aleiner, and Altshuler \cite{Basko} is that a disordered interacting localized system cannot function as its own heat bath.  In other words, if the role of delocalized states like phonons could be dispensed with then such a system would be completely insulating even at finite temperature.   Signatures of these effects have been inferred in cold gases \cite{Schreiber} and it has been proposed that if coupling to phonons is weak then there still may be some signatures of the finite temperature insulating phase \cite{Rahul}.   It has been proposed the AC conductivity may hold important signatures of these effects \cite{Sarang15a}.  There is essentially no experimental work here at all looking into this in the solid-state, but the subject is tremendously interesting.   See reviews for more details \cite{Rahul2}.}, the energy scale being probed and the amount of disorder.  See Helgren \textit{\textit{et al.}} \cite{Helgren04a} for a complete discussion on this subject.

Far away from the MIT and in regimes when electron-electron interactions are not significant, one considers a so-called Fermi glass. This is an ensemble of localized charges whose properties are primarily determined by Fermi statistics.  When the long-range Coulomb interaction is of the same order as the disorder potentials, a so-called Coulomb glass emerges, and finally, near the transition to the metallic state, fluctuations lead to a quantum critical regime.  For these various regimes, certain characteristic power law and exponential functional forms for the AC and DC transport are expected.  For instance, in a Fermi glass the AC conductivity is expected to be a power law $\sigma \propto \omega^2$ \cite{MottAC}.  In a Coulomb glass the conductivity is expected to take the form $\sigma \propto U(r) \cdot \omega + \omega^2$ \cite{ESAC1,ESAC2}.  Here, the non-interacting Fermi glass functional form is returned at high frequencies, but a linear dependence is found at low frequencies.  The crossover between these regimes is smooth and set by $U(r)$ which is the typical long range interaction strength between the charges forming the absorbing resonant pair.  There are expected to be logarithmic corrections to these power laws.  As the energy scale $U(r)$ characterizes the long range Coulomb interaction, it is expected to go to zero at the MIT.   As one approaches the quantum critical point, the behavior may also be interpreted using scaling laws (discussed below) that connect the temperature, frequency and concentration dependence of the response.  It is expected that there is a correspondence between the frequency and temperature dependent conductivity on both sides of the critical concentration. Such an analysis of the conductivity leads to a universal scaling function and defines critical exponents as discussed below.

\begin{figure}[htb]
\begin{center}
\includegraphics[width=7cm,angle=0]{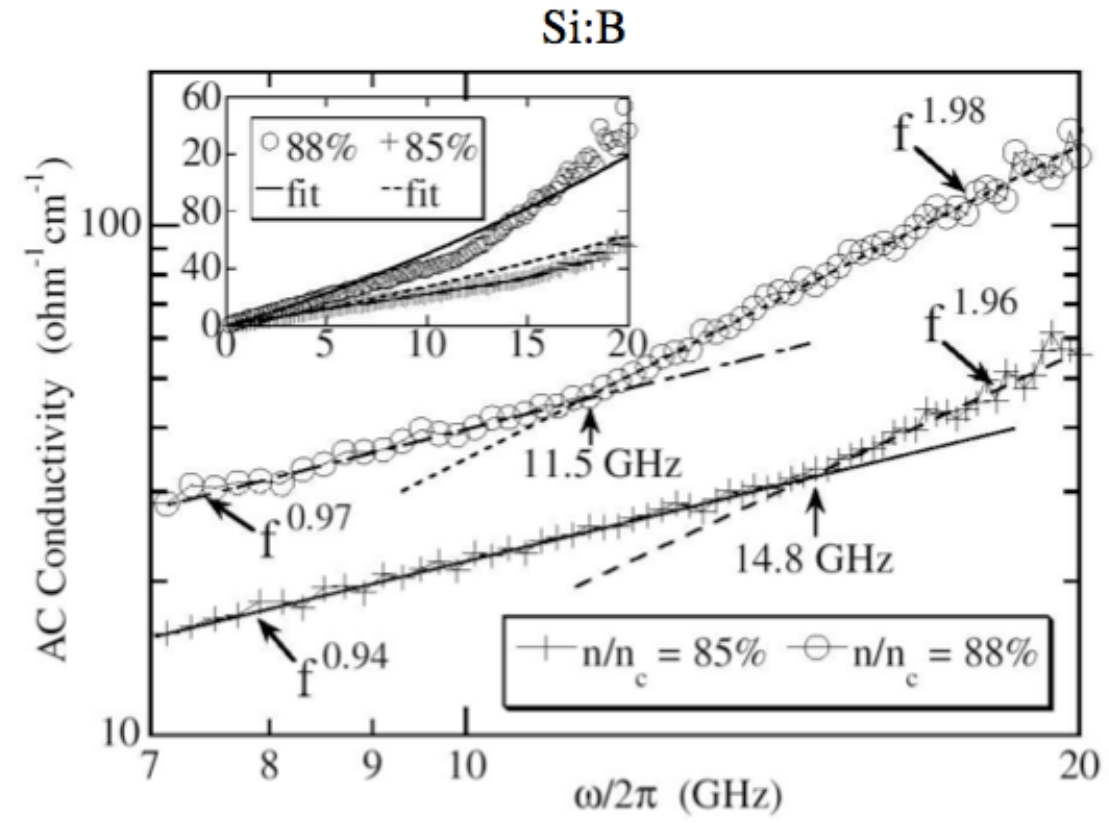}
\caption{The real part of the complex frequency dependent conductivity for two samples of Si:B at 85\% and 88\% of the way to metal-insulator transition \cite{MarkLeePRL}.  Data has been measured up to 20 GHz with a Corbino geometry microwave probe.  The crossover in the frequency dependent conductivity from linear to quadratic is qualitatively consistent with a crossover from Coulomb glass
to Fermi glass behavior.} \label{LeeandStutzmann}
\end{center}
\end{figure}

In Fig. \ref{LeeandStutzmann}, the pioneering data of Stutzmann and Lee \cite{MarkLeePRL} on Si:B  demonstrates the real part of the conductivity measured with a broad band microwave Corbino spectrometer.  The data shows a remarkable observation of conductivity linear in frequency at low frequencies and a sharp crossover at $U(r)$ to $\omega^2$ at high frequencies.   Work on Si:P on samples farther from the metal-insulator transition shows the same behavior \cite{Helgren02a,Helgren04a}, albeit the energy scale of the crossover is found at higher energies as it is set by the long range Coulomb interactions (which increases away from the MIT).    Although the proposed functional form of Efros and Shklovskii approximately fits the data of Stutzmann and Lee \cite{MarkLeePRL} (and of Helgren \textit{et al.} \cite{Helgren02a}), it fails to account for the very sharp crossover between power laws.  Stutzmann and Lee \cite{MarkLeePRL} interpreted the sharp crossover as deriving from a sharp feature in the density of states - the Coulomb gap (a depression at the Fermi energy in the density of states that is a consequence of the long range Coulomb interaction).  However, such an interpretation is at odds with theory and moreover, was not consistent with the work of Helgren \textit{et al.} \cite{Helgren02a}, where it was shown that the crossover was set by the typical interaction strength.  The sharp transition between power laws remains unexplained; although one may consider that this sharp crossover may arise from collective effects in electron motion \cite{RNBhatt}.

The theoretical functional form and experimental observation is remarkable, because the ordering of regimes is at odds with the typical situation in solid state physics.  Typically it is that non-interacting functional forms are expected at low frequency and temperature, while the explicit effects of interactions are seen at higher frequencies.  The canonical example of this ordering is the Fermi liquid, where it is the low frequency spectra which can be typically modeled in terms of quasi-free non-interacting particles (perhaps with renormalized masses).  The higher energy spectra are typically more complicated and show the non-trivial effects of electron-electron and electron-boson interactions.  Similar physics is exhibited in the heavy-fermion compounds in that it is the low frequency regime that can be understood in terms of conventional Boltzmann transport of heavy electrons.  In contrast, in electron glasses, it is the low frequency spectra which shows the effect of interactions while the high frequency spectra returns the non-interacting functional form.

This was not the original expectation.  Anderson originally coined the term ``Fermi glass" in analogy with the Fermi liquid, to describe a ensemble of localized charges whose properties where largely determined by Fermi statistics alone \cite{Anderson70a}.

\medskip

\texttt{"(The) Fermi liquid theorem is a rigorous consequence of the exclusion principle, it happens because the phase space available for real interactions decreases so rapidly (as E$^2$ or T$^2$).  The theorem is equally true for the localized case: at sufficiently  low temperatures or frequencies the non-interacting theory must be  correct, even though the interactions are not particularly small or  short range: thus the non-interacting theory is physically correct: the electrons can form a Fermi glass."}

\medskip

\texttt{- P.W. Anderson,  1970}

\medskip

It is surprising that even in a material as thoroughly studied as doped bulk silicon that there exists no clear consensus as to the ground state and the nature of the low energy excitations at low dopings.   It appears however, that Anderson's speculation is not borne out by the experimental situation.  The low lying excitations appear fundamentally changed by interactions.  Of course, existing experiments do not preclude that at even lower energies the Coulomb glass-like behavior will not break down or give way to a different response like a Fermi glass.

\subsection{Mott insulators}

Mott insulators differ from band insulators and Anderson insulators in that their insulating effects derive from correlations and not from filled bands or disorder-driven localization.  One expects to see very interesting aspects of spectral weight transfer from high energies to low as charges are doped into them and they eventually become metals.

At half-filling it is believed that such Mott insulators are characterized by upper and lower Hubbard bands, which are split by an energy $U$ that is the energy to doubly occupy a single site.  If one wants to occupy a site doubly, one must pay this energy cost $U$.  As will be discussed below the spectral weight $\int \sigma(\omega) d \omega $ is a conserved quantity which depends on the total amount of charge in a system.  It is interesting to account for how spectral weight is transfered from high to low energies by doping electrons into such a system \cite{Meinders93a}.    

\begin{figure}[htb]
\begin{center}
\includegraphics[width=8.5cm,angle=0]{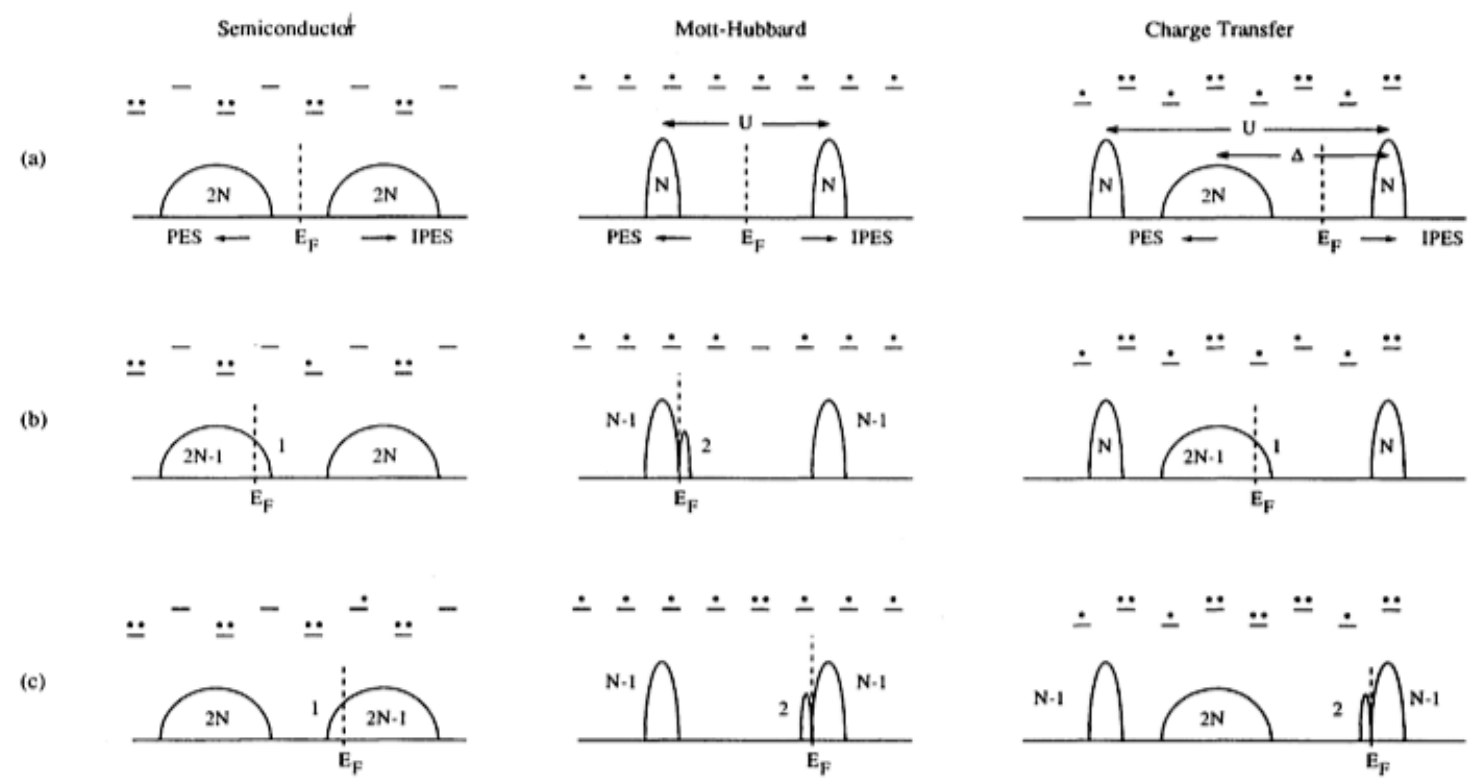}
\caption{ A schematic of the electron-removal and addition spectra for a simple semiconductor (left), a Mott-Hubbard system in the localized limit (middle) and a charge transfer system in the localized limit.  (a) Undoped (half-filled)  (b) one-electron doped, and (c) one-hole doped.  The onsite repulsion $U$ and the charge-transfer energy $\Delta$ are indicated.  From Ref. \onlinecite{Meinders93a}. } \label{SWTransfer}
\end{center}
\end{figure}

At half-filling, the lower Hubbard band is filled (occupied) and the upper Hubbard band is unfilled.  Since every site in the system has one electron, the spectral weight of upper and lower Hubbard bands are equal.  Now consider the situation after one dopes a single electron to the upper Hubbard band.  The $unoccupied$ spectral weight of the upper Hubbard band reflects the total number of sites which can have an electron added, the lower Hubbard band reflects the total number of sites $occupied$ by a single electron,  and the $occupied$ weight of the upper Hubbard band reflects the total number of sites that are doubly occupied.  These considerations then mean that doping a single electron transfers one state each from both upper and lower Hubbard bands as shown in Fig. \ref{SWTransfer} because doping changes the number of sites which are occupied/unoccupied.  It is natural to expect that a state is removed from the $unoccupied$ part of the upper Hubbard band as now a state is $occupied$.  However a state is also removed from the $occupied$ part of the lower Hubbard band as its spectral weight quantifies how many electrons sit on singly $occupied$ sites (and hence how many can be removed from singly $occupied$ sites).    The $occupied$ weight of the upper Hubbard band is then `2', and the spectral weights of $occupied$ lower Hubbard band and $unoccupied$ upper Hubbard band are both N-1.  The situation differs from doping a band semiconductor because then doping a single electron into the upper band, only takes that single state from the upper band.  There is no transfer of spectral weight across the gap.  An intermediate case is expected for the charge-transfer insulators, which the parent compounds of the high T$_c$ superconductors are believed to be.  In these compounds a charge transfer band (consisting primarily of $O$ $2p$ states) sits in the gap between upper and lower Hubbard bands and plays the role of an effective lower Hubbard band.  The energy to transfer charge from from an oxygen to copper $\Delta$ becomes the effective onsite Hubbard $U$.  Issues of spectral weight transfer in Hubbard and charge transfer insulators are discussed more completely in Ref. \onlinecite{Meinders93a}.

\begin{figure}[htb]
\begin{center}
\includegraphics[width=8cm,angle=0]{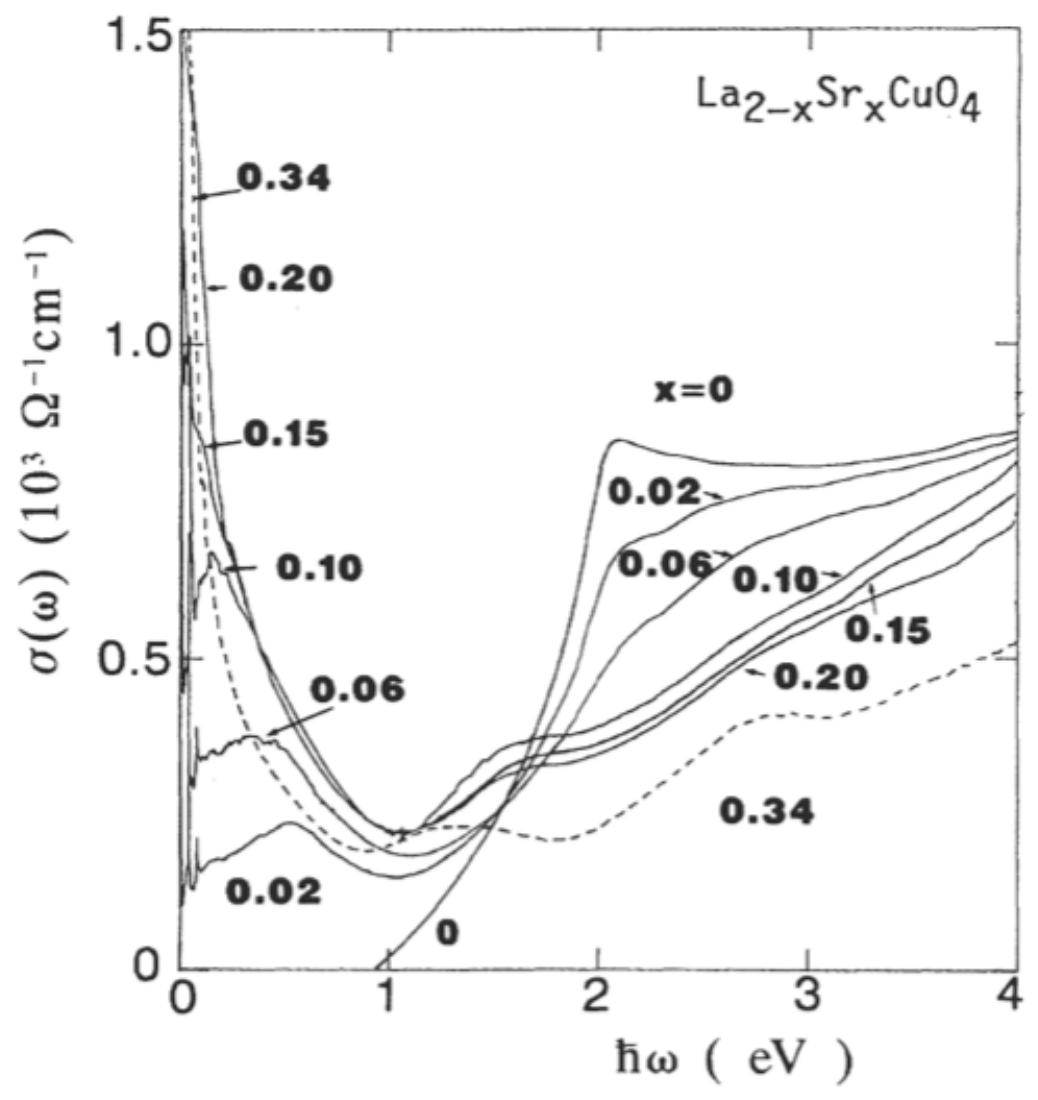}
\caption{Doping dependence of the room temperature optical conductivity of La$_{2-x}$Sr$_x$CuO$_4$.  At $x=0$ the spectra show a clear charge transfer gap of about 1.8 eV.  Upon doping with holes, spectral weight is observed to move from the high energies to low in a manner consistent with Fig. \ref{SWTransfer}.   Figure adapted from \cite{Uchida91a}.} \label{LSCO}.
\end{center}
\end{figure}

In Fig. \ref{LSCO} we show the doping dependence of the room temperature optical conductivity of La$_{2-x}$Sr$_x$CuO$_4$.  At $x=0$ the spectra show a clear charge transfer gap of about 1.8 eV.  This largely reflects excitations from O $2p^6$ states to Cu $3d^{10}$.  Upon doping with holes, spectral weight is observed to move from the high energies to low in a manner consistent with the picture for the charge transfer/Mott Hubbard insulators in Fig. \ref{SWTransfer}.   It is interesting to note that a remnant of the charge transfer band remains even for samples that have become superconductors and are known to have a large Fermi surface.

\subsection{Superconductors and other BCS-like states}

One can get a rough intuition of the electrodynamic response of superconductors for $\hbar \omega \ll  2 \Delta$ and T$\ll$T$_c$ from the relations in  Eq. \ref{zerodissp}, which give the dissipationless limit for the Drude model.  One expects a $\delta$ function peak at zero frequency in the real conductivity and a $1/\omega$ dependence in the imaginary part, whose coefficient is set by the strength of the  $\delta$  function.  At frequencies on the order of the superconducting gap $2 \Delta$, Cooper pairs can be broken and enhanced absorption should be found.  At higher temperatures, thermally excited quasiparticles are created and one expects that as they are subject to essentially normal state dissipative processes, they will give a contribution to $\sigma_1$ at finite $\omega$.

Very roughly one expects a two fluid scenario where the conductivity is approximately given by  $\sigma(\omega) = \frac{\pi}{2} \frac{Ne^2}{m} \delta(\omega = 0) + i  \frac{N(T) e^2}{m \omega}  + \sigma_{1n}(\omega,T) $ where $\sigma_{1n}$ is a normal fluid component.   At low temperatures, one expects that the superfluid density $N(T)$ is degraded as $1 - e^{- \Delta/ k _B T}$, as Cooper pairs thermally disassociate.  As the superfluid density decreases the normal fluid component increases and one may in some circumstances expect an increase in  $ \sigma_{1n}$ to go as $ e^{- \Delta/ k _B T}$.

These above considerations are only approximate however.  The explicit temperature dependence of the gap and the mutual screening effects of the superfluid and normal fluid must be taken into account.  Among other things, this means that the functional form of the contribution from `normal' electrons $\sigma_n$, although still peaked at $\omega = 0$, will be decidedly non-Lorentzian.  Within the context of the BCS theory the response of a superconductor can be calculated from the Mattis-Bardeen formalism.  I will not go into the details of the calculation here.  Interested readers should consult Tinkham \cite{Tinkham}.    One of the important results from this formalism however is the effect of the superconducting coherence factors, which depending on their sign can lead to enhanced or suppressed absorption over the normal state above and/or below the gap edge.  This means that different symmetries of the superconducting order parameter (or order parameters of other BCS-like condensates like spin- or charge-density waves), leave their signatures on the dissipative response.

\begin{figure}[htb]
\begin{center}
\includegraphics[width=7.5cm,angle=0]{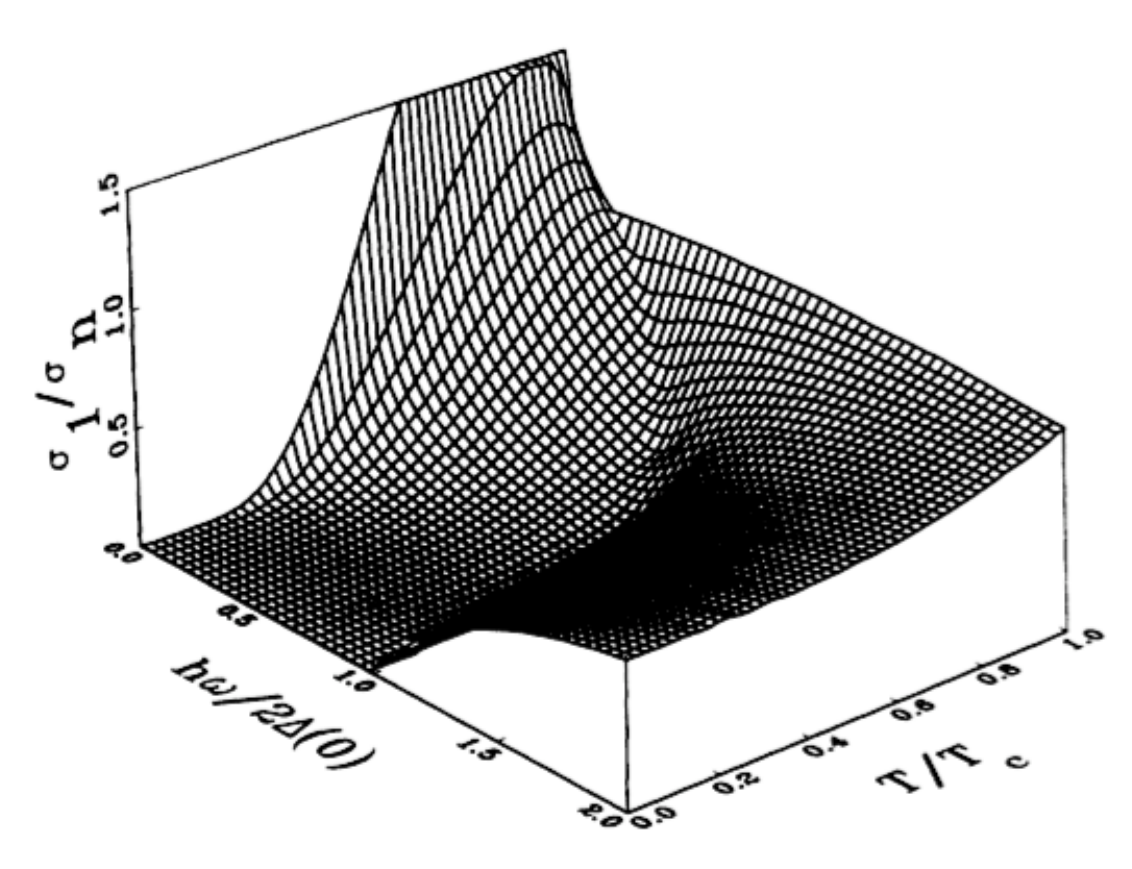}
\caption{  Temperature and frequency dependence of the conductivity $\sigma_1$ as evaluated from the Mattis-Bardeen expression.  The coherence peak exists only at low frequencies $\hbar \omega/ 2 \Delta < 0.1$.  The cusp in the surface corresponds to the energy gap.  From Ref. \onlinecite{Klein94a}.} \label{MBplot}
\end{center}
\end{figure}

Shown in Fig. \ref{MBplot} is the temperature and frequency dependence of the real part of the conductivity $\sigma_1$ for $\omega > 0$ as evaluated from the Mattis-Bardeen expression for `type II' coherence factors, which are appropriate for s-wave superconductivity.  One can see that at T=0, one has no below gap absorption, and then a gentle rise begins starting at the gap edge $2 \Delta$.  Despite the singularity in the density of states,  the conductivity is suppressed at the gap edge, as the type II coherence factors effectively cancel the enhanced density of states.  As one warms the sample, below gap excitations are possible from thermally excited quasiparticles.  Just as above the gap edge the absorption is suppressed, the absorption is enhanced over the normal state at energies below the gap.   This is again the consequence of type II coherence factors.

This anomalous behavior is more clear along the temperature axis.  If one follows the conductivity at a particular frequency as a function of temperature, one will see an enhanced absorbence at temperatures below $T_c$.  This result although surprising is direct evidence for coherence effects in the superconductor.  An analogous enhancement found in the nuclear relaxation rate $via$ NMR - the so-called Hebel-Schlichter peak -  was early evidence for the BCS theory \cite{Hebel57a}.

\begin{figure}[htb]
\begin{center}
\includegraphics[width=8cm,angle=0]{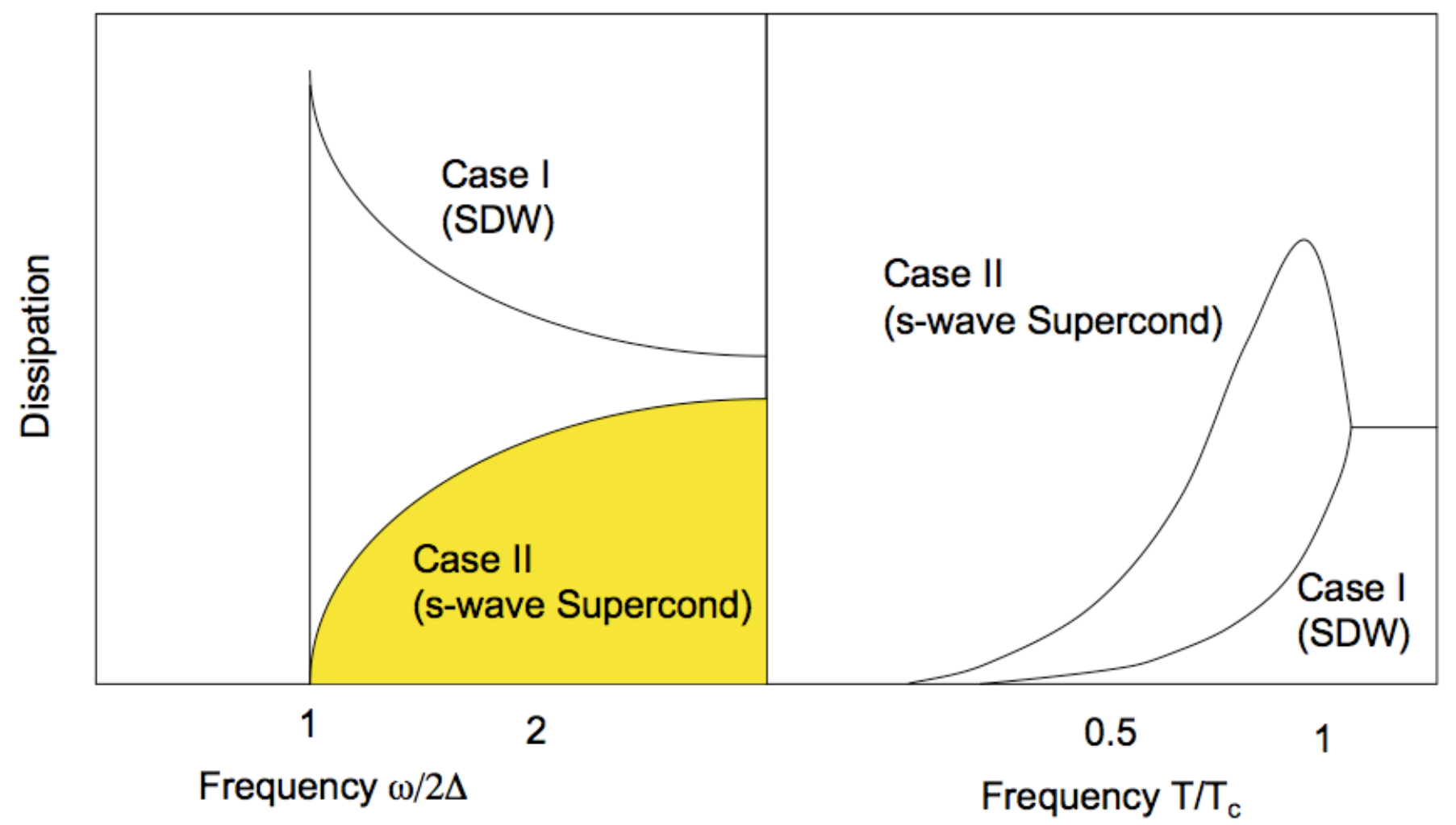}
\caption{(Color)  (left) Approximate frequency dependence of the dissipation rate from the Mattis-Bardeen theory for Type I and Type II coherence factors. (right)  Approximate temperature dependence of the dissipation rate for $\omega =   0.1 \Delta$ for both coherence factors.   } \label{CoherenceFactors}
\end{center}
\end{figure}

These features contrast with the situation in Type I coherence factors as shown in Fig. \ref{CoherenceFactors}.  In a Type I coherence factor SDW material one expects a very different response.  There the coherence factors don't cancel the singularity in the density of states.  Likewise, there is no Hebel-Slichter-like peak in the temperature dependence.

\begin{figure}[htb]
\begin{center}
\includegraphics[width=6cm,angle=0]{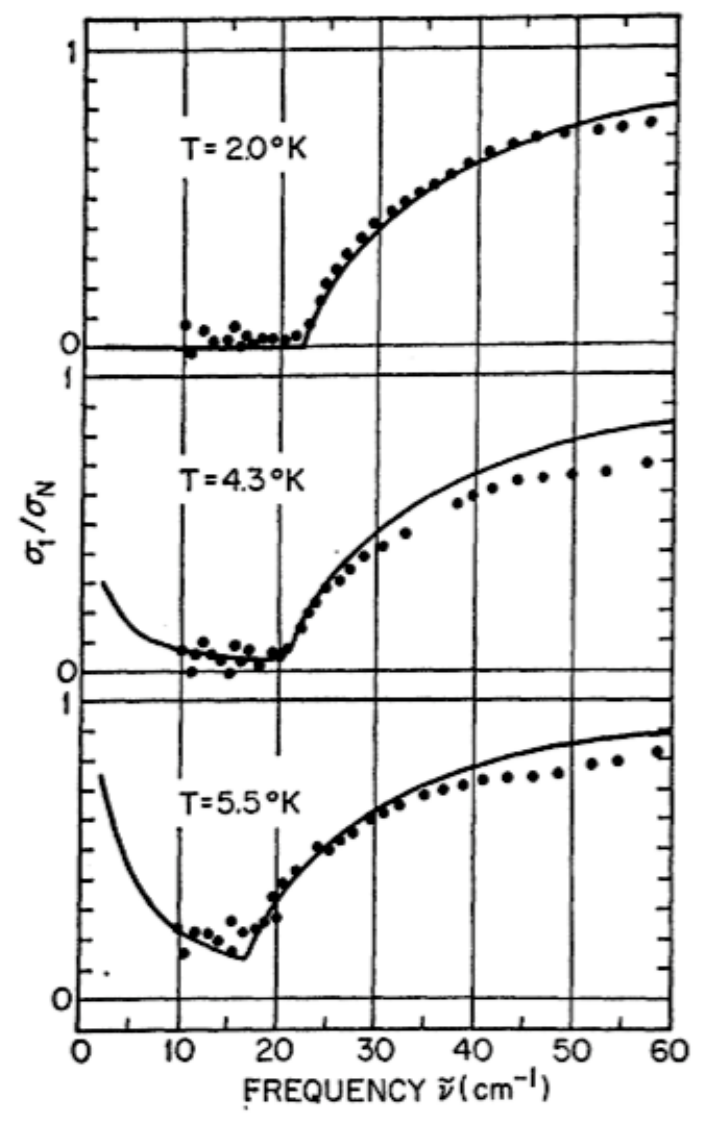}
\caption{Frequency dependence of the normalized conductivity through lead films at three different temperatures.  The results are obtained through a combination of reflection and transmission measurements.  The solid line is a calculation using the Mattis-Bardeen theory with a $2 \Delta/ \hbar$ =  22.5 cm$^{-1} $.  T$_c$ for these lead films is 7.2 $\pm$ 0.2 K.  From Ref.  \onlinecite{Palmer68a}.  } \label{PbTransmission}
\end{center}
\end{figure}

Qualitatively the behavior for Type II coherence factors is realized in conventional $s$-wave superconductors.  In Fig. \ref{PbTransmission}, the normal state normalized conductivity from some of the original measurements of Palmer and Tinkham \cite{Palmer68a} is shown, which is in good agreement with the Mattis-Bardeen prediction.  The suppression at the gap edge is consistent with type II coherence factors and hence a s-wave superconducting state.  Mattis-Bardeen predicts that subgap absorptions rise with increasing temperature and become enhanced over the normal state conductivity.

\begin{figure}[htb]
\begin{center}
\includegraphics[width=7cm,angle=0]{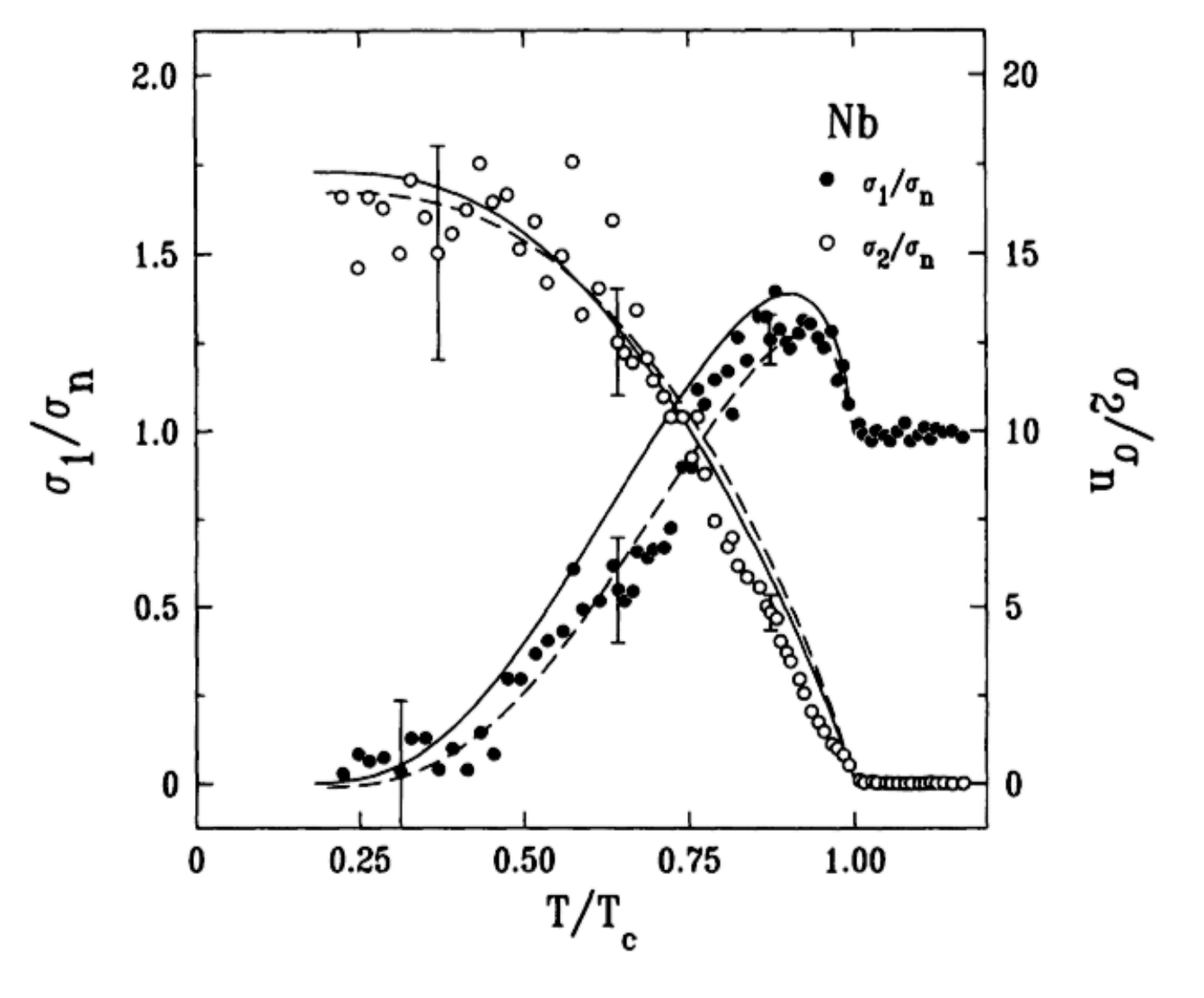}
\caption{ Temperature dependence of the complex conductivity $\sigma(\omega)$ of Nb as evaluated from surface impedance measurements at 60 GHz in cavities.  The solid curve is the weak coupling Mattis-Bardeen prediction.  The dashed curve is a strong coupling Eliashberg prediction.  From Ref.  \onlinecite{Klein94a}. } \label{NbSurface}
\end{center}
\end{figure}

In Fig. \ref{NbSurface} the conductivity of Nb at 60 GHz as measured in a microwave resonance cavity is shown as a function of temperature \cite{Klein94a}.  The peak below T$_c$ is the electromagnetic equivalent of the Hebel-Slichter peak in NMR.  This is again consistent with s-wave superconductivity.

\begin{figure}[htb]
\begin{center}
\includegraphics[width=6cm,angle=0]{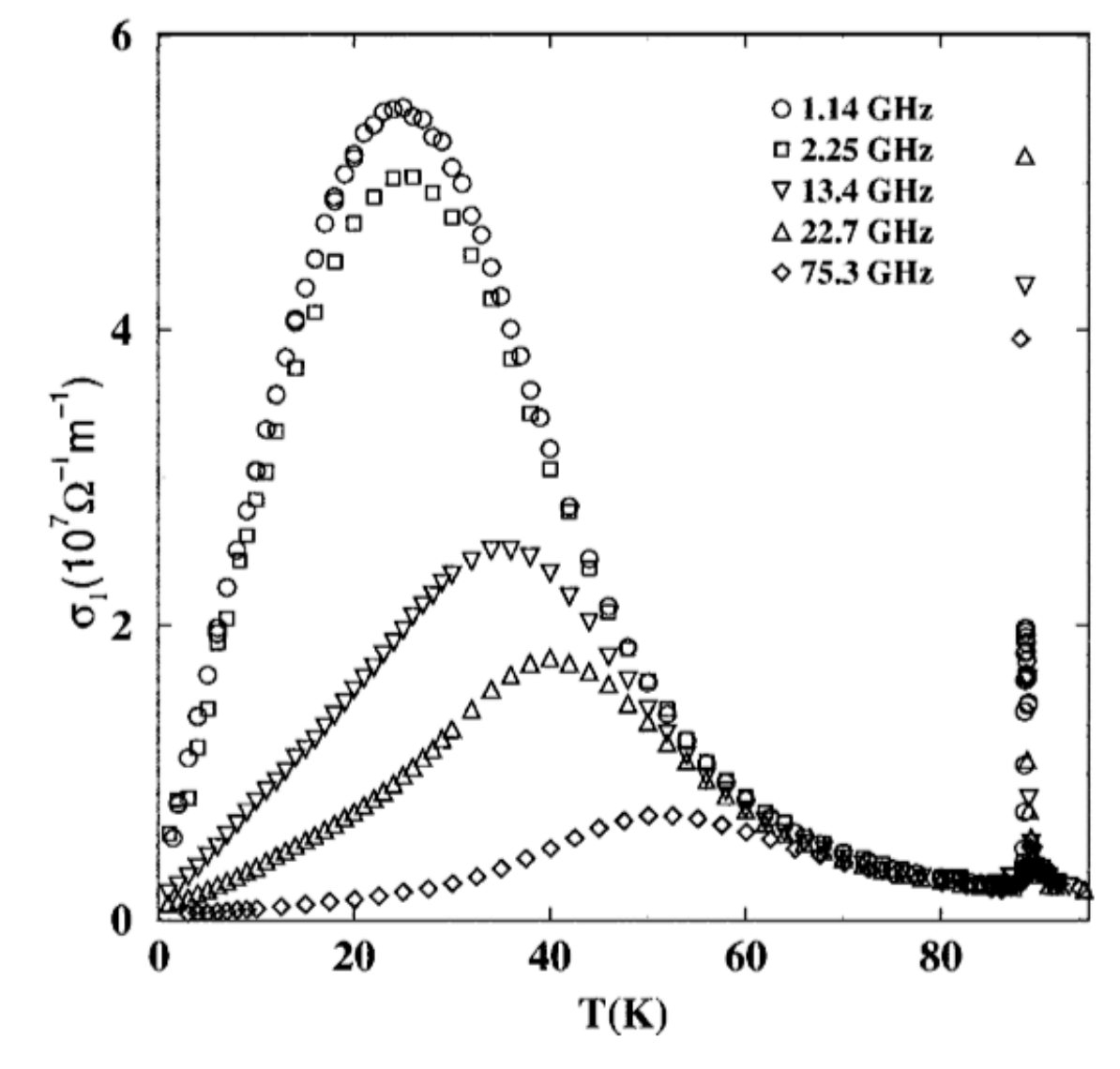}
\caption{ The real part of the conductivity extracted from the microwave surface resistance of YBCO at a number of different GHz frequencies.  From \cite{Hosseini99a}.   } \label{YBCO2}  
\end{center}
\end{figure}

\begin{figure}[htb]
\begin{center}
\includegraphics[width=7cm,angle=0]{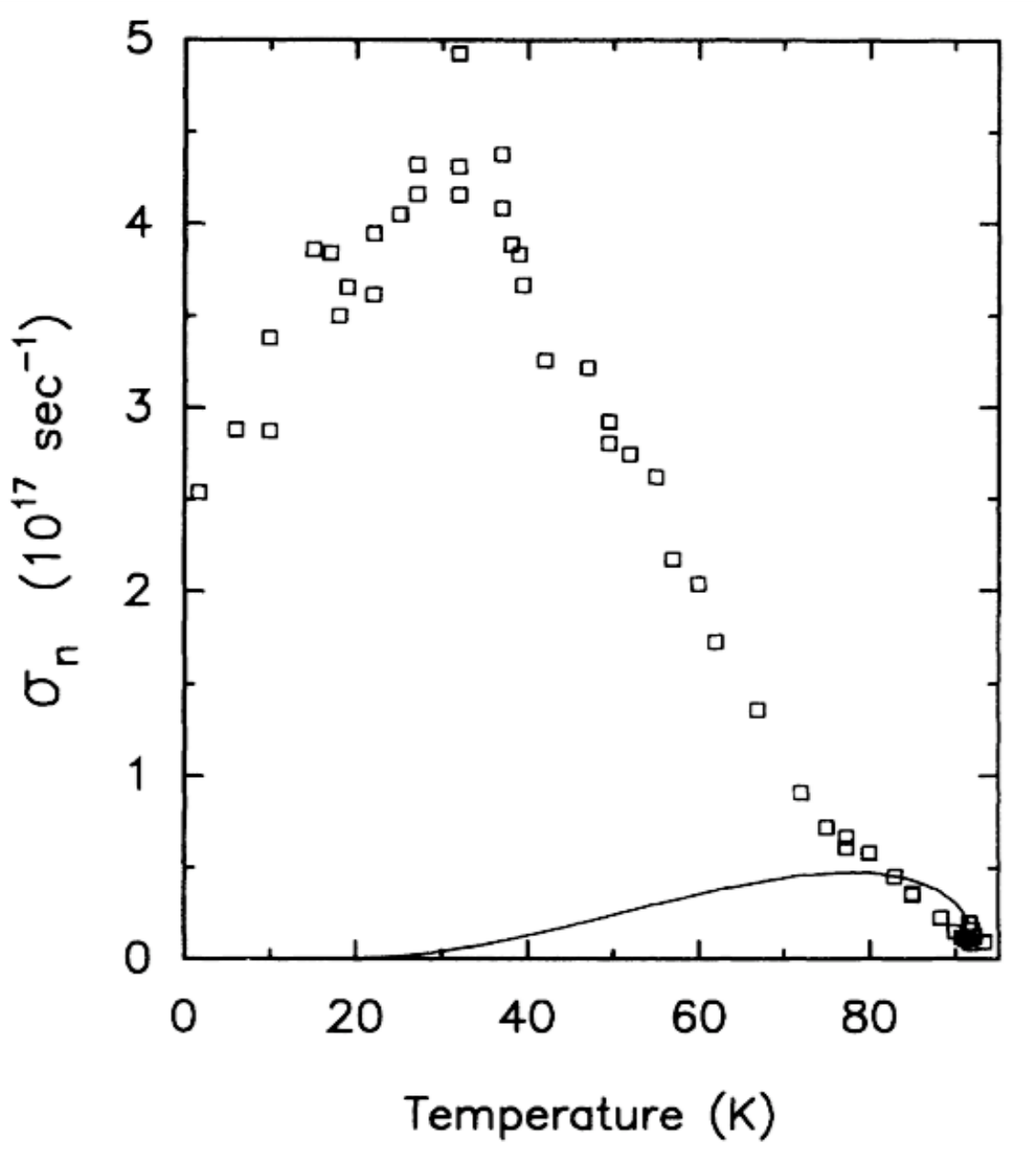}
\caption{ The real part of the conductivity extracted from the microwave surface resistance of YBCO.  The BCS conductivity (solid) is calculated using a T$_c$ of 91.8 K, a gap ratio of 3.52, and various other physical parameters such as penetration depth, coherence length, and mean free path.  It is completely incapable of describing the EM response, which shows the $d$-wave nature of these materials.  From Ref. \onlinecite{Bonn92a}.   } \label{YBCO1}
\end{center}
\end{figure}

As noted, different behavior is expected for type I coherence factors.  There one expects an absorption enhancement at the gap edge and no `Hebel-Slichter' peak.  The lack of such a peak in electromagnetic absorption of cuprate superconductors is at least partial evidence for $d$-wave superconductivity in those compounds.   It is due in part to the lack of a strong singularity in the  $d$-wave density of states, but also because the coherence factor vanishes for $q =\pi,\pi$ since $\Delta_k \approx - \Delta_{k + \pi,\pi}$ for $k$ near  $\pi,0$.  Such data is shown in Fig. \ref{YBCO2} where the conductivity measured at a number of different frequencies in the GHz range on YBCO  is displayed as a function of temperature \cite{Hosseini99a}.  The data shows a broad peak in $\sigma_1$ at around approximately 35 K, but a comparison with the Mattis-Bardeen prediction shows that it is incapable of describing it as the rise in $\sigma_1$  is much more gradual than predicted.   An explicit comparison is shown in Fig. \ref{YBCO1}.  This peak has been quite reasonably described in terms of a collapsing scattering rate below T$_c$ of quasiparticles whose number remains appreciable until low temperatures due to the $d$-wave nature of this compound.  An examination of the Drude model shows that in general one expects a peak in $\sigma_1$ when the probe frequency is approximately equal to $1/\tau$ the scattering rate.    The exhibited peak is  therefore not a Hebel-Slichter peak, and the data is consistent with $d$-wave superconductivity.  These experiments give evidence for $d$-wave superconductivity twice:  the lack of a Hebel-Slichter peak and the large $\sigma_1$ due to quasiparticle effects.   Note that the other sharp peak right at T$_c$ is believed to be due to fluctuations of superconductivity.  Its functional form is also inconsistent with it being a Hebel-Slichter peak.

For completeness, I should mention that spin density wave compounds which can be treated within the BCS formalism are expected to have a type I order parameter and a gap edge absorption enhancement.   In many cases they can also be treated within a BCS formalism.  See Ref. \onlinecite{GrunerCDW} for explicit details.

A vast literature exists on the electrodynamics of superconductors.  I have given only the most superficial treatment here. Many considerations go into how superconductivity is exhibited in optical spectra.  See Tinkham \cite{Tinkham}, for instance, for in-depth discussions on the clean and dirty limits of superconductivity and Basov and Timusk \cite{Basov05a} for the state-of-the-art on high-temperature cuprate superconductivity.

\section{Advanced Analysis and Techniques}

\subsection{Sum Rules} \label{sectionsumrules}

The optical constants satisfy a number of different sum rules \cite{Mahan90a}.   Although a number of different ones can be defined (see below), the most frequently used concerns the real part of the conductivity.  As alluded to above, there is a relation for the integral over all frequencies of the real part of the conductivity to the total number of charges $n$ and their bare masses $m_e$.  

 \begin{eqnarray}
8 \int_0^{\infty} \sigma_1(\omega) d \omega = \frac{4 \pi ne^2}{m_e}.
\label{sumrule}
\end{eqnarray}

Analysis of data in terms of sum rules provides a powerful tool that can be used to study spectral weight distributions in a relatively model-free fashion.  The above integral which extends from zero to infinity is the \textit{global oscillator strength sum rule}, which relates the integral of the $\sigma_1$ to the density of particles and their bare mass.  It is related to the oscillator strength $f_{s,s'}$ discussed above in the context of the Kubo formula.  Both are manifestations of the $f$-sum rule of elementary quantum mechanics.   Sum rules follow as consequence analyticity of the response functions and causality.   See Refs. \cite{Wooten72a,Mahan90a} for a few derivations.

\begin{figure}[htb]
\begin{center}
\includegraphics[width=8cm,angle=0]{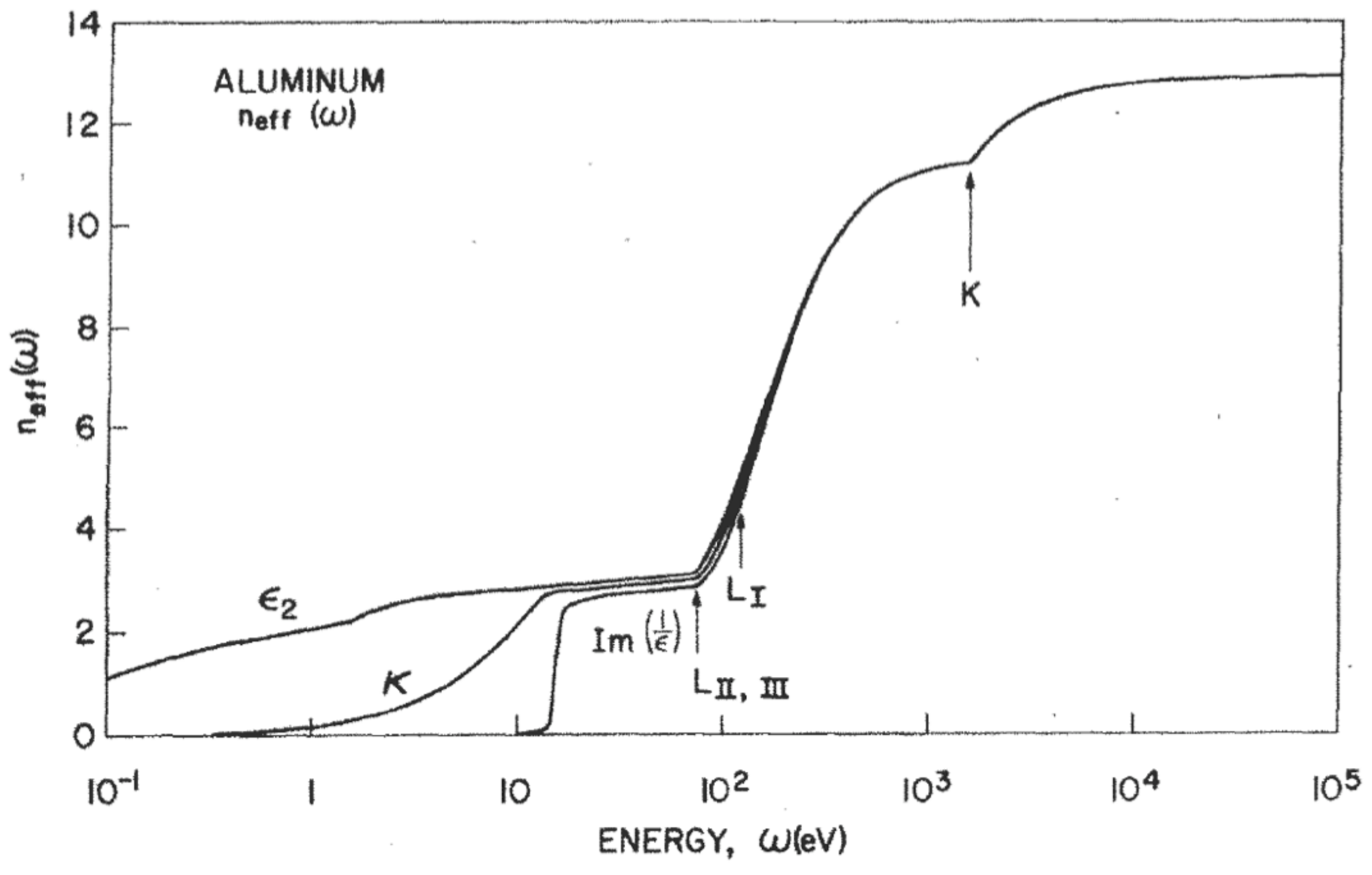}
\caption{The effective number of carriers $n_{eff}(\Omega_c)$ as a function of cutoff frequency $\Omega_c$ for aluminum.  A number of different sum rules are investigated and displayed.  The sum rule for $\epsilon_2$ is equivalent to the sum rule expressed in Eq. \ref{sumrule}.  Figure adapted from Ref. \onlinecite{Smith78a}.} \label{AlumSum}
\end{center}
\end{figure}

In real systems the integral to infinite frequencies is rarely utilized in practice and we are usually more concerned with partial sum rules such as

 \begin{eqnarray}
8 \int_0^{W} \sigma_1(\omega) d \omega = \frac{4 \pi ne^2}{m_b}.
\label{partialsumrule}
\end{eqnarray}

\noindent where $W$ is the unrenormalized electronic bandwidth and $m_b$ is the band mass.     Sum rules of this fashion can be used to get information about occupation and correlation in a subset of bands.    Of course if the integral cut-off $W$ is extended to high enough energies the full sum rule Eq. \ref{sumrule} must be fufilled  This puts a fundamental constraint on the various values that the individual sub-band band masses can assume.  A partial sum rule analysis of the equivalent quantity $\epsilon_2$ applied to the simple metal case of aluminum is shown in Fig. \ref{AlumSum}.   A number of different sum rules are also displayed.  Aluminum's nominal electronic configuration is $1s^22s^22p^63s^23p^1$.  Depending on the cutoff of the integral, the partial   sum rule for different orbitals is satisfied.  Integrating up to approximately the plasma frequency (15.5 eV)   returns the number of electrons in the $n=3$ valence band.  At the x-ray $L$ edge at around 100 eV, one begins to reveal spectral weight of the $n=2$ formed bands.  One can see that if the integral is performed to high enough energies (well above the $K$ edge of aluminum) one recovers the aluminum's atomic number `13' if one assumes the free-electron mass in Eq. \ref{partialsumrule}.  It is also interesting to note that the rate of convergence is different for the different sum rules plotted.   As will be discussed briefly below, this is related to how spectral weight is distributed in the spectra and can be related to quantities like the strength of interactions.

In correlated systems one frequently makes use of an even more limited sum rule and only performs the integral over an energy only several times the Drude width to determine the mass renormalized due to interactions.  For instance the small spectral weight  in the very narrow Drude peak as shown in Fig. \ref{SCInsAC2} gives a very large mass ($\approx 10^2 m_e$).  An integral as such over a small renormalized Drude peak with small spectral weight is how the very large masses in Fig. \ref{SpecificHeatOpticalMass} were generated.   One can view this small mass as coming from a very narrow interaction derived band where the integral only has to be performed over a small fraction of $W$ to capture its spectral weight.   However, the sum rules of Eqs. \ref{sumrule} and \ref{partialsumrule} must ultimately be satisfied at high enough energies.   This means that there must be be some higher energy satellite in $\sigma_1$ that contains enough spectral weight to return the band mass $m_b$ when the  integral in Eq. \ref{partialsumrule} is performed out to the unrenormalized band width $W$.  Note that similar information can be gained from applying the extended Drude model analysis, which is outlined in the next section (Sec. \ref{DLsection}).

Sum rules have been used extensively in the analysis of the data for correlated electron systems because, in some circumstances, they allow a relatively model independent way to analyze the data.  For instance, one can show that for a tight binding model with nearest neighbor hopping only that 

\begin{eqnarray}
8 \int_0^{W} \sigma_1(\omega) d \omega =  -  \frac{\pi e^2 a_r}{2 \hbar^2} K_r.
\end{eqnarray}

\noindent where $a_r$ is the lattice constance in the incident $E$ field's polarization direction and $K_r$ is the effective kinetic energy.  This relation has been used extensively in the cuprate superconductors to attempt to show evidence for or against a novel lowering of the kinetic energy-driven mechanism for superconductivity.  This is in contrast to the usual mechanism in the BCS theory where it is the potential energy that drops when falling into the superconducting state.  Such analysis is not trivial in these materials, as there is evidence that the spectral weight that goes into the condensate is derived not just from the near E$_F$ electronic band over a width much smaller than $W$, but over an energy range many times the band width $W$\footnote{It may be that this extreme mixing of low and high energy scales between, which prohibits a true low energy description is an alternative definition of a correlated material.}.

\begin{figure}[htb]
\begin{center}
\includegraphics[width=6cm,angle=0]{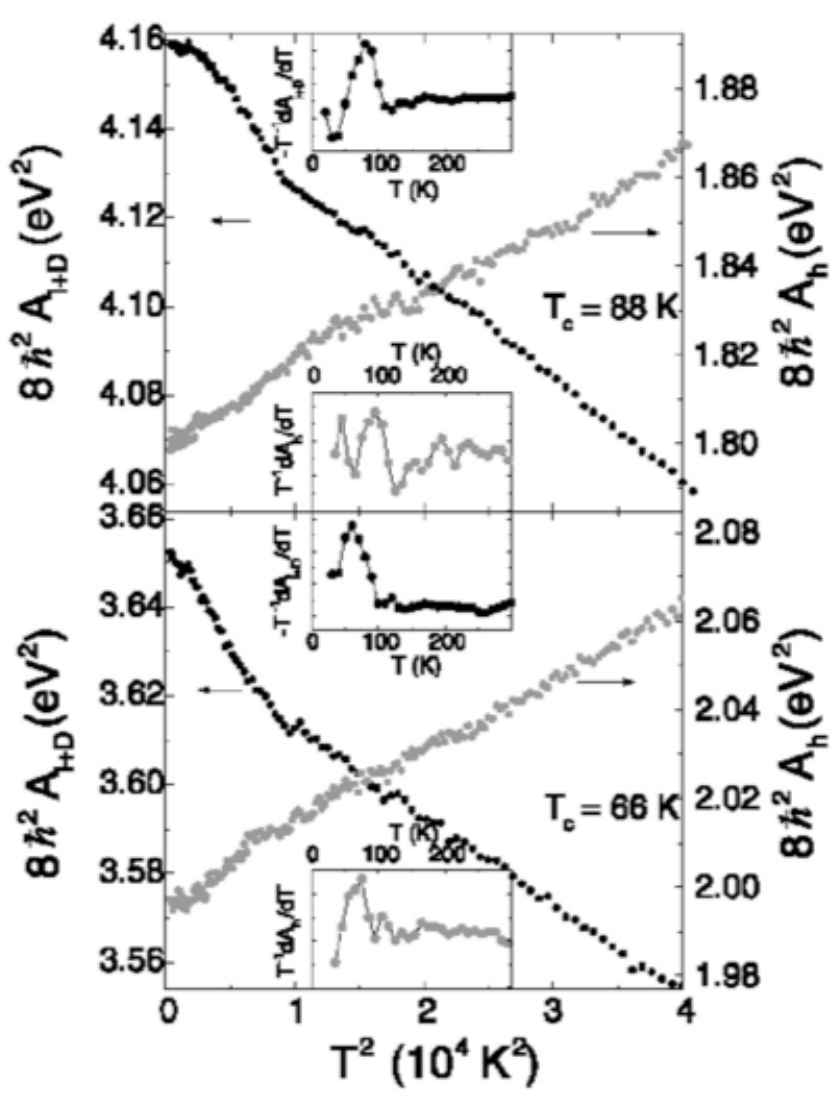}
\caption{Temperature dependence of the low-frequency spectral weight A$_{1+D}(T)$ and the high-frequency spectral weight $A_h(T)$, for optimally doped and underdoped Bi2212. Insets: derivatives $-T^{-1}dA_{1+D} / dT$ and  $T^{-1}dA_{1h} / dT$. From Ref. \onlinecite{Molegraaf02a}.} \label{hajo}
\end{center}
\end{figure}

It is believed that the usual sum rule for superconductors (Ferrell-Glover-Tinkham) \cite{Tinkham}  is satisfied for the $ab$ plane conductivity of the cuprate superconductors up to about $10\%$ accuracy \cite{Basov05a}.  Molegraaf \textit{et al.} \cite{Molegraaf02a} used a combination of reflectivity and spectroscopic ellipsometry in Bi2212  (Fig. \ref{hajo}) to conclude that about 0.2 - 0.3$\%$ of the total strength of the superconducting $\delta$ function is collected from an energy range beyond 10,000 cm$^{-1}$ to 20,000 cm$^{-1}$ \footnote{10,000 cm$^{-1}$ is approximately equivalent to 1.24 eV and 14,390 K!  It is remarkable that such energy scales are explicitly relevant for a 100 K phenomenon.   In contrast conventional superconductors satisfy the `Ferrell-Glover-Tinkham' sum rule at energies approximately 10 - 20 $\Delta$ \cite{Tinkham}.} and that the kinetic energy decreases in the superconducting state.  This is a very small effect, but still large enough to account for the condensation energy.   Such analysis of total spectral weight transfer can be tricky as it requires the analysis of spectral weight beyond the measurement regime, which is only constrained through the Kramers-Kronning relations.  These results or rather the analysis of the data has been disputed by Boris \textit{et al.},  \cite{Boris04a} who form a nearly opposite conclusion regarding spectral weight transfer based on nearly identical experimental data.

The c-axis response has been investigated in a similar manner and a similar transfer of spectral weight from very high energies to the $\delta$ function has been observed, although in this case the magnitude of the effect is definitively insufficient to account for T$_c$ \cite{Basov05a}.

As mentioned above, there are a number of sum rules in addition to the most commonly applied one for $\sigma_1$ (or equivalently the first moment sum of $\epsilon_2$).   Probably next most commonly applied is the sum rule for the first moment of the loss function, whose integral is equal to that for the first moment sum of $\epsilon_2$ \cite{Mahan90a}.   This has been called the longitudinal $f$-sum rule.  A similar sum rule concerns the first moment sum of the absorption coefficient $\kappa(\omega)$ \cite{Smith78a}.  There is also a sum rule on the loss function itself, that can be related to the Coulomb correlation energy of the ground state wavefunction \cite{Nozieres58a,Nozieres59a}.  Similarly, there is second moment conductivity sum rule and a third moment loss function sum rule that in certain circumstances allows one to evaluate the strength of umklapp scattering and the static electron-lattice energy \cite{Turlakov03a,Turlakov03b}.   There are also first $inverse$ moment sum rules for the conductivity and loss function, which are equal to various combinations of pure numbers (For instance $\int^\infty_{-\infty} - \textrm{Im} \frac{1}{\omega \epsilon(\omega)} d \omega = \pi$).   These allow the calibration of, for example, the absolute intensity of an electron energy loss measurement since these sums have no dependence on any material parameters.  Other related sum rules include the inertial sum rule on the index of refraction ($\int_0^\infty [n(\omega) - 1] d\omega = 0 $)  \cite{Shiles80a,Altarelli72a,Altarelli74a} and the ``DC sum rule" on the dielectric function  ($\int_0^\infty [\epsilon_1(\omega)- 1] d\omega = -2\pi^2 \sigma_{DC}$)  \cite{Shiles80a,Altarelli72a}.  There is also a sum rule on the frequency dependent scattering rate \cite{Marsiglio01a}, which is discussed below.  For further details on sum rules, see the very extensive literature on this subject \cite{Mahan90a,vanderMarel03b,Marsiglio01a,Smith78a,Nozieres58a,Nozieres59a,Turlakov03a,Turlakov03b,Shiles80a,Altarelli72a,Altarelli74a}.

\subsection{Extended Drude Model}\label{DLsection}

Within the Drude-Lorentz (Sec. \ref{SectionDL} )  model we consider that conduction electrons are quasi-free.  The scattering rate of the Drude intraband contribution is considered to be characterized by a single frequency independent relaxation time $\tau$.  It is well known however that various inelastic channels can contribute to relaxation in solids, each with a characteristic frequency dependence.  For instance, in 3D the fermion-fermion scattering rate is expected to exhibit an $\omega^2$ dependence at low frequency.   An electron scattering off an Einstein boson (a single spectral mode with a well-defined frequency $\omega_0$) is expected to show an onset in increased scattering at the boson frequency  $\omega_0$ which is the threshold to make a real (as opposed to virtual) excitation.

A number of authors have proposed various treatments that can capture these frequency dependences in optical spectra.   G\"otze and W\"olfle introduced a memory function $ M(\omega)$ approach to capture correlations effects where the optical conductivity could be written in a modified Drude form as

 \begin{eqnarray}
\sigma(\omega) =  \frac{\omega_p^2}{4 \pi (M(\omega) - i \omega  )}.
\end{eqnarray}

Similar ideas around the same time for a frequency dependent scattering rate were introduced by P.B. Allen in the context of the electron-phonon problem \cite{Allen71a}.  J. W. Allen and J. C. Mikkelsen \cite{Allen77a} proposed that one could capture frequency dependent effects experimentally through an $extended$ $Drude$ $model$ (EDM) where the mass and scattering rate are explicitly frequency dependent.  Inverting the complex conductivity one gets
 
 \begin{eqnarray}
\frac{m^{\ast}(\omega)}{m_b}= - \frac{\omega^{2}_{p}}{4\pi\omega}Im\left[\frac{1}{\sigma(\omega)}\right],
\label{EDMmass} \\
\frac{1}{\tau(\omega)}= \frac{\omega^{2}_{p}}{4\pi}Re\left[\frac{1}{\sigma(\omega)}\right],
\label{EDM}
\end{eqnarray}
 
\noindent where $m_b$ is the band mass. 
 
Both in spirit and formalism this is reminiscent of standard treatments in many-body physics where one posits that the effects of interactions can be captured by shifting the energy of an added electron from the bare non-interacting energy  $\epsilon(\textbf{k})$ by a complex $self$-$energy$   $\Sigma(\omega,\textbf{k}) = \Sigma_1(\omega,\textbf{k})   + i \Sigma_2(\omega,\textbf{k}) $.  One can express the optical response in terms of the complex self-energy $\Sigma^{op}(\omega) = \Sigma^{op}_1(\omega) + i \Sigma^{op}_2(\omega)$ as 

 \begin{eqnarray}
\sigma(\omega) = - i \frac{\omega_p^2}{4 \pi (2\Sigma^{op}(\omega) - \omega  )}.
\label{mstar2}
\end{eqnarray}

and in terms of previously defined quantities 

 \begin{eqnarray}
\Sigma_1^{op}  =  \frac{\omega}{2} (1 - m^*/m_b), \nonumber \\
\Sigma_2^{op} = - \frac{1}{2 \tau (\omega)}.
\label{opticalselfenergy}
\end{eqnarray}

Note that as $\frac{1}{\tau(\omega)}$ and $m^*(\omega)$ are a particular parameterization of the real and imaginary parts of a complex response function, they are Kramers-Kronig related.  Analogous to the case of $\sigma(\omega)$ detailed above, if one knows  $\frac{1}{\tau(\omega)}$ for all $\omega$, then $m^*(\omega)$ can be calculated and vice versa.   In a related fashion, if there are frequency dependent structures in $\frac{1}{\tau(\omega)}$ then the effective mass must be enhanced in some frequency regions.

A caveat should be given about the interpretation of optical self-energies.  Although an analogy can be made to a momentum averaged version of the quasi-particle self-energies that appear in one-particle spectral functions, the two self-energies are not exactly the same.  While quasi-particle scattering depends only on the total charge scattering rate to all final states,  transport and optics experiments have a much larger contribution from backward scattering than forward scattering as it degrades the momentum much more efficiently.  In general this means that the quantities derived $via$ optics emphasize backward scattering over forward scattering and so the optically derived quantities can contain a vertex correction not included in the full quasi-particle interaction.  In this sense the two self-energies are formally different although they can contain much of the same information.   The connection in so far as it is understood explicitly is discussed in Ref. \onlinecite{Allen15a}.  A number of other different methods exist to calculate the finite frequency scattering rate \cite{gotze_72,shulga_91,mitrovic_85,sharapov_05} and a nice comparative treatment of a number of different approaches has been given \cite{Bhalla17a}.

\begin{figure*}[htb]
\begin{center}
\includegraphics[width=18cm,angle=0]{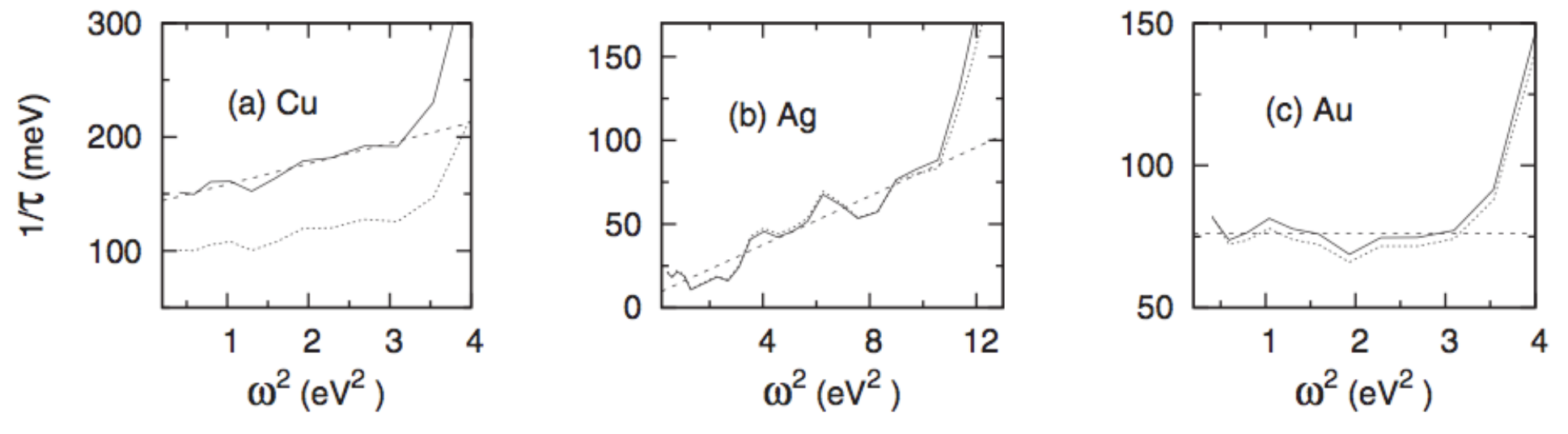}
\caption{Scattering rate vs. $\omega^2$ of (a) Cu, (b) Ag and (c) Au. Solid and dotted lines are the bare scattering rate $1/\tau$  and the dressed scattering rate $1/\tau^*$ respectively as detailed in the text.  From \cite{Youn06a}} \label{ScatRateMetals}
\end{center}
\end{figure*}

\begin{figure}[htb]
\begin{center}
\includegraphics[width=6cm,angle=0]{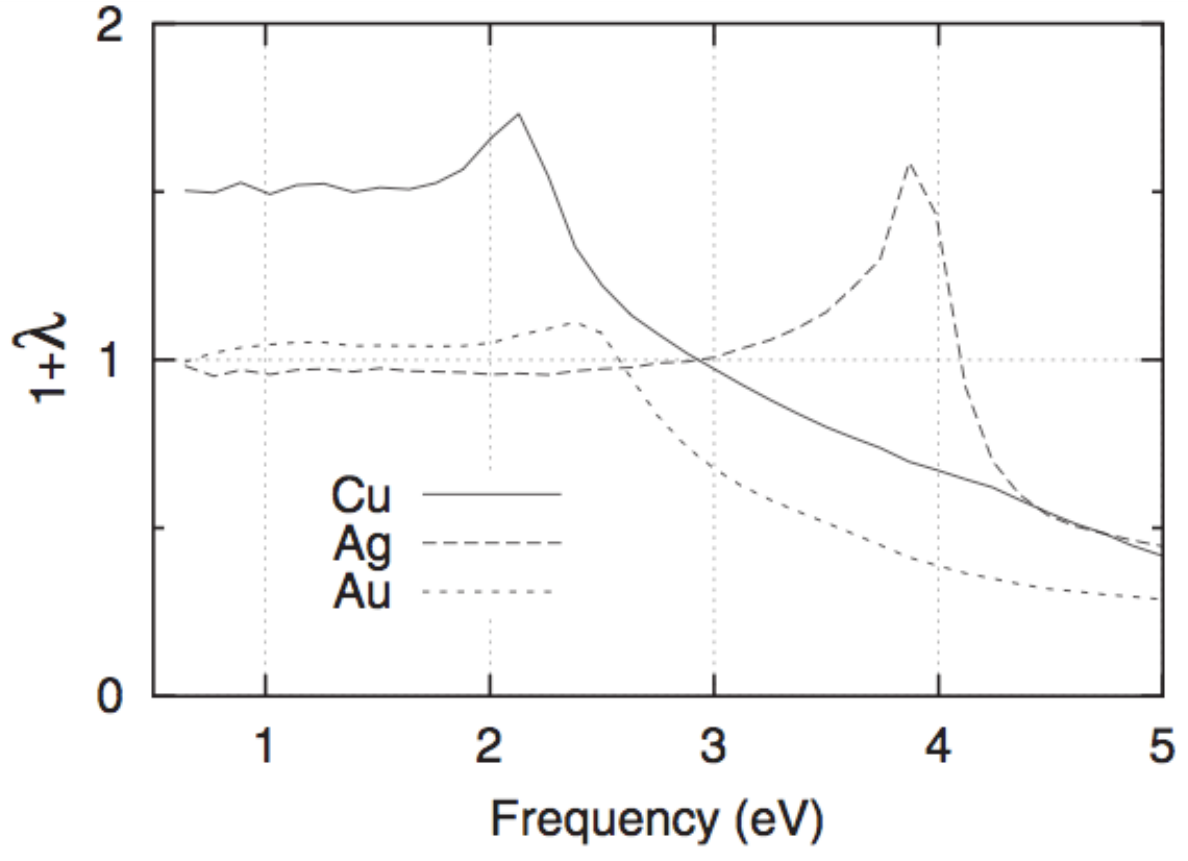}
\caption{Mass enhancement factor $1 + \lambda$ of noble metals. From \cite{Youn06a}} \label{MassMetals}
\end{center}
\end{figure}

\begin{figure}[htb]
\begin{center}
\includegraphics[width=6cm,angle=0]{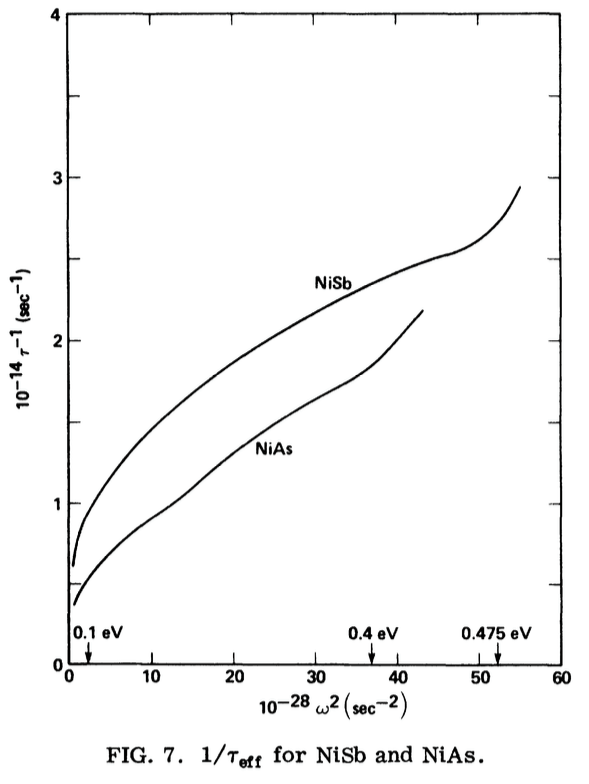}
\caption{The quantity $1/\tau^*(\omega) $ from Eq. \ref{Memoryfunction} for NiAs and NiSb \cite{Allen77a}.   One can see that below the energy scale of the interband transitions (0.4 eV and 0.475 eV) an approximately $\omega^2$ dependence is found. } \label{NiCompounds}
\end{center}
\end{figure}

As mentioned above, frequently the extended Drude model can be introduced in terms of a complex frequency dependent $Memory$ function, $M(\omega)$

 \begin{eqnarray}
\sigma(\omega) =  \frac{\omega_p^2}{4 \pi (M(\omega) - i \omega  )}
\end{eqnarray}

The memory function  $M(\omega) $ can be seen defined in terms of an effective scattering rate and effective mass  $ 1/\tau(\omega) - i \omega \lambda(\omega)$, where $\tau$ is the lifetime and $1 + \lambda  =  m^*/m_b $ and hence

 \begin{eqnarray}
\sigma(\omega) =  \frac{\omega_p^2}{4 \pi (1/\tau(\omega) - i \omega [1 + \lambda(\omega)] )} .
\label{Memoryfunction}
\end{eqnarray}

One can put Eq. \ref{Memoryfunction} in the form of the standard Drude model with the substitutions  $\tau^*(\omega) = [1 + \lambda(\omega) ] \tau(\omega)$ and $\omega_p^{*2} = \omega_p^{2} /[1 + \lambda(\omega) ]  $ to get

 \begin{eqnarray}
\sigma(\omega) =  \frac{\omega_p^{*2}(\omega)}{4 \pi} \frac{ \tau^*(\omega) } {1- i \omega  \tau^*(\omega)  } .
\label{sigmastar}
\end{eqnarray}

This equation describes the frequency dependence of a particle with renormalized plasma frequency (and hence renormalized mass) and renormalized scattering rate.     Compare Eq. \ref{sigmastar} with Eq. \ref{Drudeplasmafreq}.  Note that the quantity $1/\tau^*$ is not equivalent to the quantity defined in Eq. \ref{EDM} above, as $1/\tau^*$ includes the renormalization effects of the lifetime as well as the mass, whereas  $1/\tau$ includes only lifetime effects.  The two quantities differ by a factor of $1 + \lambda(\omega)$.  In this sense one may argue that $1/\tau(\omega)$ is the more intrinsic quantity, as also evidence by its direct proportionality to the imaginary part of the self-energy, but this point is essentially a semantic one.    $1/\tau^*$  is the effective scattering rate (or dressed scattering rate) however, because for weak frequency dependence it is the actual half width of the Drude peak in the optical conductivity.

Various contributions to the total scattering rate such as  electron-electron (ee), electron-phonon (ep), electron magnon (em), etc. add within the prescription set by Matthiesen's rule as $\frac{1}{\tau(\omega)} =  \frac{1}{\tau_{0}} + \frac{1}{\tau_{ee}} +  \frac{1}{\tau_{ep}}  +  \frac{1}{\tau_{em}} $.  Note that there is typically also a contribution $\frac{1}{\tau_{0}}$  to the elastic scattering which is frequency independent that comes from static disorder.  As mentioned above this can be seen as a measure of the degree to which the translational symmetry is broken by disorder and hence the degree to which strict momentum conservation in the optical selection rule $q = 0$ can be violated.

The near-E$_F$ electrons of the noble metals, like Cu, Ag, and Au have free-electron-like $s$-states, which form a Fermi surface that deviates from almost spherical symmetry only near the Brillouin zone necks in the $<111>$ directions.  The above analysis applied to such metals is shown in Fig. \ref{ScatRateMetals} and Fig. \ref{MassMetals}.  For Cu and Ag the scattering rate shows an approximately $\omega^2$ dependence of the scattering rate, which is in accord with the expectation for electron-electron scattering \footnote{One should interpret the reported $\omega^2$ dependence here for simple metals carefully.   Indeed one may expect that electron-phonon scattering would be more dominant in simple metals.}.  It is also interesting that electron-electron scattering is observed up to such high energies as we may have expected other contributions such as electron-phonon scattering to give an observable contribution.  We should reiterate that a purely translational invariant electron gas, cannot dissipate momentum by electron-electron collisions alone.  In this regard effects like umklapp or interband (Baber) scattering are essential to see the effects of such collisions in the optical response.  Also note here the difference between the bare scattering rate $1/\tau$ and the dressed scattering rate $1/\tau^*$.  The difference is only significant when there is a large renormalization in the effective mass.
 
In Fig. \ref{NiCompounds} are shown extended Drude model data for the quantity $1/\tau^*(\omega) $ from Eq. \ref{Memoryfunction} for NiSb and NiAs \cite{Allen77a}.   These antimonides are more strongly interacting $3d$ systems.   The room-temperature optical reflectivity of metallic NiSb and NiAs were measured between 0.05 and 5.0 eV, and the optical constants have been deduced by a Kramers-Kronig analysis.    Unlike the good metals discussed immediately above, the color of these systems is more dull with NiAs appearing yellow, and NiSb appearing light pink due roughly to the same mechanism as gives copper and gold its colors.   One can see in Fig. \ref{NiCompounds} that below the energy scale of the interband transitions (0.4 eV and 0.475 eV) an approximately $\omega^2$ dependence is found.   This is consistent with the Fermi liquid expectation.   However notice the vastly larger scale of the coefficient of the $\omega^2$ term in as compared to the simple metals.

Although the dual quadratic dependencies of the $ \omega^2$ and $T^2$ forms of the electron-electron scattering rates in a Fermi liquid follows in a relatively model independent fashion, more consideration needs to go into their relative magnitudes.   Indeed, there has been interest in the relative prefactors of the $ \omega^2$ and $T^2$ terms.   The prediction from canonical Fermi liquid theory is that the frequency and temperature dependent scattering rate goes as

 \begin{eqnarray}
1/\tau\propto (\hbar\omega)^2 + (p\pi k_B T)^2
\end{eqnarray}
where $p=2$ \cite{MaslovA,Gurzhi,MaslovB,Berthod}.   It appears that in many cases this simplest prediction has not been observed.    Different values have been observed:  $p\sim 1$ in URu$_2$Si$_2$ \cite{Nagel}, $p\sim 2.4$ in the organic material BEDT-TTF \cite{Dressel}, and $p\sim 1.5$ in underdoped HgBa$_2$CuO$_{4+\delta}$ \cite{Mirzaei}.  In contrast, the claim for the very clean Fermi liquid Sr$_2$RuO$_4$ is that scaling of $\omega$ and $T$ are most consistent with $p=2$ \cite{Stricker}.   However lower temperature and frequency data should be taken.   It has been proposed that the deviations from $p=2$ can be explained by elastic energy-dependent scattering, which can decrease the value of $p$ below 2  \cite{MaslovA,MaslovB}. Such a contribution depends on energy, but not on temperature and therefore contributes a ${{\omega}^{2}}$  but not a T$^2$ term to the self-energy.   See Ref. \onlinecite{MaslovA} for further discussion of these details and summary of the experimental situation.

Note that in Eqs.  \ref{EDMmass} and \ref{EDM} there is the matter of how to define the plasma frequency $\omega_p$\footnote{Note that this is only important for the overall scale factor of these expressions.   Their functional dependence and form is independent of the choice of the plasma frequency.}.    For the electron gas the plasma frequency is given by $4 \pi n e^2/m$, however the precise definition is less clear in a real material with interactions, interband transitions, and orbital mixing. In the above analysis $\omega_p$ must come from the spectral weight of the full intra-band contribution.   Rigorously the plasma frequency should be defined from the sum rule Eq. \ref{sumrule} as

 \begin{eqnarray}
\int_0^\infty \sigma_1^{intra}(\omega) =  \frac{\omega_p^2}{8}.
\label{SW}
\end{eqnarray}

\noindent where  $\sigma_1^{intra}$ is contribution to the conductivity coming from all (and only) $intraband$ spectral weight\footnote{Note that in anything other than a simple Drude system the plasma frequency here is not given by the plasma minimum in the reflectivity as that quantity is renormalized by a factor of $\frac{1}{\sqrt{\epsilon_\infty}}$.  As mentioned above, this later quantity is known as the screened plasma frequency or plasmon frequency.} .  The integral of this intraband contribution extends all the way to infinity, which gives of course a practical difficulty in real systems because the value of the spectral weight will be contaminated by $interband$ contributions.  Note that for a complicated interacting system the integral and $\omega_p$ must  include not only the contributions of the weight of the low frequency Drude peak, but must also include higher energy parts of the spectra (satellites) if they derive from the same non-interacting band.  It can be difficult experimentally to estimate $\omega_p$ accurately.  For instance, in the case of a system with strong electron-phonon coupling the spectra may be approximately modeled as the usual Drude model with a  simple $\omega = 0$ Lorenztian and frequency independent scattering rate at low frequencies, but with a higher frequency satellite at the characteristic phonon frequency which results physically from the excitation of a $real$ phonon as well as an electron-hole pair.  The high frequency satellite manifests in the scattering rate as an onset at the phonon frequency as an additional decay channel becomes available at that energy.  In this sense the plasma frequency which goes into an extended Drude analysis is the plasma frequency associated with the spectral weight of both the low frequency `Drude' contribution $and$ the satellite.  The spectral weight of the low frequency part may be clear, but it may be a difficult practical matter to clearly identify the spectral weight of the satellite, due to the overlapping contributions of it with true interband transitions etc.  In practice, a high frequency cutoff to the integral in Eq. \ref{SW} is usually set at some value which is believed to capture most of the intraband contribution, while minimizing contamination by interband transitions.  A number of different criteria can be used:  the requirement that the integral is temperature independent, an appeal to reliable band structure calculations etc. etc.

There can also be a difficulty in definitions of the plasma frequency for systems described by the Mott-Hubbard model, where a strong onsite repulsion splits a single metallic band into upper and lower Hubbard `bands'.  In such cases, one has one band or two depending on definitions and so the matter of the true intraband spectral weight can be poorly defined.  In such cases it is important to remember that in this and other cases that the Drude model is a classical model applicable only to weakly interacting mobile particles.  Although the EDM can be put on more rigorous ground expressing its quantities as optical self-energies, one still is always making an analogy and connection to a non-interacting system.  To the extent that the EDM is just a parameterization of optical spectra, it is always valid.    However, to the extent that its output can be interpreted as a real mass and scattering rate of $something$ (i.e. a particle), it is important that that something exists.  In other words, it must be valid to discuss the existence of well-defined electronic excitations in the Fermi-liquid sense in the energy regimes of interest.

In this regard, one must also be aware of the contribution of any interband terms in the evaluation  of Eqs. \ref{EDMmass} and \ref{EDM}.   These expressions for the EDM are only valid for excitations that are nominally $\omega = 0$.  In principle finite frequency interband excitations (classically Drude-Lorentz) can overlap with a broad Drude intraband term giving artifacts that should not be considered as contributions to the scattering rate and effective mass.   The sharp onset in $1/\tau(\omega)$ in Fig. \ref{ScatRateMetals} and the cusp in Fig. \ref{MassMetals} are due to a onset of strong interband excitations in this spectral region (Fig. \ref{NobleDielectric}).

\begin{figure}[htb]
\begin{center}
\includegraphics[width=7cm,angle=0]{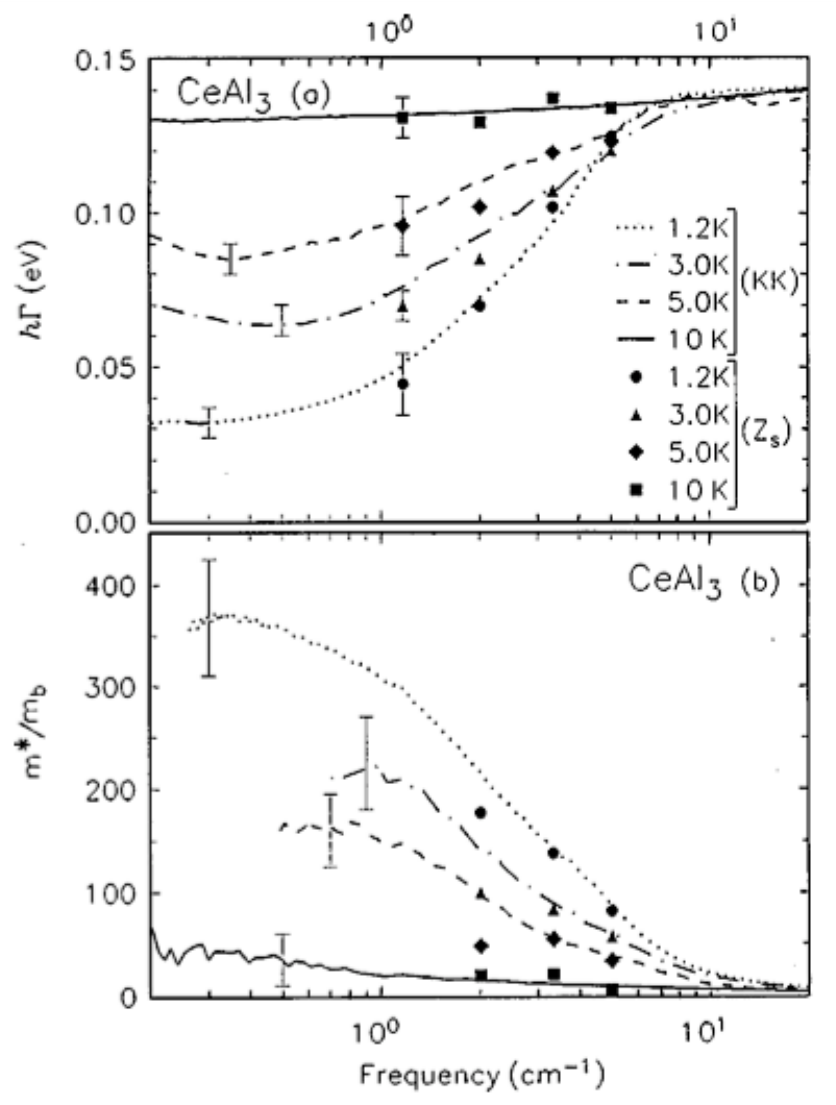}
\caption{Scattering rate and effective mass in CeAl$_3$. (a)  Frequency-dependent optical scattering rate for CeAl$_3$ at four temperatures, compiled using both infrared reflectivity and microwave cavity data. (b) Frequency dependence of m$^*$ obtained from Eq. \ref{EDM} \cite{Awasthi93a}.} \label{EDMCeAl3}
\end{center}
\end{figure}

The EDM has been used in many different contexts in strongly correlated systems.  An ideal example is afforded by its application to heavy fermion systems.  As mentioned above, due to interaction of conduction electrons with localized moments the conduction electrons can undergo extremely large mass renormalizations of a factor of a few hundred over the free electron mass.  These mass renormalizations develop below a coherence temperature $T^*$.   Such effects are reflected in the optical spectra.  As seen in Fig. \ref{EDMCeAl3}, at high temperatures an EDM analysis of CeAl$_3$ reveals a pure Drude-like, essentially frequency independent mass and scattering rate.  The conventionally sized mass and large scattering rate at high temperatures reflect the effects of essentially normal electrons scattering strongly and incoherently with the localized moments\footnote{Note that the conventional EDM analysis expresses $m^*$ as a ratio with the band mass $m_b$ if $\omega_p$ is treated correctly as discussed above.   This is in contrast to other techniques like, for instance, de Haas- van Alphen that expresses $m^*$ as a ratio with the free electron mass $m_e$.}.   The spectra show no particularly interesting frequency dependence.  Below the coherence temperature at low frequencies significant renormalizations to the mass and scattering rate are seen.  The mass develops a significant enhancement and the scattering rate drops.  The coherence temperature $T^*$ and the frequency above which the unrenormalized values are recovered $\omega_c$ are equivalent to each other to within factors of order unity (with $\hbar = k_B = 1$).  Remarkably the optical mass as compared to the band mass is increased by a factor of almost 350.  Note that although both the mass and the low frequency scattering rate undergo dramatic changes, the DC Drude conductivity is not changed significantly, as $\sigma_{DC} \propto \tau /m $ and the respective changes approximately cancel.  This is reflected in the fact that the relation $m^*(\omega \rightarrow 0)/m_b$ is expected to be equivalent to $ \tau/ \tau^*(\omega \rightarrow 0)$ to within factors of order unity \cite{Millis87a}.   Note that these changes in scattering rate and mass are reflected in the optical conductivity itself by the formation of a very sharp zero frequency mode (essentially a Drude peak) which rides on top of the normal Drude peak \cite{Millis87a}.

\begin{figure*}[htb]
\begin{center}
\includegraphics[width=8.5cm,angle=0]{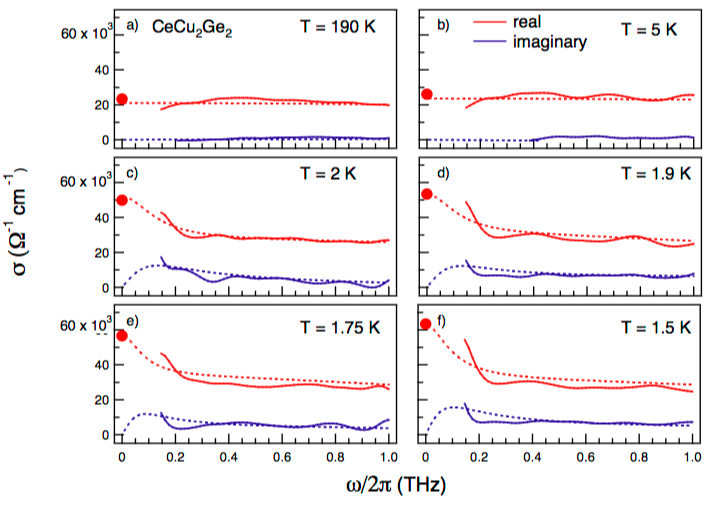}
\includegraphics[width=8.5cm,angle=0]{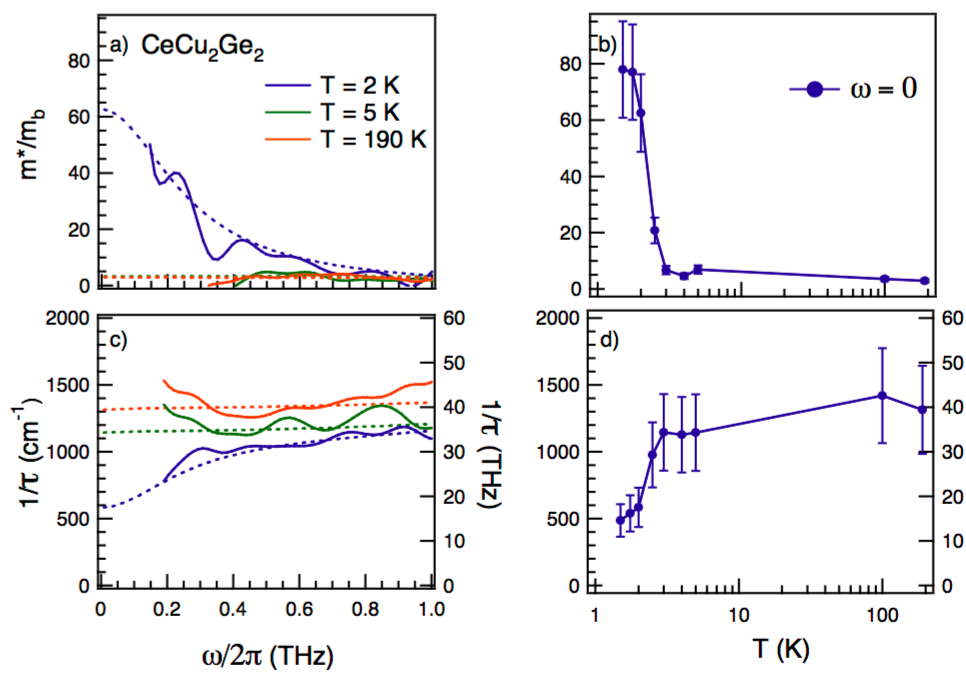}
\caption{(Left)  a-f) Real and imaginary parts of the complex conductivity of CeCu$_2$Ge$_2$ as a function of frequency for six temperatures in the frequency range of 0-1 THz. Solid lines indicate experimental data, while dashed lines show results of a fit described in the text. The values of the DC conductivity are also shown as a solid circle.  (Right) The renormalized mass of CeCu$_2$Ge$_2$  as a function of frequency derived from the extended Drude model for three temperatures. Solid lines indicate experimental data with dashed lines showing results of a fit described in the text. b) Renormalized mass of CeCu$_2$Ge$_2$  at zero frequency as a function of temperature based on the zero frequency extrapolation of the extended Drude model fits. Note the log scale on temperature axis. c) Scattering rate as a function of frequency with numerical fits shown in dashed lines for three temperatures. d) Scattering rate at zero frequency as a function of temperature. The error bars indicate uncertainty, which is primarily due to the ambiguity in determining the plasma frequency from FTIR measurements.   From Ref. \onlinecite{Bosse12a}
} \label{GraceTHz}
\end{center}
\end{figure*}

Data from my group \cite{Bosse12a} using THz spectroscopy on the Kondo-lattice antiferromagnet CeCu$_2$Ge$_2$ shows that a narrow Drude-like peak forms (Fig. \ref{GraceTHz} Left) at low temperatures, which is similar to ones found in other heavy fermion compounds that do not exhibit magnetic order.  Using this data in conjunction with DC resistivity measurements, we obtained the frequency dependence of the scattering rate and effective mass through an extended Drude model analysis (Fig. \ref{GraceTHz} Right). The zero frequency limit of this analysis yields evidence for large mass renormalization even in the magnetic state, the scale of which agrees closely with that obtained from thermodynamic measurements.  This mass renormalization in the low frequency and temperature limit is of order 50-100 times the band mass mb, which is consistent with the enhancement inferred from specific heat measurements.  In other systems, the expectation is that non-Fermi liquid behavior may reveal itself in frequency dependent scattering with dependencies slower than quadratic.   Please see Ref. \onlinecite{MaslovA} for a theoretical discussion and Ref. \onlinecite{Bosse16a} for a representative experiment on CeFe$_2$Ge$_2$.

Evidence that the renormalizations in the optical mass reflect the existence of real heavy particles in these systems and not just a convenient parametrization of the optical spectral can be seen in the above discussed Fig.  \ref{SpecificHeatOpticalMass} which shows a proportionality of the optically measured mass to the linear coefficient of the specific heat (which is proportional to inverse of the near E$_F$ density of states and hence the mass).  It is remarkable that a parameter determined dynamically - essentially by shaking charge with an oscillating E-field - is precisely related to a quantity which is determined $thermodynamically$ by quantifying the amount of heat absorbed.  This graph provides a remarkable demonstration of the quasiparticle concept.

As mentioned, in addition to electron-electron scattering, the effects of electron-boson interactions also reveal themselves in the optical spectra.  Spectral features in the optical conductivity deriving from electron-boson coupling were originally discussed in the context of Holstein processes - the creation of a real phonon and an electron-hole pair with the absorption of a photon -  in superconducting Pb films \cite{Allen71a,Joyce70a}.  The EDM analysis can reveal significant information about the interaction of electrons with various bosonic modes.     Allen derived an expression for the optical scattering rate at T=0K \cite{Allen71a}

 \begin{eqnarray}
\frac{1}{\tau(\omega)} = \frac{2 \pi}{ \omega}  \int^\omega_0   d\Omega (\omega - \Omega) \alpha^2 F(\Omega).
\end{eqnarray}
where $\alpha^2 F(\Omega)$ is the phonon coupling constant multiplied by the phonon density of states.

\begin{figure}[htb]
\begin{center}
\includegraphics[width=8cm,angle=0]{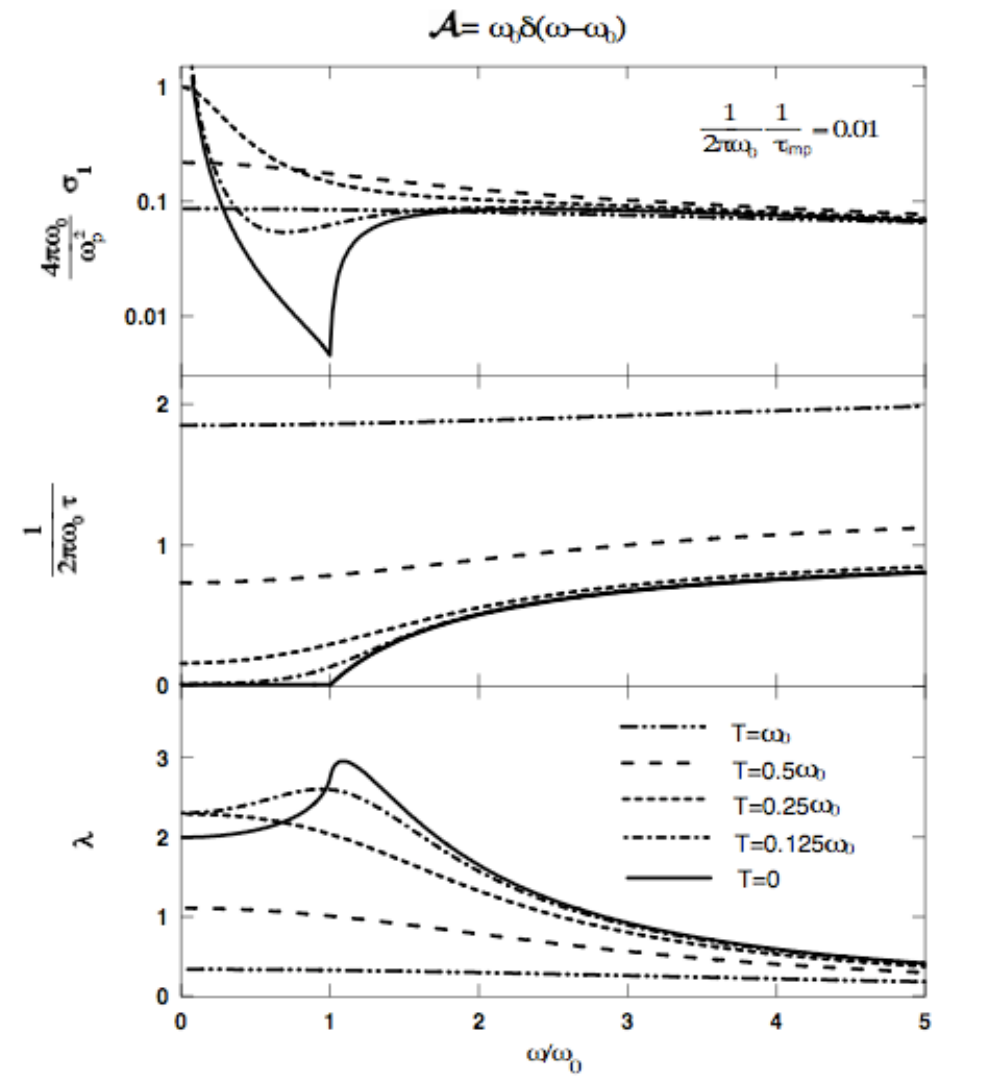}
\caption{Electron-boson model calculations with a boson spectral density $A(\Omega) = \omega_0 \delta(\omega -\omega_0)$. The top panel gives the optical conductivity and the lower two panels show the corresponding scattering rate and mass enhancement.  The coupling constant in this calculation is set to 1.  From \cite{Puchkov96a}. } \label{PuchhkovSim}
\end{center}
\end{figure}

In Fig. \ref{PuchhkovSim} the optical conductivity calculated for an electron with impurity scattering coupled to a single Einstein phonon is shown for a few different temperatures.   (See Ref. \onlinecite{Puchkov96a} for details regarding the calculation of the effects of electron-boson scattering in optical spectral.) The lower two panels show the scattering rate and mass enhancement generated from the optical conductivity.  At low temperatures the optical conductivity (Fig. \ref{PuchhkovSim} (top)) reveals a sharp Drude peak at low frequencies, which results from the usual electron-hole excitations.  The sharp onset in absorption at the phonon frequency is the threshold for excitation of a $real$ phonon.  At threshold, electron-hole pairs are excited additionally, but it is the phonon which carries away most of the energy.  This is reflected in the scattering rate, which is flat and small at low $\omega$, but shows a cusp and sudden increase at $\omega_0$ as the phase space for scattering increases discontinuously  (Fig. \ref{PuchhkovSim} (middle)).

Although real phonons can only be excited above threshold, electron-phonon coupling manifests itself at energies below threshold by an increase in the effective mass (Fig. \ref{PuchhkovSim} (bottom)).  Physically, this can be understood as a renormalization of the electron-hole pair energy modified by the excitation of $virtual$ phonons.  This renormalization of the energy manifests itself as an increased mass.  The electron-hole pair can be viewed as dragging a cloud of virtual phonons with it.  One can get an approximate measure of the boson coupling function as  \cite{Marsiglio01a}

 \begin{eqnarray}
\alpha^2 F(\omega) =   \frac{1}{2 \pi } \frac{d^2}{d \omega^2}[\omega Re \frac{1}{\sigma(\omega)}].
\end{eqnarray}

\begin{figure}[htb]
\begin{center}
\includegraphics[width=8cm,angle=0]{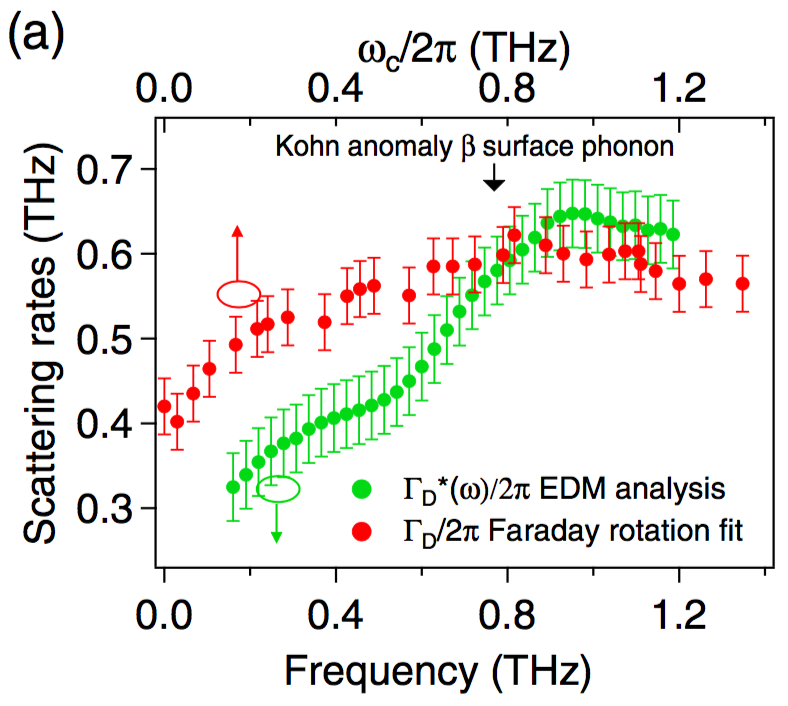}
\caption{(Color) The scattering rate as a function of cyclotron frequency (red). The fully renormalized scattering rate by mass through extended Drude analysis as a function of frequency (green). The black arrow indicates $\beta$ surface phonon frequency at the Kohn anomaly.   From Ref. \onlinecite{Wu15a}.} \label{TILiangphonons}
\end{center}
\end{figure}

Extended Drude model analysis was applied to the scattering rate of the Drude transport of Bi$_2$Se$_3$ topological insulator surface states in Ref. \onlinecite{Wu15a}.   A shown in Fig. \ref{TILiangphonons} an sudden increase of the scattering rate as a function of frequency is seen.   The frequency of scale of the increase is close to the scale of the previously observed Kohn anomaly of the surface $\beta$ phonon, 0.75 THz at $2k_F$.    When magnetic field was applied in the Drude peak shifted to finite frequency in a fashion expected for a cyclotron resonance and it was observed that as it moved above this $\beta$ phonon frequency it also broadened (Fig. \ref{TILiangphonons}).   Both of these effects can be understood from the perspective that when the excitation energy is larger than the energy of the phonon, a possible decay mechanism of the electron-hole pair is through exciting the phonon.

\begin{figure}[htb]
\begin{center}
\includegraphics[width=9cm,angle=0]{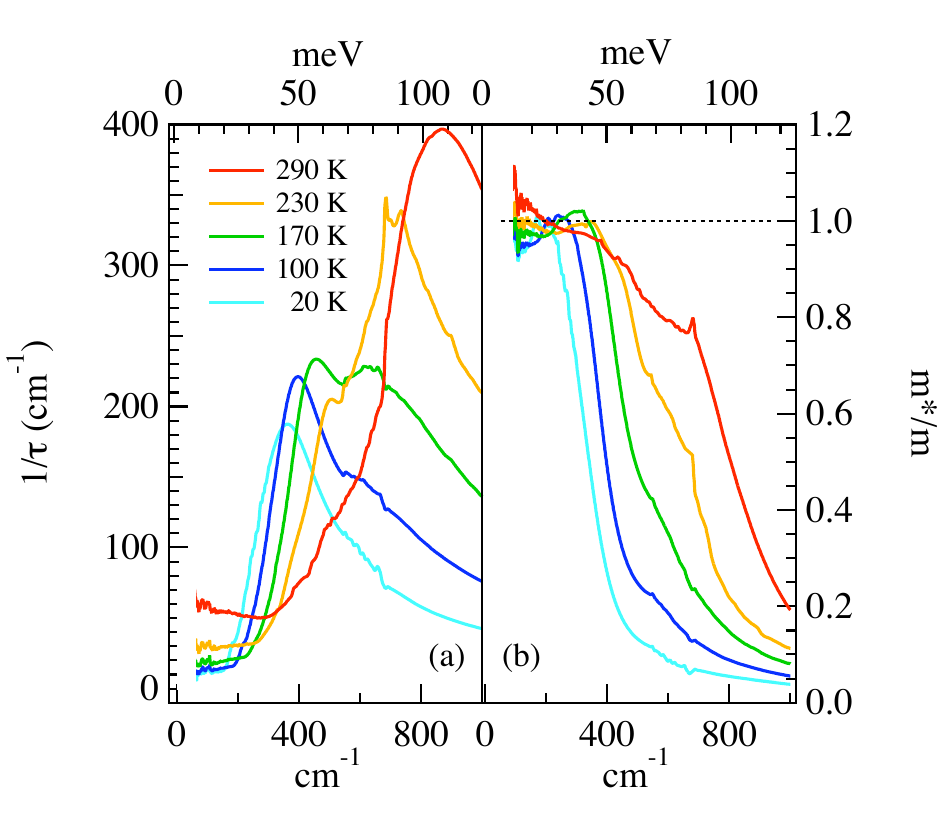}
\caption{(Color) Frequency dependent scattering rate (a) and effective mass (b), calculated from the optical data of elemental bismuth using $\epsilon_\infty$.   The low frequency scattering rate $\tau^{-1}(\omega)$ progressively falls as the temperature is lowered.  An approximately
frequency independent region is interrupted by a sharp onset in scattering and a decrease of the effective mass, which was associated with plasmon scattering.} \label{BiEDM}
\end{center}
\end{figure}

\begin{figure}[htb]
\begin{center}
\includegraphics[width=8.5cm,angle=0]{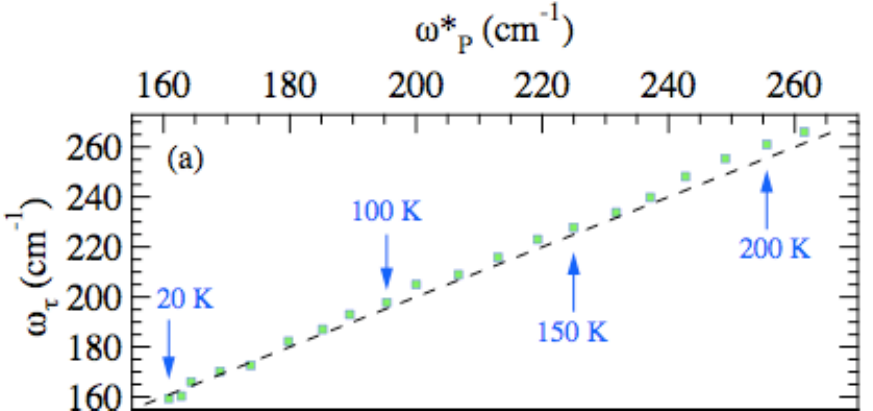}
\vspace{4mm}
\caption{(Color) A parametric plot of the sharp onset in scattering plotted versus the independently measured plasmon frequency in elemental bismuth. The plasmon frequency changes monotonically as the sample is cooled from room temperature to 15K.  } \label{Parametric}
\end{center}
\end{figure}

Although the results of the calculation displayed in Fig. \ref{PuchhkovSim} is for phonons, the general idea holds for electron-boson coupling  generically \cite{Marsiglio01a, Carbotte99a} although, of course, the bosonic density of states is generally more complicated than a single sharp mode.  For instance Tediosi \textit{et al.} \cite{Tediosi07a} found a remarkable example of this concept in semi-metal bismuth, where the bosonic excitation is formed out of the collective excitations of the electronic ensemble itself i.e. the plasmon.  This is an effect that should be present in all metals, but is enhanced in a semi-metal like bismuth due to its very low carrier density and high $\epsilon_\infty \approx 100 $.   At low frequencies Tediosi \textit{et al.} \cite{Tediosi07a} found  in their EDM analysis (Fig. \ref{BiEDM}) that the mass and scattering rate are approximately frequency independent as expected for the Drude model.  However, at higher frequencies there is sharp onset in the scattering rate at a strongly temperature dependent position.  As one cools bismuth, the charge density changes dramatically and the plasmon frequency (as given by the zero crossing of $\epsilon_1$) drops by  a factor of almost two.  It was found that the energy scale of this sharp onset in scattering tracks the independently measured plasmon frequency as shown in Fig. \ref{Parametric}.  One can note the strong resemblance of the data to the calculation of an electron interacting with sharp mode in Fig. \ref{PuchhkovSim}.  A strong coupling as such between electrons and plasmonic electron degrees of freedom may in fact be captured within the same Holstein Hamiltonian that is used to treat the electron - longitudinal phonon coupling and describe polarons.  For this reason this collective composite excitation has been called a $plasmaron$ \cite{Lundqvist1,Lundqvist2} when exhibited in the single particle spectral function.  The observation of Tediosi \textit{et al.} was the first such observation optically.

\begin{figure}[htb]
\begin{center}
\includegraphics[width=8.5cm,angle=0]{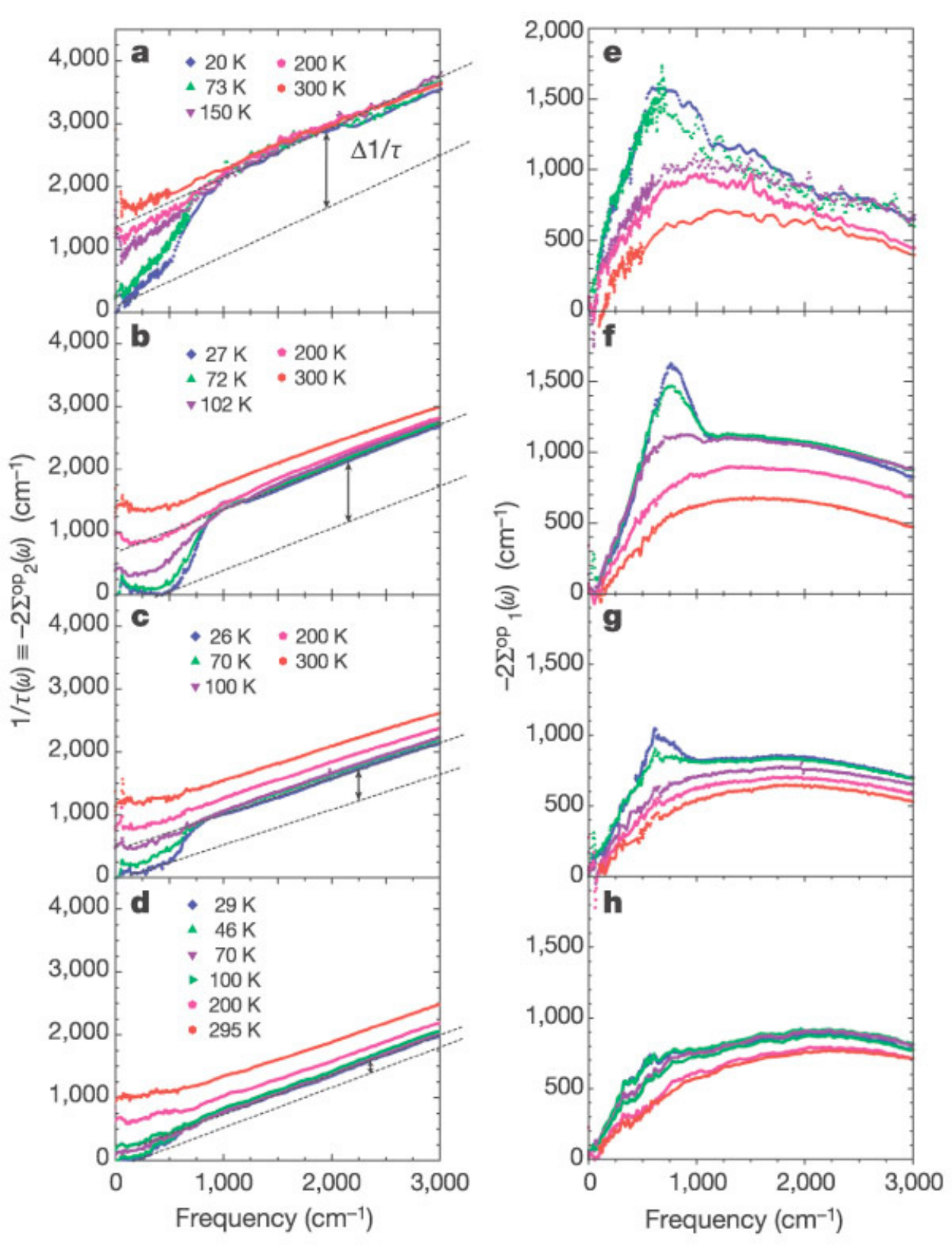}
\caption{(Color) a-d, The frequency and temperature dependent optical scattering rate, $1/\tau(\omega) $ for four doping levels of Bi$_2$Sr$_2$CaCu$_2$O$_{8+\delta}$.  a, T$_c$ = 67 K (underdoped); b, 96 K (optimal); c, 82 K (overdoped); d, 60 K (overdoped). e-h, The real part of the optical self-energy, - 2$\Sigma_1^{op}(\omega)$ as defined in Eq. \ref{opticalselfenergy}.  From Ref. \onlinecite{Hwang04a}.} \label{ScatRateCuprates}
\end{center}
\end{figure}

EDM analysis has been applied extensively to the high-T$_c$ cuprate superconductors \cite{Basov05a}.  In Fig. \ref{ScatRateCuprates} I show the imaginary and real parts of the optical self-energy as defined in Eq. \ref{opticalselfenergy} above for a series of four doping levels of the compound Bi$_2$Sr$_2$CaCu$_2$O$_{8+\delta}$ \cite{Hwang04a}.  The displayed samples span the doping range from the underdoped to the severely overdoped.    Again one can notice the resemblance of the data to that of the model calculation in Fig. \ref{PuchhkovSim} and indeed data of this kind, along with similar signatures in angle-resolved photoemission was interpreted as a generic signatures of electron-boson interaction in the cuprates.

The identity of this boson is a matter of intense current debate in the community.  Some groups have pointed to a many-body electronic source \cite{Carbotte99a, Hwang04a,Kaminski01a,Johnson01a} related to the `41 meV' magnetic resonance mode discovered via neutron scattering \cite{Rossat91a,Mook93a}.  Others have argued that these feature's presence above T$_{c}$, the relative doping independence of its energy scale, and its universality among material classes demonstrate a phononic origin and indicates the strong role that lattice effects have on the low-energy physics \cite{Lanzara01a} 
\footnote{My (NPA's) opinion is that these are by and large straw-man arguments and perhaps not the right way to approach the problem of the origin of high-T$_c$.  In a strongly correlated system like the cuprates, it is clear that everything is strongly coupled to everything else.  Magnetism is strongly connected to lattice effects and vice versa.  In this regard, it is not surprising that experiments sensitive to the lattice show lattice effects and experiments sensitive to magnetism show magnetic effects.  One should not expect in such materials that various degrees of freedom neatly partition in separate distinct subsystems.}.  We note that whatever the origin of this effect, it is almost certainly the same as that which causes the upturn in the optical scattering rate  \cite{Hwang04a,Carbotte99a,Basov05a}.  We should also mention that much of the discussion regarding these self-energies and the like has implicitly assumed that the only possibility of understanding such features are a consequence of some sort of electron-boson coupling, but this is not necessarily the case.   There are proposals that purely electronic interactions and proximity to a Mott insulator can give such features \cite{Chakraborty08a,Byczuk07a}.

\subsection{Frequency dependent scaling near quantum critical points.}

A quantum phase transition (QPT) is a zero-temperature transition between two distinct ground states as a function of a non-thermal parameter, such as magnetic field, pressure, charge density etc. \cite{Sondhi97a,SachdevBook}.  Unlike conventional phase transitions, which are driven by thermal fluctuations, they are driven essentially by diverging fluctuations of the system's zero point motion.
  Once thought to be only of academic interest,  the existence of a QPT nearby in a material's parameter space is believed to influence a whole host of finite temperature properties.  QPTs may hold the key to understanding the unusual behavior of many systems at the forefront of condensed matter physics  \cite{Sondhi97a}. 

Just as in the case of conventional phase transitions, QPTs are believed to be characterized by diverging length and time scales if second order.  In the disordered state, one envisions an order parameter that fluctuates slower and slower over longer and longer length scales as the transition is approached until at the transition these length and time scales diverge.  These length and time scales are the correlation length $\xi_c$ and time $\tau_c$.  The divergence of the fluctuation time scales, which in the case of conventional transitions is called $critical$ $slowing$ $down$ implies the existence of a characteristic frequency $\omega_c$ which vanishes at the transition.  

In a QPT, the transition occurs as a function of some non-thermal parameter $K$.  It is usually considered that close to the transition at $K_c$,   $\xi_c$ diverges in a manner $\propto |K- K_c|^{-\nu}$ and $\tau_c$ diverges $\propto |K- K_c|^{-z \nu } \sim \xi^{z}$ where $\nu$ and $z$ are called the correlation length and dynamic exponent respectively.

From the classical case a well established formalism exists for the understanding of various physical quantities close to continuous (2nd order) transitions.  Widom's scaling hypothesis assumes that close to the transition the $only$ relevant length and time scales are those associated with diverging correlations in the order parameter \cite{Stanley71a}.  Such considerations hold for QPTs also and means that physical observables such as magnetization or conductivity are expected to have scaling forms, in which independent variables such as the temperature, probe frequency or wave vector appear in the argument only as a product with quantities like $\xi_c$ and $\tau_c$.  This is expressed formally in terms of $finite-size$ scaling.  For an incident wave vector $q=0$ such a scaling form can be written

 \begin{eqnarray}
O(K,\omega,T)= \frac{1}{T^{d_o/z}} F(\omega/T , T \tau_c).
\label{scaling}
\end{eqnarray}

Here $O$ is some observable and $F$ is the scaling function.  $d_o$ is the scaling dimension, which determines how various physical quantities change under a renormalization group transformation.  It is often close to the ``engineering" dimension of the system (for instance 0 for 2D conductivity), but can acquire an anomalous dimension in certain circumstances.   Generally one expects frequency and temperature dominated regimes of these response functions for $\omega \gg T,\omega_c$  and $T \gg \omega,\omega_c$ respectively.  For further discussion on scaling forms and motivation on this and other related points see the excellent review by Sondhi \textit{et al.} \cite{Sondhi97a}.  

Scaling functions written as a function of $\xi_c$ and $\tau_c$ can be very powerful as they allow the analysis of experimental data independent of microscopic theory.  For instance, the scaling exponents $\nu$ and $z$ only depend on certain global properties of the system such as symmetry, dimensionality, and the nature of the dominant interactions.  

There are many studies where the existence of a QPT is inferred from frequency dependent scaling (see  for example \cite{vanderMarel03a}) in a regime where frequency effects dominate over temperature ones, but there are far fewer studies where scaling is investigated as one passes through a known QPT.  I know of only two where scaling in the frequency dependent conductivity has been investigated.  This is unfortunate because, as mentioned, such studies could potentially be tremendously powerful for the investigation of correlated systems where the appropriate underlying physical model is not clear.  This shows the extreme experimental challenges inherent in  accessing the experimentally relevant frequency and temperature ranges where $\hbar \omega \gg k_B T$ while both $\hbar \omega$ and $k_B T$ are as low as possible, but also below any higher energy scales that are not relevant to the ordered state.  Hopefully this situation will change with the increasing prominence and feasiility of broadband microwave and THz techniques.

\begin{figure}[htb]
\begin{center}
\includegraphics[width=8cm,angle=0]{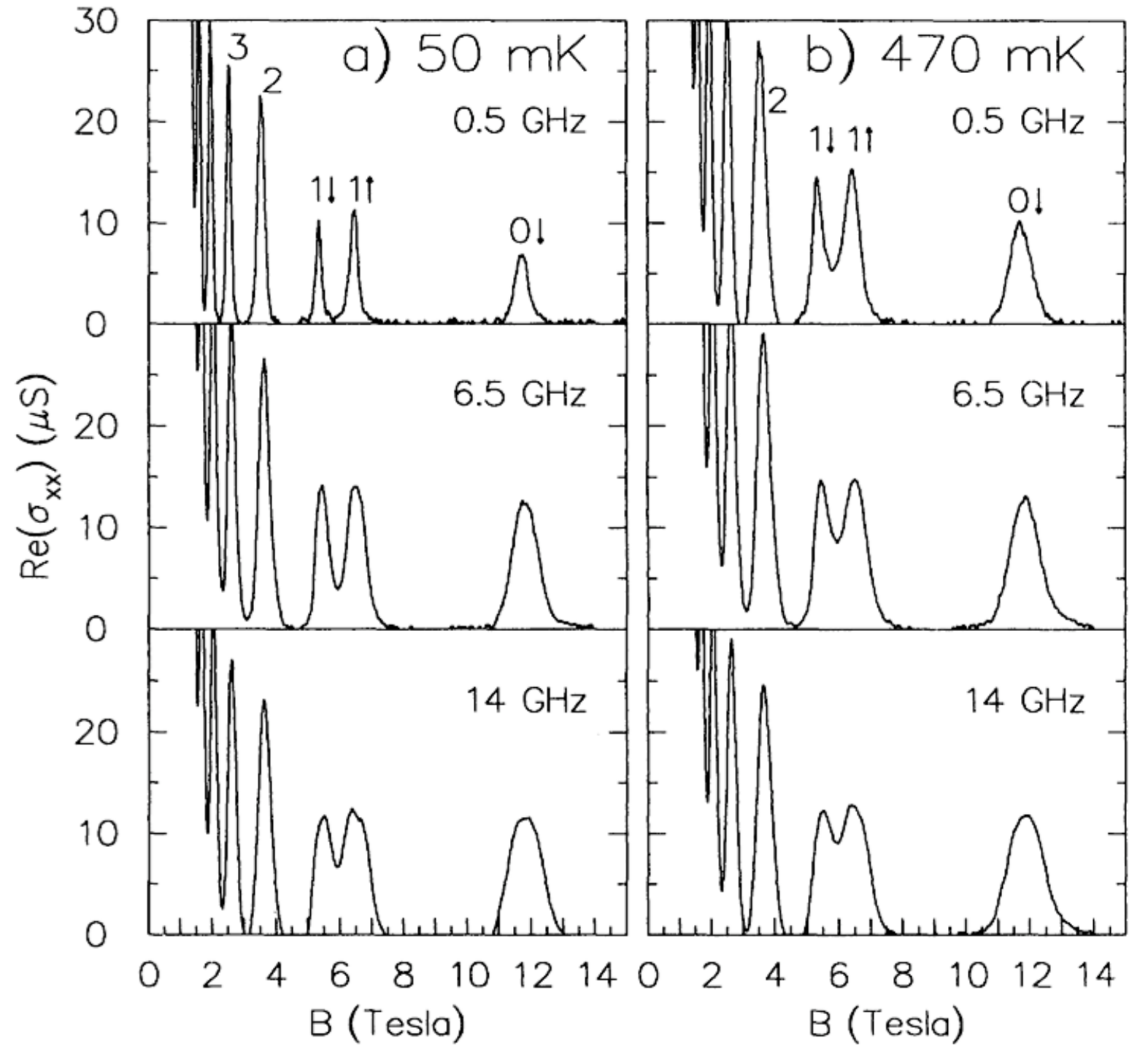}
\caption{Re $\sigma_{xx}$ vs. B at three frequencies and two temperatures.  Peaks are marked with Landau level $N$ and spin index. } \label{Engel1}
\end{center}
\end{figure}

The first experimental study to investigate frequency dependent scaling near a QPT was that of Engel \textit{\textit{et al.}} \cite{Engel93a}, who used a waveguide coplanar transmission setup to perform broadband microwave spectroscopy on quantum Hall systems.  In quantum Hall systems the longitudinal resistivity goes to zero at field values where the transverse resistivity assumes perfectly quantized values.  Although the value of the Hall resistivity is perfectly quantized, the width of the transition region depends on temperature \cite{Wei88a} and on measurement frequency \cite{Engel93a}.  In Fig. \ref{Engel1} the Re $\sigma_{xx}$ vs. B is shown at three different frequencies and two different temperatures.  One can see that at 50 mK, the width of the transitions regions ($\Delta B$) is strongly frequency dependent, while at 470 mK the widths are almost frequency independent.

\begin{figure}[htb]
\begin{center}
\includegraphics[width=8cm,angle=0]{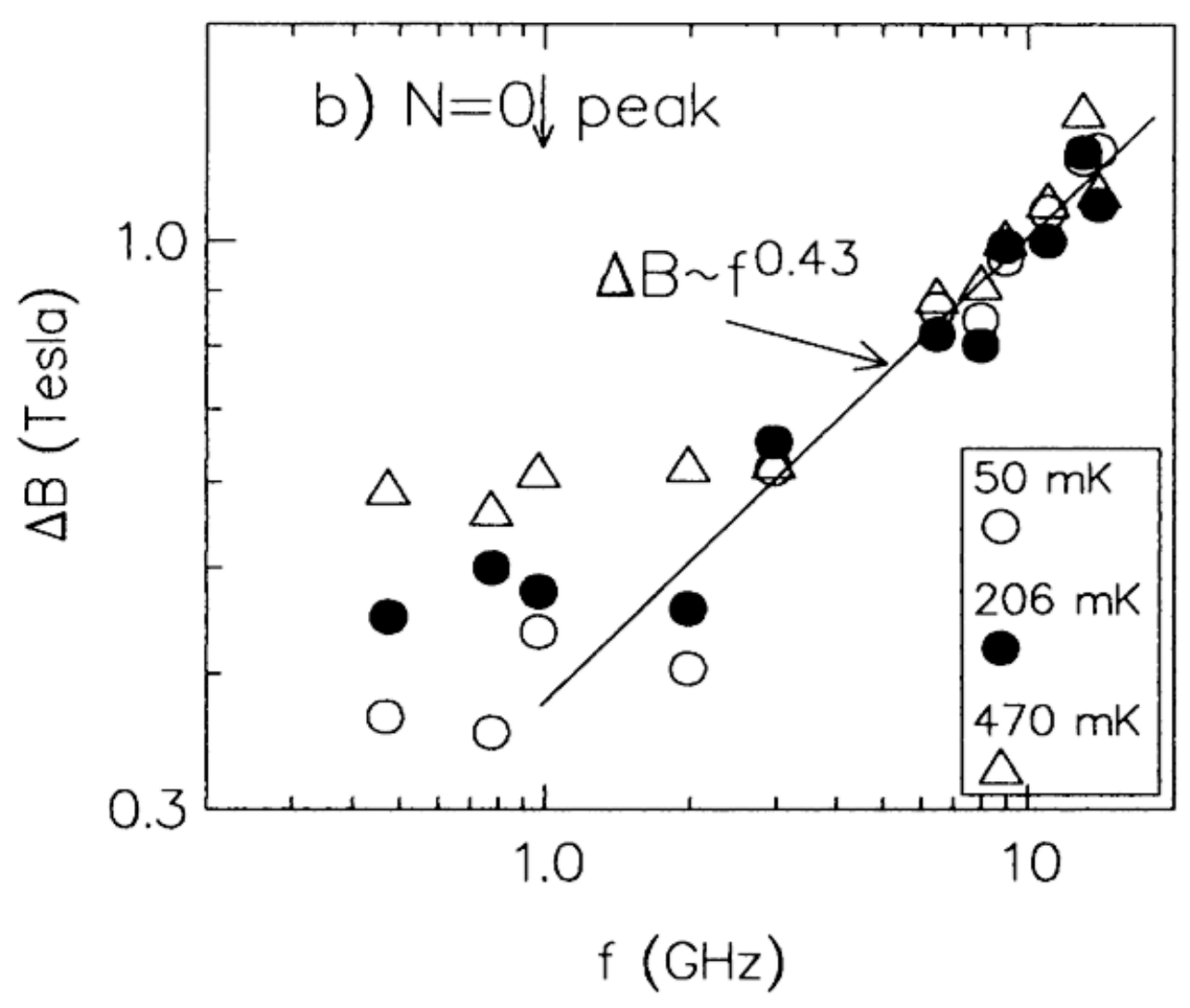}
\caption{Peak width between extremal points in longitundinal resistance  $d$ Re $\sigma_{xx}/ d B$ $\Delta B$ vs. measurement frequency at three different temperatures.  Data are shown for the $N=0 \downarrow$ Landau level.} \label{Engel2}
\end{center}
\end{figure}

This frequency and temperature dependence of  $\Delta B$ (as defined as the extrema of the derivatives of  Re $\sigma_{xx}$) is summarized in Fig. \ref{Engel2} for one of the spin-split levels.   Similar data are found for other levels.  One sees that for all such levels at low frequencies, the data are frequency independent and temperature dependent, whereas for low temperatures the data are temperature independent and frequency dependent.  The crossover between regimes takes place when $3 \hbar \omega \sim  k_B T$.  Such data are consistent with scaling theories of quantum criticality, where the only relevant frequency scale at the QPT is set by the temperature itself.  It is also significant that the crossover condition between frequency and temperature dominated regimes involves $\hbar$, showing that the physics is essentially quantum mechanical.  In the high frequency data it is observed that spin split levels give a $\Delta B$ which is approximately $\propto \omega^{\gamma}$, with $\gamma =  0.43$.  This is consistent with scaling theories that give an exponent for quantum percolation of $1/\gamma = z \nu = 7/3$ and also consistent with the temperature dependence of these transition widths as measured by DC transport that show a $T^\frac{1}{7/3}$ dependence.  It was claimed that through an analysis of both the temperature and frequency dependent exponents, one can extract the dynamic exponent $z$, which  yields $z=1$.  This is consistent with long range Coulomb interaction.  

Both the DC data \cite{Wei88a} and the high frequency data have transition widths which exhibit a power law dependence on temperature or frequency with an exponent of $3/7$ at all spin-split transitions, showing that all such transitions between quantum Hall levels fall into the same universality class.   It should be possible in such a study to incorporate both temperature and frequency dependence into a single scaling function although this has not been done yet.  It is expected that a general scaling function of the form

 \begin{eqnarray}
\rho(B,T,\omega) =  F(\omega/T,\delta/T^{\frac{1}{z\nu}})
\label{OmegaTScaling}
\end{eqnarray}

\noindent applies where $\delta = |B - B_c|/B_c$ measures the distance to the quantum critical point.  This function is equivalent to Eq. \ref{scaling} using the fact that the scaling dimension of the resistivity vanishes in $d=2$.  In the high frequency and high temperature limits,  Eq. \ref{OmegaTScaling} reduces to one where frequency or temperature appear in the argument as $\frac{\delta}{\omega^{1/zv}}$ and $\frac{\delta}{T^{1/zv}}$ respectively.

\begin{figure}[htb]
\begin{center}
\includegraphics[width=8cm,angle=0]{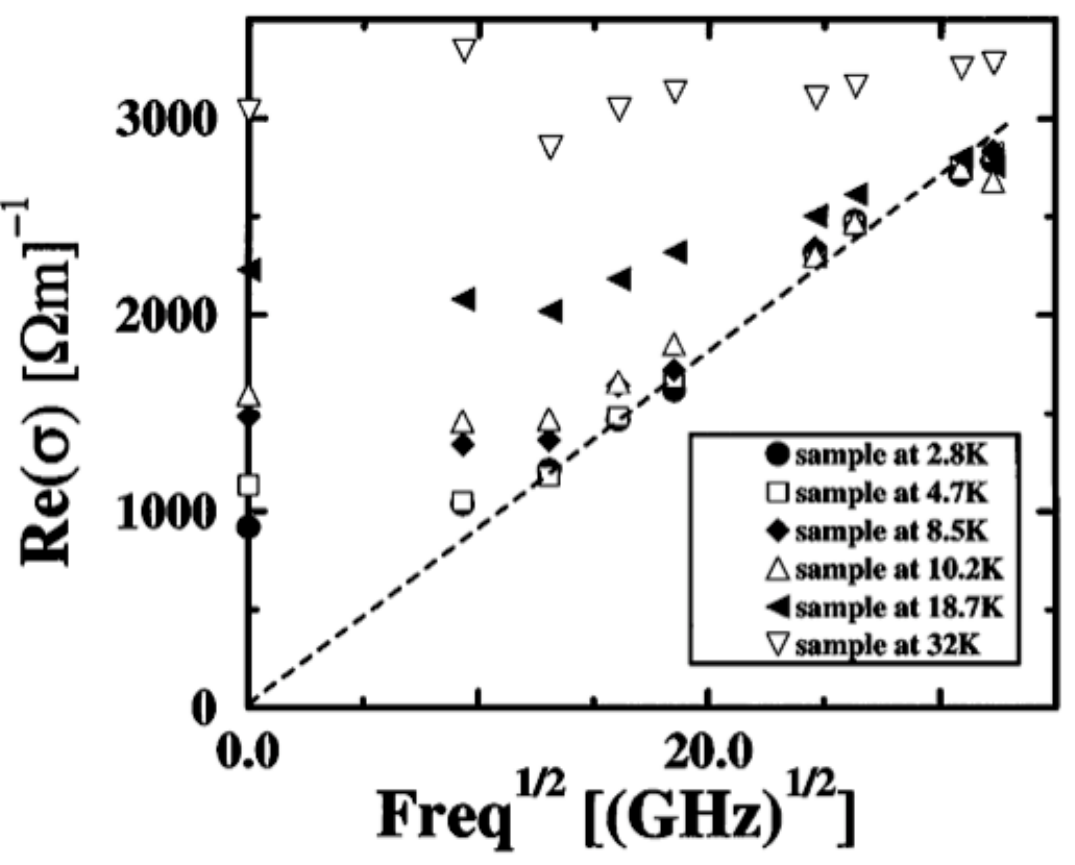}
\caption{Re$\sigma(\omega)$ plotted against $\sqrt{\omega}$ for temperatures 2.8 to 32K and frequencies 87-1040 GHz. For the lowest temperatures and when $\hbar \omega > k_B T$, the data follow a $\sqrt{\omega}$.  The data approaches the DC values plotted on the vertical axis for $\hbar \omega \ll k_B T$.  } \label{Carini1}
\end{center}
\end{figure}

\begin{figure}[htb]
\begin{center}
\includegraphics[width=8cm,angle=0]{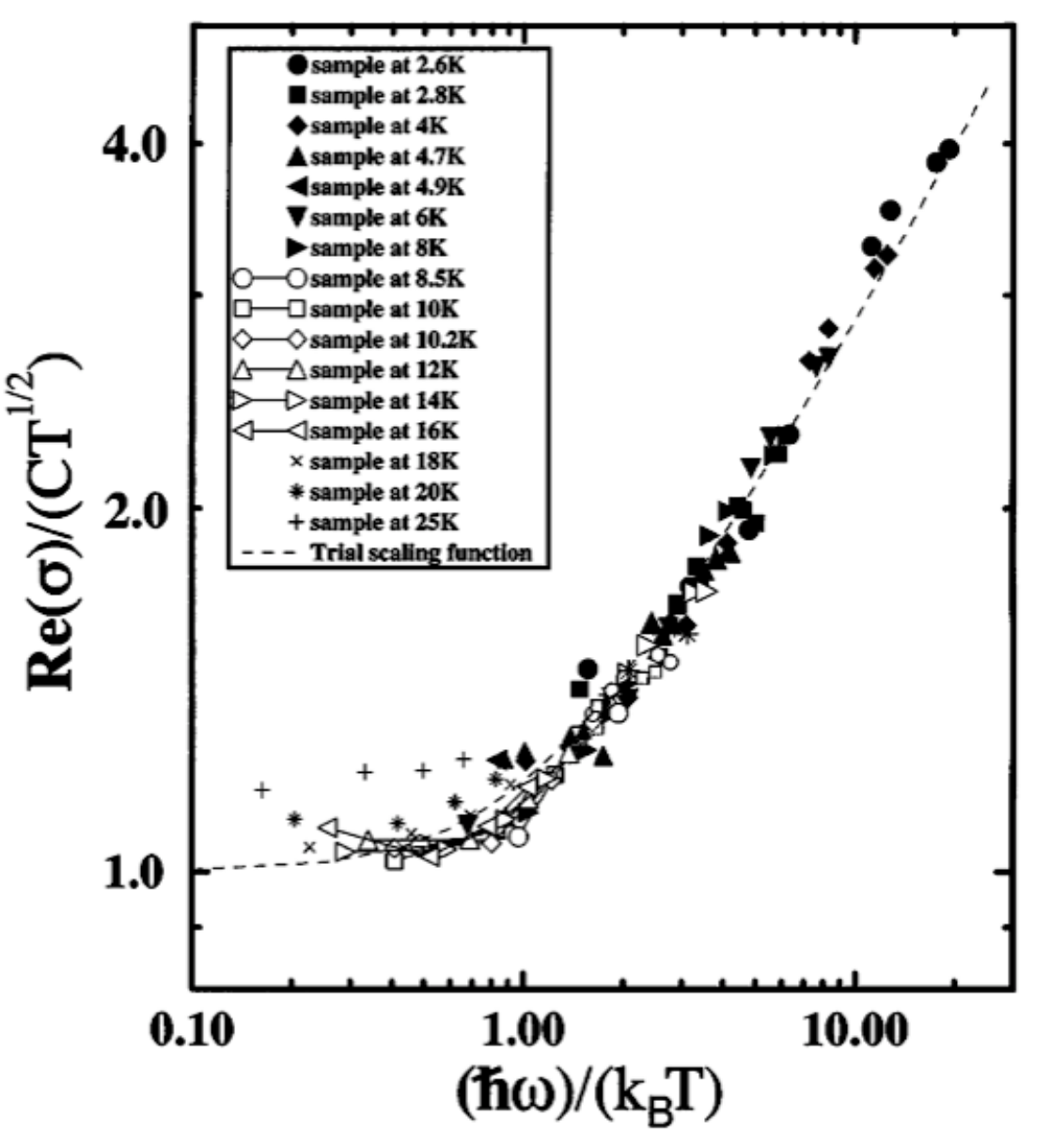}
\caption{Log-log plot of conductivity scaled by the factor  $C=475 (\Omega m K^{1/2})^{-1}$ vs. $\omega/T$.  For temperatures 16 K and below, the data for the entire experimental frequency range collapse onto a single curve within the experimental noise. Higher temperature data begin to rise above the collapsed data systematically for low scaled frequencies. The dashed line is a trial scaling function as described in the text.} \label{Carini2}
\end{center}
\end{figure}

The other frequency dependent study was that of Lee \textit{et al.} \cite{Lee98a,Lee00a} who applied the ideas of finite size scaling to the 3D metal-insulator (MI) transition in Nb$_{1-x}$Si$_x$.  They measured the frequency dependent conductivity $via$ millimeter wave transmission through thin films whose relative concentrations of Nb to Si were tuned exactly to the MI QPT point at $x_c$.  As in the study of Engel \textit{et al.} \cite{Engel93a}, for samples tuned right to quantum criticality it is observed that for the $\hbar \omega \gg k_B T$ data the conductivity is temperature independent and that for  $\hbar \omega \ll k_B T$, the data are frequency independent (Fig. \ref{Carini1}).  A crossover from a frequency dominated regime to a temperature dominated regime implies the existence of a scaling function $\sum$ of a form that can be inferred from above  Eq. \ref{scaling}

 \begin{eqnarray}
\sigma(x_c,T,\omega) = C T^{\frac{1}{z}}\sum(\frac{\hbar \omega}{k_B T}).
\label{CariniScaling}
\end{eqnarray}

In previous work on similar samples it was found that for samples tuned to the critical concentration the DC conductivity followed a T$^{1/2}$ relation implying $z=2$.  As shown in Fig. \ref{Carini2}, successful scaling can be found by plotting the scaled conductivity Re $\sigma(T,\omega)/C T^{1/z}$ using this $z$.  This procedure collapses the data over the entire measured frequency range for temperatures 16 K and below, implying the applicability of a scaling function that depends $only$ on the scaled frequency.  Data for temperatures higher than 16 K starts to rise above the other collapsed curves, indicating the appearance of other mechanisms (electron-phonon scattering for example) which limits the size of quantum fluctuations.  This is consistent with the DC result where the conductivity starts to deviate from the T$^{\frac{1}{2}}$ above 16 K.

As already inferred from Fig. \ref{Carini1}, the scaling function shows both temperature and frequency dominated regimes, with a crossover at approximately $\hbar \omega \sim k_B T$.  Again, this shows that at the QPT the only energy scale is set by the temperature itself.  Note, that this crossover happens at a slightly different $\omega/T$ ratio than the quantum Hall case, but is still of order unity as expected.   For high frequencies the scaling function follows a power law dependence with the same exponent as was used to scale the vertical axis with $z=2$.  This means that an equivalent scaling function could have been found by dividing the conductivity data by $\omega^\frac{1}{2}$ and plotting the data as a function of $T/\omega$.  Lee \textit{et al.} \cite{Lee98a} attempted to guess the form of the scaling function with the expression $\sum = Re(1 - i b \hbar \omega/k_B T)^{1/2})$.  The function successfully captures the high and low $\omega/T$, but misses the sharp crossover near $\hbar \omega/k_B T \sim 1 $.

The value of the dynamic exponent  $z=2$, which was found in this study is interesting.  It is different than what is expected from models without a density of states singularity, where $z=d$ or for those with straight Coulomb interaction ($z=1$ observed in the quantum Hall case).  However, it is consistent with several field-theoretic scenarios which give $z=2$ \cite{Belitz94a}.

\begin{figure}[htb]
\begin{center}
\includegraphics[width=7cm,angle=0]{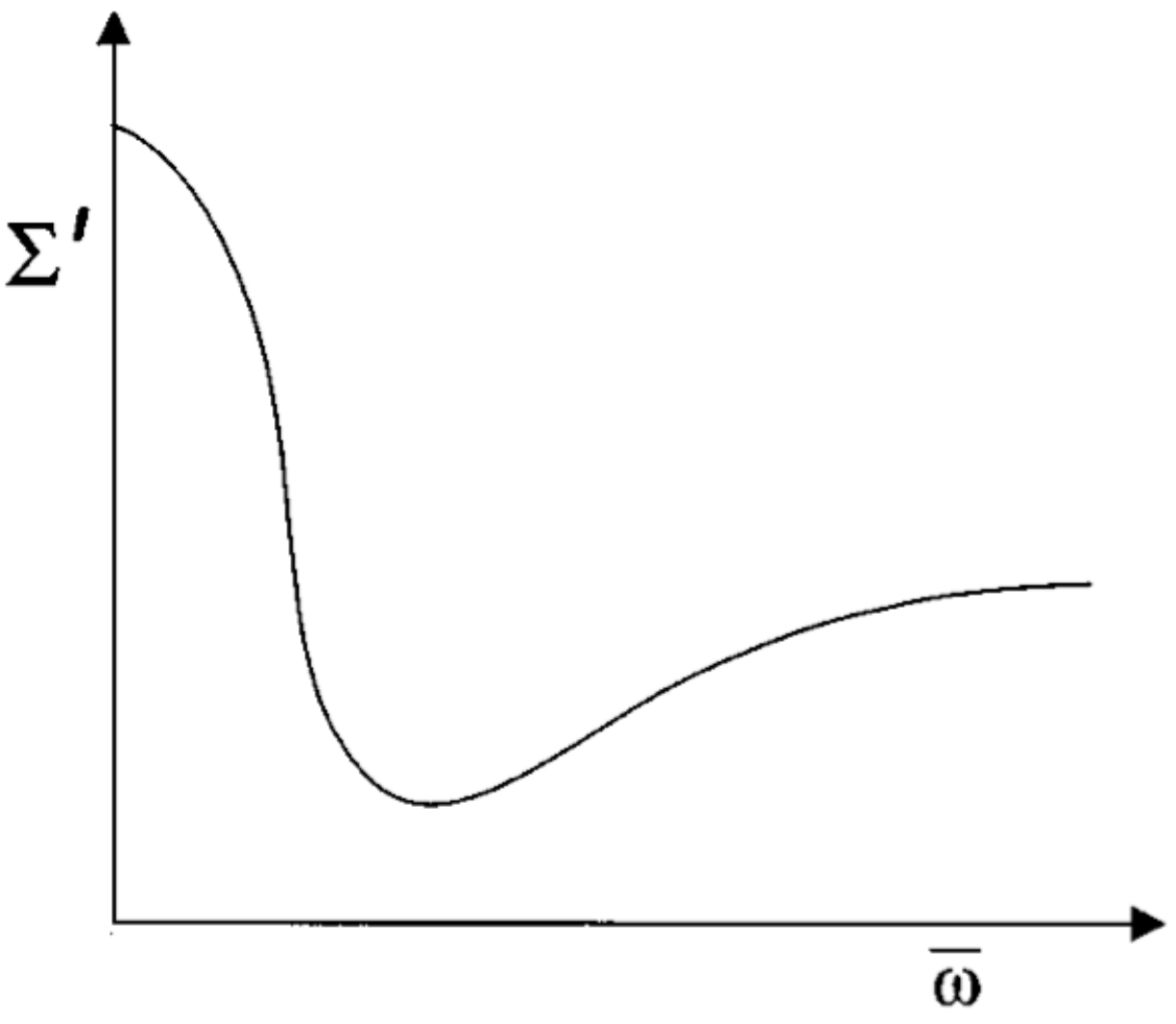}
\caption{The real part of universal scaling function $ \Sigma$, as a function of $\bar{\omega} = \omega /T $.    There is a Drude-like peak from the inelastic scattering between thermally exciting carrier, which falls off at order  $\bar{\omega} \sim 1$.  At larger  $\bar{\omega}$ there is a crossover to a collisionless regime.   Importantly the function gives different values in the  $\bar{\omega} \rightarrow 0$ and  $\bar{\omega} \rightarrow \infty$ limits.} \label{Damle}
\end{center}
\end{figure}

I should mention that as presented above, frequency scaling does not necessarily give us any information on exponents etc. that we could not also get from the temperature scaling. Here the primary importance of frequency scaling was in its ability to demonstrate how characteristic time scales diverge at the QPT, which leaves the temperature itself as the only energy scale in the problem.  However, as discussed by Damle and Sachdev \cite{Damle97a}, the fact that at the QPT itself, response functions can written as a universal function of $\omega /T$, means that one does not necessarily expect the same behavior in the $\omega =0$, $T \rightarrow 0$ (incoherent) limit as in the $\omega \rightarrow 0$, $T = 0$ (phase-coherent) limit  .  Since all DC experiments are in the former limit, while the vast majority of theoretical predictions are in the latter, finite frequency measurements can in fact given unique insight.   It was predicted by Damle and Sachdev that at the 2D superfluid-insulator transition the conductivity is equal to two different universal numbers of order $e^2/h$ in the $\omega/T \rightarrow 0$ and  $\omega/T \rightarrow \infty$ limits as shown in Fig. \ref{Damle}.  This is true even if $\omega$ and $T$ are both asymptotically small.

\subsection{Using light to probe broken symmetries}
\label{Symmetries}

Optical spectroscopies are most often used to probe dynamical correlations in materials, but they are also a probe of symmetry.   Polarization anisotropies are of course sensitive to structural anisotropies, but have been much less used as a probe of more exotic symmetry breakings in ordered states.  All physical properties of a solid must respect its underlying symmetries.  As mentioned in Sec. \ref{Responsefunctions}, formally this is codified as Neumann's principle that states that the symmetry transformations of any intrinsic physical property of a crystal (such as a response function) must include $at$ $least$ the symmetry transformations of the point group of that crystal \cite{Nye85a}.  The symmetries of a material typically manifest in the form of an algebra relating matrix elements or overall constraints (transposition, unitarity, hermiticity, normality, etc.) on the tensor elements of a response function.


Traditionally one regards Neumann's principle as constraining the form of response functions directly.   However, in determining the outcome of an optical experiment, it is not enough to know just the response function.  One must apply the appropriate boundary conditions at the vacuum-material interface.  However boundary conditions can have notorious subtleties in magnetoelectric or chiral systems that have a non-local response. Improperly applied this approach has given incorrect answers~\cite{Arfi,Bungay,Mineev10a,Mineev13a,Hosur,Pershoguba} to questions like whether $\mathcal{T}$ symmetric but inversion symmetry breaking systems can have a Kerr rotation (it cannot~\cite{ArmitageJones,Cho2016}).  

Therefore a powerful alternative approach is to constrain the ``scattering functions" directly by symmetry as motivated in Ref.~\onlinecite{ArmitageJones}.  Then one appeals only to the symmetry of the experimental geometry, which means that the possible discrete symmetries of solid such as reflections, rotations, inversion, rotation-reflections and  time-reversal symmetry may be probed by selecting and measuring the incoming and outgoing far field polarization states of light.

In linear response the optical transmission or reflection properties of a material can be captured by a $2 \times 2$ ``Jones" matrix e.g. a second rank tensor of complex elements.  Just like with the response functions, the symmetries of a material typically manifest in the form of an algebra relating matrix elements or overall constraints (transposition, unitarity, hermiticity, normality, etc.) on the form of Jones matrix.  This was worked out in gratuitous detail by me in Ref. \onlinecite{ArmitageJones}.   Jones matrices are particular useful for time-domain THz spectroscopy as one typically measures the complex transmission function in such experiments.   In the most general case, the Jones matrix is comprised of 4 independent complex values.  In the basis of $x-y$ linear polarization it is

\begin{equation}
\hat{T} = \left[\begin{array}{cc}T_{xx} & T_{xy} \\T_{yx} & T_{yy}\end{array}\right] ,
\label{Tmatrix}
\end{equation}
that acts on an $x-y$ vector in the following fashion

\begin{equation}
 \left[\begin{array}{cc}T_{xx} & T_{xy} \\T_{yx} & T_{yy}\end{array}\right] 
\left[\begin{array}{c}E_{x}^i \\E_{y}^i\end{array}\right]  = \left[\begin{array}{c}E_{x}^t \\E_{y}^t\end{array}\right].
\label{Tmatrix}
\end{equation}

The above equation forms an eigenvalue-eigenvector problem.    If light is sent through the system under test with a well-defined linear polarization state in, for instance, the $x$ direction, then the Faraday rotation (rotation of transmitted light) may be measured through the relation tan$(\theta_F) = T_{xy} / T_{xx}$.  Going forward, I will replace the entries $T_{ij}$ in Eq. \ref{Tmatrix} with entries $A,B,C,D$ for convenience e.g. 

\begin{equation}
\left[\begin{array}{cc}T_{xx} & T_{xy} \\T_{yx} & T_{yy}\end{array}\right] =
\left[\begin{array}{cc}A & B \\C & D \end{array}\right] .
\label{ABCD}
\end{equation}

Consider the simplest examples coming from rotation symmetry.  If a material possesses a $\hat{R}_z(\pi/ 2)$  rotation symmetry ($C_4$) around the $z$ axis, its Jones matrix will be invariant under a $\frac{\pi}{2}$ rotation.   Rotation matrices have the form

  \begin{align}  
\hat{R}_z(\phi) &= \left[\begin{array}{cc} \mathrm{cos} \phi &  \mathrm{sin} \phi \\  \mathrm{-sin} \phi &  \mathrm{cos} \phi \end{array}\right].
\label{Rotationmatrix}
\end{align}

The rotation matrix is applied to a Jones matrix by sandwiching it between the rotation matrix and the inverse of the rotation matrix.  This procedure often causes confusion. It can be motivated as follows.  Consider a situation where we consider incident light falling on a material, being transmitted, and then detected in the manner of Eq.  \ref{Tmatrix}.   However note that one never measures the transmitted vector directly.   One must always measure a {\it number} that is the projection of the electric field vector on a reference vector.  So the outcome of a measurement where one wants to know the $x$ electric field of a transmitted wave is expressed as

\begin{equation}
[ 1  \; \; 0] \left[\begin{array}{c}E_{x}^t \\E_{y}^t\end{array}\right] = E_x^t,
\end{equation}
e.g. we measure the electric field by its projection on the $x$-direction.  Now consider a situation where we choose a different rotated coordinate frame to {\it describe} the experiment, but we leave the experiment itself untouched.  Since nothing changes in the experiment, nothing in the experimental signal changes.  The experimental signal could be for instance the voltage on a gauge that is completely oblivious to the fact that the experimenter has changed their reference frame.  This is enshrined in the ``covariance principle" which -- for our purposes -- states that physical laws (which manifest in response functions) are independent of the coordinate system used to describe them.   This is represented mathematically as

\begin{equation}
[ 1  \; 0] \; \hat{R}_z^{-1}(\phi) \; \hat{R}_z(\phi) \; \hat{T}   \; \hat{R}_z^{-1} \;  \hat{R}_z  \; \left[\begin{array}{c}E_{x}^i \\E_{y}^i\end{array}\right] = E_x^i.
\end{equation}
It corresponds to choosing a rotated coordinate frame by rotating the coordinates of the initial vector (by $ \hat{R}_z(\phi)$), the coordinates describing the transmission matrix (by $ \hat{R}_z(\phi) \; \hat{T}  \; \hat{R}_z^{-1} $), and the coordinate of the reference vector (by $ \hat{R}_z^{-1} $ operating from the right).  These transformations clearly leaves the experimental signal unchanged as the rotation matrices are multiplied in pairs with their inverses.  Now one must recognize that rotating the coordinates to describe the system is the same as counter rotating the entire system (reference vector, sample, and initial field).   Again the same experimental signal would be obtained.  {\it Now} consider rotating only the sample while leaving all else invariant.   This operation clearly corresponds to  $ \hat{R}_z(\phi) \; \hat{T}  \; \hat{R}_z^{-1}(\phi) $, from which one can infer the general form of a point group transformation applied to a transmission matrix.

Applying the relevant rotation matrix for $\hat{R}_z(\pi/ 2)$ to Eq. \ref{ABCD} gives
 
  \begin{align}  
\hat{R}_z(\pi/ 2) &= \left[\begin{array}{cc} 0 &  1 \\  -1 &  0 \end{array}\right],  \nonumber \\    \tilde{\hat{T}} = \hat{R}_z (\pi/2)  \hat{T}  \hat{R}_z^{-1}(\pi/2) &= \left[\begin{array}{cc}D & -C \\-B & A \end{array}\right] .
\label{Piover2Rotation}
\end{align}

\noindent If the material has $\hat{R}_z(\pi/ 2)$ symmetry then this matrix must be equal to the initial one in Eq. \ref{ABCD} and hence the Jones matrix for a material with $\hat{R}_z(\pi/ 2)$ symmetry must have the form

\begin{equation}
C_4(z)  \;\;  \Longrightarrow \; \; \hat{T} = 
\left[\begin{array}{cc}A & B \\-B & A \end{array}\right].
\label{C4}
\end{equation}

Other symmetries (and even time-reversal symmetry) can be handled in a somewhat similar fashion as shown in Ref. \onlinecite{ArmitageJones}.   TDTS measurements probing and related ones probing circular polarization anisotropies have played an important role in determining symmetry in the cuprates \cite{Xia08a,Lubashevsky14a}.

Despite the utility in applying Neumann's principle to the scattering matrices directly, in some circumstances insight can be gained from applying symmetry considerations directly to response functions  e.g. to the conductivity and not only the scattering functions, which are themselves functions of the response functions.  A simple example due to J. Orenstein\footnote{From his 2013 Cargese lectures.} showing the power of this approach is as follows.   Consider the possibility of a ``bizzare" Ohm's law that reflects a current driven by magnetic field $J_i = \beta_{ij} B_j$, which can be contrasted to the usual Ohm's law $J_i = \sigma_{ij} E_j$.   The above mentioned covariance with respect to the inversion of the coordinate system gives that  $- J_i =  \mathcal{I}\{ \beta_{ij} \} B_j$ and hence $  \mathcal{I}\{ \beta \} = - \beta$.  $\mathcal{I}\{ \beta_{ij} \}$  represents the inversion operation acting on the response function.  Note that inversion does not change the axial vector magnetic field.  However if the material has inversion symmetry
the aforementioned Neumann's principle determines that $  \mathcal{I}\{ \beta \} =  \beta$.  For both Neumann's principle and the covariance principle to apply, $\beta =0$ and hence an Ohm's law of this kind cannot exist.

However, applying symmetry operations to response functions can result in complications and not just with regards to the boundary conditions discussed above.   A simple extension of the above treatment to the normal Ohm's law with respect to time-reversal symmetry leads to an erroneous conclusion.   A time-reversed Ohm's law is $- J = \mathcal{T}\{ \sigma \} E$.   Here time-reversal affects the direction of current, but not the electric field $E$.  Covariance with regards to time implies that $ \mathcal{T}\{ \sigma \} = - \sigma$ and with Neumann's principle this seems to indicate that the conductivity would be zero in a system with time-reversal symmetry.  This is obviously not true. The essential problem is that a non-zero dissipative conductivity implies an entropy that is increasing in time, which gives a direction to the arrow of time.  In order to do this more properly one has to formulate a condition similar to the de Hoop reciprocity that I had used in Ref. \onlinecite{ArmitageJones} for scattering matrices that allows one to distinguish between the breaking of time-reversal symmetry in the material and dissipation.   Applying this idea to response functions directly has not been done to the best of my knowledge, but would be interesting.

Even more information about symmetries can be determined by probing the {\it non-linear} optical response of materials.  As shown in Eqs. \ref{PolarizationExpansion} and \ref{OhmsLawExpansion}, frequently we will want to expand the response of materials in powers of the driving field.  The non-linear response is determined by tensors of higher rank and hence are constrained by symmetries in an even more explicit fashion.    For instance the second harmonic generation (SHG) $\chi^{(2)}$ response is particularly sensitive to the presence of global inversion symmetry because unlike the linear electric-dipole susceptibility tensor, which is allowed in all crystal systems, the SHG electric dipole susceptibility vanishes in centrosymmetric point groups, leaving the much weaker electric-quadrupole susceptibility as the 
primary source of second harmonic signal.  Measurements of the nonlinear susceptibility have made important contributions in the detection of subtle broken symmetry states of matter in cuprates and iridates \cite{Zhao16a,Zhao17a}.   Although some aspects of these symmetries can -- in principle -- be measured through x-ray or neutron diffraction experiments displacements of oxygen or bond currents (as proposed for the pseudogap state of the cuprates) may not be very sensitive to x-rays.  Polarization experiments also have the unique advantage of being sensitive to true long-range order e.g. slow fluctuations of an order will not give a polarization anisotropy.   This is unlike scattering style experiments that always have a finite frequency window of the scattered particle and can in principle pick up a signal from a slowing fluctuating order.

\begin{figure}[htb]
\begin{center}
\includegraphics[width=7cm,angle=0]{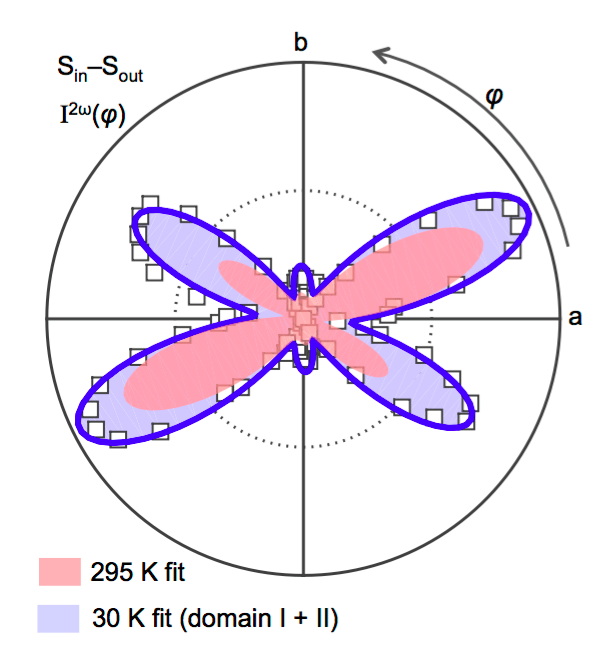}
\caption{Angular anisotropy of the second harmonic generation signal of YBa$_2$Cu$_3$O$_y$ above and below  the pseudogap temperature $T^*$.  Polar plot of  $I^{2\omega} (\varphi) $ measured at T = 30 K in $S_{in}-S_{out}$ geometry from YBa$_2$Cu$_3$O$_{6.92}$ (squares). The blue curve is a best fit to the average of two 180° rotated domains with $2'/m$ point group symmetry as described in Ref. \onlinecite{Zhao17a}. The fit to the T = 295 K data (pink shaded area) is overlaid for comparison.  From Ref. \onlinecite{Zhao17a}.} \label{RotationalAniso}
\end{center}
\end{figure}

\begin{figure}[htb]
\begin{center}
\includegraphics[width=7cm,angle=0]{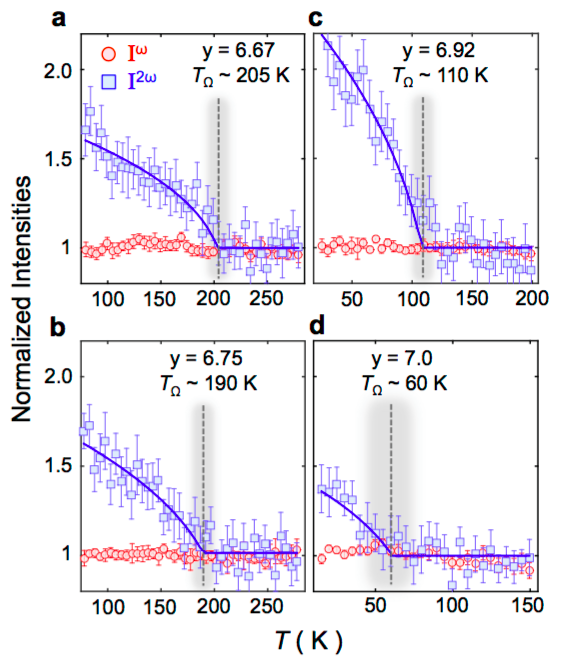}
\caption{Inversion symmetry breaking transition across the phase diagram of YBa$_2$Cu$_3$O$_y$. Temperature dependence of the normalized linear and SH response of YBa$_2$Cu$_3$O$_y$ with a) $y = 6.67$, b) 6.75 c) 6.92 and d) 7.0  to their room temperature values. Data were taken in  $S_{in}-S_{out}$  geometry at angles corresponding to the smaller SHG lobe maxima at T = 295 K.  Blue curves overlaid on the data are guides to the eye. The inversion symmetry breaking transition temperatures T$_{\Omega}$ determined from the SHG are marked by the dashed lines. The width of the shaded gray intervals represent the uncertainty in T$_{\Omega}$.  From Ref. \onlinecite{Zhao17a}.} \label{TempDepend}
\end{center}
\end{figure}

In Fig. \ref{RotationalAniso}, I show the data of Zhao et al. who used second harmonic generation to reveal the symmetries broken at the pseudogap temperature in the cuprates \cite{Zhao17a}.   They show that spatial inversion and two-fold rotational symmetries are broken at a temperature close to T$^*$ (Fig. \ref{TempDepend}) while mirror symmetries perpendicular to the Cu-O plane are absent at all temperatures. This transition occurs over a wide doping range and persists inside the superconducting dome, with no detectable coupling to either charge ordering or superconductivity. The results suggest that the pseudogap region coincides with an odd-parity order that does not arise from a competing Fermi surface instability and exhibits a quantum phase transition inside the superconducting dome.

Other work has uncovered a odd-parity hidden order in a spin-orbit coupled correlated iridate \cite{Zhao16a} and a multipolar nematic phase of matter in the metallic pyrochlore Cd$_2$Re$_2$O$_7$ \cite{Harter17a}.  This latter phase spontaneously breaks rotational symmetry while preserving translational invariance, but is odd under inversion symmetry.   Such a multipolar nematic order cannot be identified even in principle using using conventional charge transport anisotropy measurements because the loss of inversion symmetry is manifested in the spin texture of the Fermi surface.

\subsection{Pump-probe spectroscopy and non-equilibrium effects}
\label{PumpProbe}

Thermal equilibrium is one of the cornerstones of modern statistical mechanics and condensed-matter physics. By considering and averaging over ensembles that do not evolve in time, microscopic physical phenomena can be related to macroscopic physical laws, in a general way that is independent of the specific model used to describe them. Indeed, the idea of thermal equilibrium even allows us to define concepts as basic as that of temperature itself.  However to the increased ease of use of ultrafast femtosecond lasers, non-equilibrium measurements of solids have become of very prominent and really represent a wide open area of inquiry.

The study of materials in a non-equilibrium configuration present numerous challenges.   Obviously if one takes a strongly correlated system (which is hard to understand by itself) and non-equilibrium effects (which are had to understand by themselves) and combines them, simplicity is not the natural outcome.  But there can be reasons for wanting to push physical systems out of equilibrium and study their behavior.  At least three classes of phenomena and motivations for studying a condensed-matter system out of equilibrium come to mind.

1) Experiments are typically performed by pumping the system at one time and probing it at a later stage. After pumping, one hopes to learn something about the relaxation mechanisms and timescales of the equilibrium phase by watching the system relax and return to equilibrium through the decay of its elementary excitations.   There is also the possibility of using multiple pulses of light to induce not just absorption but also emission.   This requires reasonably long lived excitations, but is the foundation for the incredibly powerful technique of 2D spectroscopy that is used in in NMR (to excite nuclei) and in the optical range (to study atomic vibrations) \cite{Cundiff}.   The technique has been developed in the THz range and applied to insulating spin systems \cite{Lu17a}.  I consider this a tremendously promising area.

2) One may drive the system in such a way as to allow access to material configurations (for example, time averaged structure or free charge density) that cannot be accessed in equilibrium. Intense pump pulses may therefore change the free-energy landscape and allow a competing phase to be stabilized in a transient fashion. The hope is that the transient phase reflects equilibrium possibilities in a larger generalized parameter space.

3) One may drive a system to achieve a non-equilibrium phase that cannot exist or is not stabilized without a time - dependent driving field. In this sense, the time- dependent pump should be considered a term in the Hamiltonian.  Work that claims that a light-illumination-driven Floquet topological insulator can be stabilized is of this variety \cite{Cayssol13a,Wang13a,Mahmood16a}.   There has also been a tremendous amount of interest in the possibility of light-driven superconducting phases \cite{Hu14a,Mitrano16a} by the Cavalleri group.

This has become a huge area of inquiry and aside from the above, I won't go into more detail.   Please see classic \cite{Averitt02a} and more recent reviews \cite{Cayssol13a,Mankowsky16a,Orenstein12a,Nicoletti16a} for further details.

\subsection{Topological materials and Berry's phase}

Ordered states of matter have been traditionally categorized by the symmetries they break. Upon the ordering of spins in a ferromagnet or the freezing of a liquid into a solid, the loss of symmetry distinguishes the ordered state from the disordered one.  However, it was discovered that materials can also be distinguished from each other by specific \textit{topological} properties that are encoded in their quantum mechanical wave functions \cite{Armitage,Hasan,Xiao}.  This discovery that topological properties of quantum mechanical wave functions are important to a wide class of quantum materials has heralded a revolution in condensed matter physics. Topological states are now known to play a critical role in a wide range of condensed matter phenomena, from the well known quantum Hall effect, 3D topological insulators (TIs), and Weyl semimetals (WSMs) to the fundamental properties of polarization and magnetization \cite{Xiao}.

In many cases, these topological properties are characterized by ``topological invariants", which are quantum numbers sensitive only to topological properties and insensitive to system specific details. At the level of non-interacting electrons many of these topological properties can be captured through the concept of the Berry phase incurred by an electron as it moves through the Brillouin zone in momentum ($k$) space \cite{Xiao}. The related quantities of the Berry curvature and Berry connection act like a pseudo-magnetic field and pseudo-vector potential, respectively, in $k$-space. For instance, a 2D topological invariant, the Chern number, is an integral of the Berry curvature over a 2D Brillouin zone and accounts for the net Berry ``flux" through the surface \cite{Xiao}. These topological invariants underlie topological states, such as topologically protected surface states in topological insulators and bulk states in topological semimetals. The exotic phenomena exhibited by topological materials have resulted in a host of studies aimed at elucidating their properties over the past decade. Furthermore, new materials are continually being discovered, including type-II Weyl semimetals (which exhibit a tilted Weyl cone in $k$-space) \cite{Armitage}, nodal line semimetals (which have lines or rings in $k$-space over which bands touch), and quadratic band touching systems \cite{Kim,Emmanouilidou,Wang}.   In the vast majority of cases the studied materials can be understood in the context of weakly interacting fermions and do not properly fall under the heading of strongly correlated electrons.   Still, the fields have very much progressed together intellectually and there has been much cross-fertilization so I cover these issues here\footnote{These materials collectively have been referred to as ``Quantum Materials".   As all materials are ``quantum", its not clear what this terminology means.  That has not stopped many (myself included) from using it.  Perhaps the best we can do is fall back on Justice Potter Stewart's famous dictum ``I know it when i see it".}.

\begin{figure}[htp]
\includegraphics[width=0.95\columnwidth]{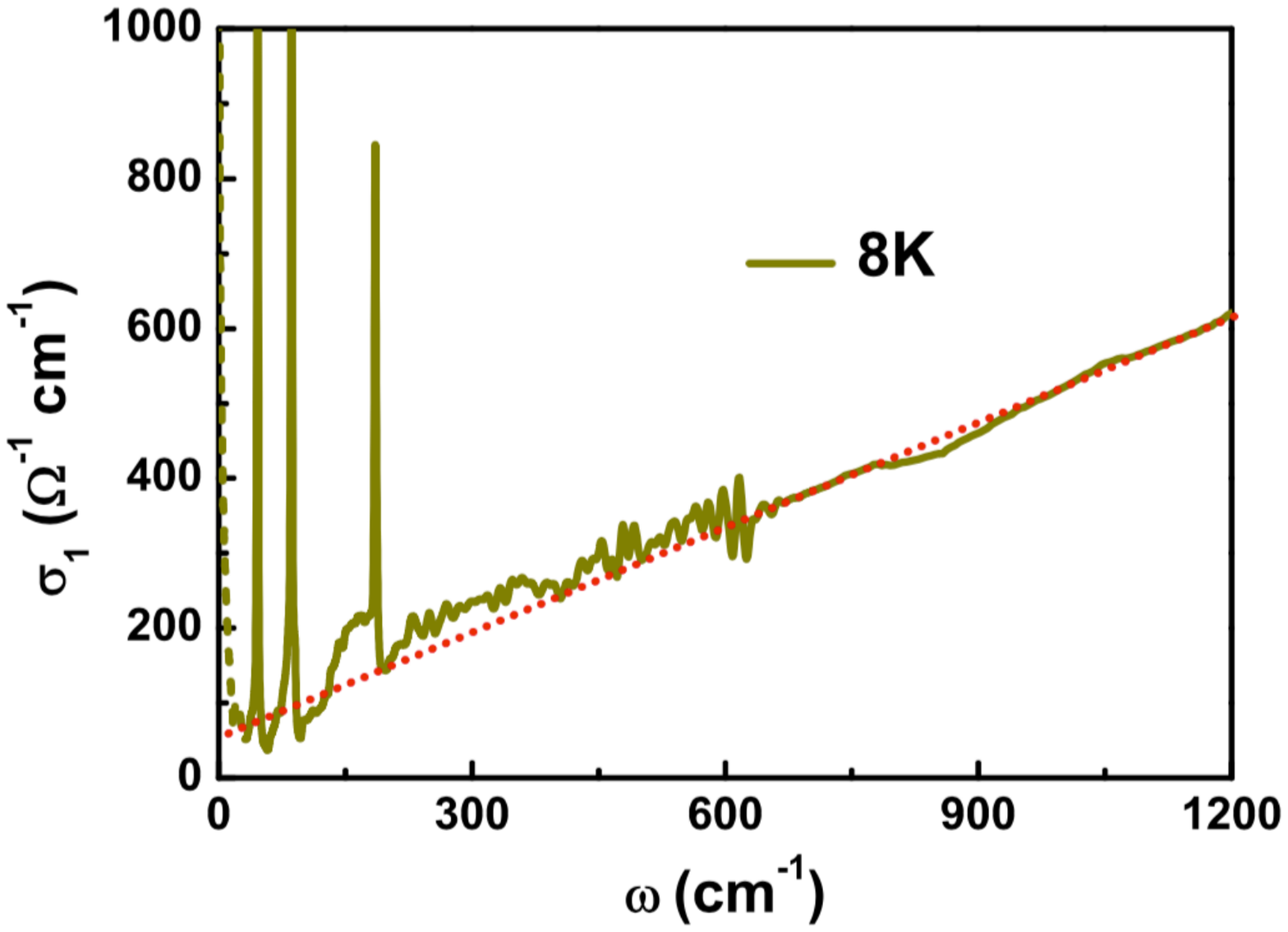}
\includegraphics[width=0.9\columnwidth]{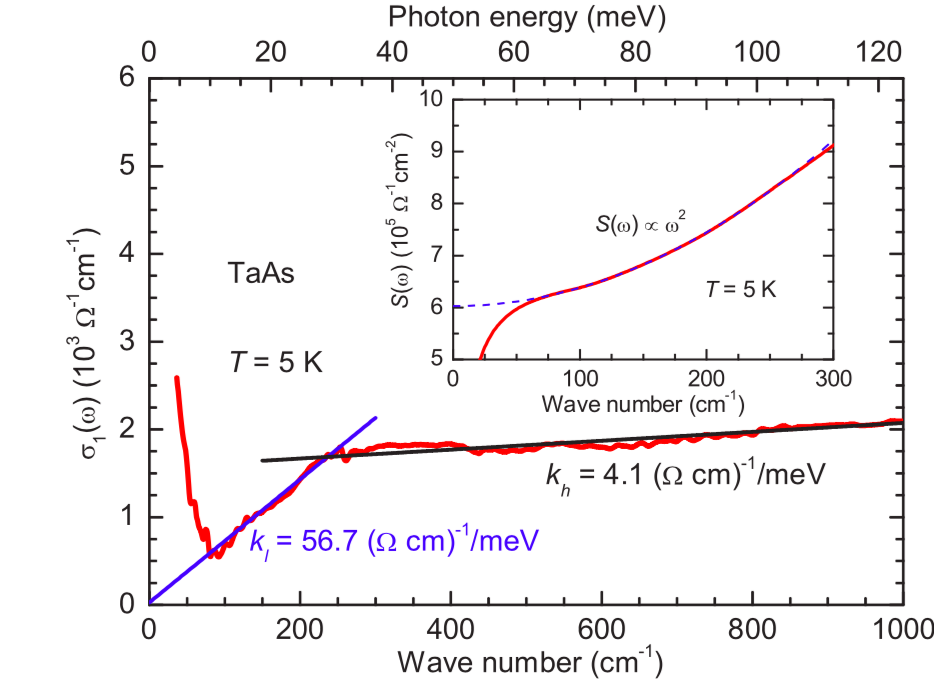}
\caption{(Top) The optical conductivity of ZrTe$_5$ at 8 K at frequencies below 1200 cm$^{-1}$. The red (dotted) line is the linear fitting of $\sigma_1(\omega). $ From Ref. \onlinecite{Chen15a}.  (Bottom) Optical conductivity for TaAs at 5 K. The blue (steep) and black (shallow) solid lines through the data are linear guides to the eye. The blue (steep) line show the Weyl part of the spectrum, while the black (shallow) likely line comes from higher energy non-Weyl states.  The inset shows the spectral weight as a function of frequency at 5 K (red solid curve), which follows an $\omega^2$ behavior (blue dashed line).   From \cite{Xu16a}.}
 \label{ZrTe5Optics}
\end{figure}

In many topological materials -- in at least some regime -- the electron dynamics are expected to be semi-classical.   For instance in topological insulator surface states in weak applied magnetic fields, the motion is captured by the semi-classical expressions for cyclotron resonance \cite{Wu15a}.   See Fig. \ref{TILiangphonons}.  But in other cases, there can be features directly ascribable to novel aspects of the band structure.   For instance, 3D Dirac and Weyl systems are predicted to have a number of interesting semi-classical transport and optical effects that are diagnostic for this state of matter \cite{Hosur12a,Burkov11a}.   With small modifications, most of these results apply equally to Weyl and Dirac systems.  In the absence of impurities and interactions the free fermion result for the conductivity in the low energy limit (where quadratic or higher order terms in the dispersion can be neglected), which arises from interband transitions across the Weyl or Dirac node when the chemical potential (E$_F$) is at the Weyl or Dirac point is

\begin{equation}
\sigma_1(\omega) = N \frac{e^2}{12 h} \frac{|\omega|}{v_F},
\label{LinearConduct}
\end{equation}
where $v_F$ is the Fermi velocity and $N$ is number of nodes \cite{Hosur12a,Burkov11a,Hosur13a,Tabert16b}.  This prediction is closely related to the prediction and observation in 2D Dirac system of single layer graphene that for interband transitions the optical conductance should be  $G_1 (\omega) = \frac{e^2}{\hbar}$, giving a frequency independent transmission that is quantized in terms of the fine structure constant $\alpha$ as $T(\omega) = 1 - \pi \alpha$ \cite{Ando02a,Kuzmenko08a,Nair08a}.  In a 2D material like graphene the Kubo-Greenwood expression for the conductance from interband transitions can be written (for the chemical potential at the Dirac point  and T=0) as $G_1 (\omega) = \frac{\pi e^2}{\omega}  |\textbf{v}(\omega)|^2 D(\omega) $ where $ \textbf{v}(\omega)$ is the velocity matrix element between states with energies   $ \pm \hbar \omega / 2$ and $g(\omega)$ is the 2D joint density of states.  The universal conductance arises because the Fermi velocity factors that come into the matrix element are canceled by their inverse dependence in the density of states.  In these 3D Dirac systems, another factor of $\omega$ comes in in the density of states yielding Eq. \ref{LinearConduct}.   Please see Fig. \ref{ZrTe5Optics} for results showing this linearity for the possible Dirac system ZrTe$_5$ and WSM TaAs \cite{Chen15a,Xu16a}.

Note that Eq. \ref{LinearConduct} implies a logarithmic divergence of the real part of the dielectric constant through Kramers-Kronig considerations \cite{Rosenstein13a,Jenkins16a}.   The corresponding imaginary conductivity is

\begin{equation}
\sigma_2(\omega) = - \frac{2}{\pi}N \frac{e^2}{12 h} \frac{|\omega|}{v_F}  \mathrm{log} \frac{2 \Lambda v_F}{\omega}.
\label{LinearConductImaginary}
\end{equation}
\noindent where $ \Lambda $ is a UV momentum cutoff.  

Quite generally, in noninteracting electron systems consisting of two symmetric bands that touch each other at the Fermi energy the optical conductivity generically has power-law frequency dependence with exponent $(d-2)/z$ where $d$ is the dimensionality of the system and $z$ is the power-law of the band dispersion \cite{Basci13a}.  Such power-law behavior is a consequence of the scale-free nature of such systems.   It has been argued \cite{Fisher91a} that at a conventional continuous transition the optical conductivity should scale as $(d-z-2)/z$.   Due to their scale free nature one may regard Dirac systems as intrinsically quantum critical with a dynamic exponent equal to the band dispersion power-law.   With the usual substitution for the effective dimensionality of a quantum critical system $d_{eff} = d + z$ the generic power-law expression for the Dirac conductivity follows.   Please see Ref. \onlinecite{Armitage} for further details on the optical response of WSMs and DSMs.

\begin{figure}[htb]
\begin{center}
\includegraphics[width=7cm,angle=0]{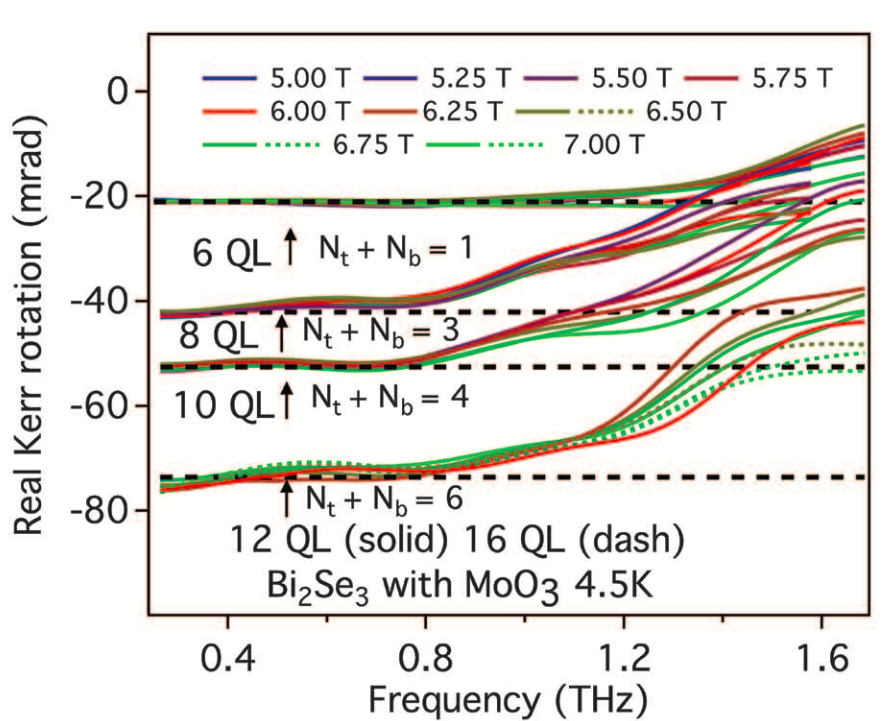}
\caption{Faraday rotation for a number of different Bi$_2$Se$_3$ samples of different thicknesses.   Different thicknesses give different charge densities at the surface. From Ref. \onlinecite{Wu16b}.  Dashed lines represent values for the universal values of the Faraday rotation.  From Ref. \onlinecite{Wu16b}.} \label{Axions}
\end{center}
\end{figure}

Of course, even better than the above sort of experiments would be the ability to identify and measure response functions that directly reflect the topological invariants of the system. There are two known cases where this has been done. The first is the classic observation of the quantized Hall conductance in 2D electron gases under a DC magnetic field, which directly reflects the topological Chern number of the system \cite{vonKlitzing}. The second example is the quantized magnetoelectric response of a topological insulator, measured by my group using TDTS \cite{Wu16b}.  This is a realization of the much celebrated axion electrodynamics and quantized electromagnetic response proposed in the early days in the TI field \cite{Qi,Essin,Maciejko} and represents the second time in the annals of condensed matter physics when a topological invariant has been measured in a solid state system by charge transport.   As shown in Fig. \ref{Axions}, for fields higher than 5 T, we observed a quantized Faraday (and Kerr) rotations, but in a regime where the dc transport is still semiclassical as it gets shorted by the non-chiral side surfaces.  A nontrivial Berry's phase offset to these values gives evidence for axion electrodynamics and the
topological magnetoelectric effect.

In both of the aforementioned examples where a topological invariant has been measured, the response is exhibited in the linear response regime.  One may ask whether other topological states of matter, such as Weyl or Dirac semimetals (DSMs) or nodal line systems, also have quantized observables in response to electromagnetic fields, and/or if specific signatures of Berry phase effects in general can be deduced.  In this regard, it has been pointed out that it may be possible to address these questions by going beyond the linear optical response into a regime where the material response depends nonlinearly on the incident electric field \cite{Morimoto}.  Intense, ultrashort optical pulses (from THz to x-ray frequencies) provide a new way to not only probe the novel phenomena displayed by topological matter, but also to drive these materials into new regimes that were inaccessible via conventional approaches.  For the latter, the advantages of nonlinear optical techniques have been demonstrated in a number of condensed matter systems. For example, as discussed in Sec. \ref{PumpProbe} in high-temperature superconductors, optical SHG has been used to reveal the broken symmetries characterizing the pseudogap phase and intense mid-IR pulses have been used to photoinduce superconductivity.
 
Theoretical studies have shown that the nonlinear optical response of solids is intimately related to their topological properties \cite{Morimoto}. It was demonstrated using the Floquet formalism that various nonlinear optical effects, such as the shift current in non-centrosymmetric materials and the photovoltaic Hall response, can quite generally be described in a unified fashion by quantities involving the Berry connection and Berry curvature.  For instance, it has been shown that the SHG signal can be related to the shift vector $\mathbf{R}$, which in turn can be directly linked to the Berry connection \cite{Morimoto}.  In addition, under a DC magnetic ($\mathbf{B}$) field, SHG is intrinsically sensitive to the chiral anomaly in WSMs,  an effect unique to these materials that transfers particles between Weyl points with opposite chiralities under parallel electric and magnetic fields. Finally, the circular photogalvanic effect (CPGE) in WSMs is also sensitive to the Berry phase as well as the chirality of Weyl fermions \cite{Juan16a}.

One of the few nonlinear optical effects that has been measured in these materials to date is SHG on 3D Dirac materials.   Theoretical studies indicate that SHG can yield signatures of several quantum effects that in turn shed light on the underlying topology. For example, Ref. \onlinecite{Morimoto16b} demonstrates that transitions arising near Weyl nodes between linearly dispersing bands in a WSM should give a nearly universal prediction in the low frequency limit for the non-linear susceptibility ($\chi^{(2)}$)

\begin{equation}
\chi^{(2)} = \frac{g(\omega) <v^2\mathbf{R}>}{2 i \omega^3 \epsilon_0}.
\end{equation}

With the density of states $g(\omega)$ proportional to $\omega^2$, the SHG signal is predicted to diverge as $1/\omega$.     Although this result is reminiscent of the $\omega$ dependence of the optical conductivity discussed above for these systems, the 1/$\omega$ divergence is a unique signatures for inversion-breaking WSMs in particular because it vanishes in DSMs.  However, similar to the case of the linear in $\omega$ conductivity, the SHG divergence will be cutoff by disorder and nonzero Fermi energy in real materials.    Wu et al. \cite{Wu16a} has found a giant, anisotropic $\chi^{(2)} $ at 800 nm (1.55 eV) in TaAs, TaP, and NbAs, which may be related to a high frequency limit of this physics.  In the spectral range measured, the effect is of order 7000 $pm/V$, which is an order magnitude larger than in GaAs, which is the material with the next largest coefficient.  It is important to probe SHG and the shift current at even lower frequency to look for the dependence of $\omega$.    Important efforts in this direction have been made by Patankar et al. \cite{Patankar}, who show that the effect is even bigger at even lower frequency and who also derive a new theorem that relates the spectral weight of the nonlinear conductivity to the third cumulant of the ground state, which is a quantity related to the skewness or intrinsic asymmetry of the ground-state polarization distribution.  Other effects have been proposed such as a nonlinear Hall effect arising from an effective dipole moment of the Berry curvature in momentum space  \cite{Sodemann15a} and a photoinduced anomalous Hall effect\cite{Chan16b} for WSMs and photogalvanic effects \cite{Cortijo16b} in DSMs.

In DSM systems there is no direct photocurrent without driving electric field \cite{Shao15a}, but because of their spin selective transitions the photoconductivity is anisotropic for polarized radiation.  Ref. \onlinecite{Chan16a} proposes that inversion symmetry breaking WSMs with tilted Weyl cones (Type II most effectively) and doped away from the Weyl point, will be efficient generators of photocurrent and can be used as low frequency IR detectors.  Such a photocurrent has  been demonstrated via the circular photogalvanic effect (CPGE)  \cite{Ma17a} in TaAs.  The CPGE is the part of a photocurrent that switches its direction with changes to the handedness of incident circular polarization.   It can be shown \cite{Moore10a} to be sensitive to the anomalous velocity derived by Karplus and Luttinger \cite{Karplus54a} that was later interpreted as a Berry-phase effect \cite{Sundaram99a,Jungwirth02a}.  \onlinecite{Ma17a} also pointed out that such experiments can also measure uniquely the distribution of Weyl fermion chirality in the BZ.

An interesting proposal of Ref. \onlinecite{Juan16a} was that of a quantized response also in the CPGE of inversion symmetry broken WSMs that possess no mirror planes or four-fold improper rotation symmetries (e.g. structurally chiral).  The CPGE usually depends on non-universal material details.  Refs.\cite{Juan16a,Morimoto16b} predict that in Weyl semimetals and three-dimensional Rashba materials without inversion and mirror symmetries, that the trace of the CPGE is quantized (modulo multi-band effects that were argued to be small) in units of the fundamental physical constants.  It is proposed that the currents obey the relation

\begin{equation}
\frac{1}{2}\left[{dj_{\circlearrowright} \over dt}- {dj_{\circlearrowleft} \over dt} \right]= C {2\pi e^3 \over h^2 c \epsilon_0} I 
\label{eqinjection}
\end{equation}
where $C$ is the integer-valued topological charge of Weyl point and $I$ is the applied intensity.    Alternatively, the right hand side of the equation can be expressed as $ C {4\pi \alpha e \over h} I $ where $\alpha$ is the fine-structure constant.  In this expression, the currents for left and right circular polarization are perpendicular to the polarization plane.  Alternatively the quantity $\left[j^{\mathrm{sat}}_{\circlearrowleft}-j^{\mathrm{sat}}_{\circlearrowright}\right] $ may be measured if the relaxation time $\tau$ is sufficiently long and known independently.  An attractive property of this response is that it is related to the chiral charge on a single node.  The total node chirality in the BZ must of course be zero, however this does not prevent a CPGE.  In an $\mathcal{P}$ breaking material with no mirror planes, the Weyl nodes of opposite chirality do not need to be at the same energy.   One node can be Pauli blocked rendering it inert and giving a quantized response for some finite range in frequencies.   The proposed double Weyl system SrSi$_2$ \cite{Huang16b} that has no mirror planes (unlike TaAs) or RhSi \cite{Chang2017}  (which is predicted to have  six-fold-degenerate double spin-1 Weyl nodes and a four-fold-degenerate node) may be good candidates for this effect.  Its predicted magnitude is well within the range of current experiments.  Using the universal coefficient $ \frac{e^3}{\hbar^2 c \epsilon_0}$ = 22.2 $\frac{A}{W \cdot ps},$ \onlinecite{Juan16a} predict for $\tau \sim 1$ ps a steady state photocurrent of $\sim 2 \frac{nA}{W/cm^2}$, which is approximately 100 times that found in the topological insulator films \cite{Okada16a}.

SHG can potentially also give insight into other effects, such as the quantum nonlinear Hall effect, which arises from an effective dipole moment of the Berry curvature in k-space \cite{Jia16a}.   Adding a $\mathbf{B}$  field makes SHG also sensitive to the chiral magnetic effect (directly linked to the chiral anomaly), the magnetochiral effect (a difference in the material response to left and right circularly polarized light (LCP and RCP, respectively), and the Berry curvature \cite{Cortijo16b,Morimoto16b}. The relative strengths of these responses depend on the orientation of $\mathbf{B}$ with respect to the vector $\mathbf{b}$ in momentum space that connects the Weyl nodes, indicating that SHG measurements will also shed light on this important parameter.  Second harmonic generation may be a relatively simple way to gain substantial insight into the intrinsic topological properties of 3D nodal semimetals. 

Even more insight into topological invariants may be given by high-order THz sideband generation (HSG), which has been proposed  as a probe of the Berry phase in materials like MoS$_2$ or bilayer graphene \cite{Fan13a,Yang14a}.  In this scheme a linearly polarized laser is used to excite charge across the bandgap, after which elliptical or circularly polarized radiation accelerates the charge around a closed loop in the Brillouin zone. Upon recombination, the electron-hole pair emits polarized light, the plane of which is predicted to be rotated by exactly the accumulated Berry's phase in the loop \cite{Yang14a}. One may map out the momentum dependence of the Berry phase by using successively higher THz pulse amplitudes.  A related experiment has been performed by Banks et al. \cite{Banks17a} who find that a linear birifrigence in their high-order sideband generation (HSG) in semiconductors is affected by Berry phases.    Using a combination of a near-band gap laser beam, which excites a semiconductor that is being driven simultaneously by sufficiently strong terahertz (THz)-frequency electric fields, the highest-order sidebands are associated with electron-hole pairs being driven coherently across roughly 10$\%$ of the Brillouin zone around the $ \Gamma$ point.    The THz pulse adiabatically drives an electron coherently across a large portion of the Brillouin zone and picks up a Berry's phase before recombination.

\section{Acknowledgments}

I would like to thank the organizers of the 2008 Boulder School for Condensed Matter Physics for the opportunity to first talk about these topics.   These lecture notes were then updated for the 2018 QS3 Summer School at Cornell University and the 2018 Princeton Summer School on Condensed Matter Physics.   I'd also like to thank Luke Bilbro, Vladimir Cvetkovic, Natalia Drichko, Amit Keren, and Wei Liu for helpful suggestions and careful reading of the original version of these notes.

\end{document}